\documentclass[12pt,aps,preprintnumbers,superscriptaddress,nofootinbib,prd,onecolumn,floatfix]{revtex4-2}
\pdfoutput=1

\usepackage{multirow,array}
\usepackage{amsmath,amssymb}
\usepackage{graphicx}
\usepackage{color}
\usepackage{url}
\usepackage{slashed}
\usepackage{footmisc}
\usepackage{hyperref}
\usepackage{braket}
\usepackage{colortbl}
\usepackage{orcidlink}
\usepackage{physics}
\usepackage{mathrsfs}
\usepackage{feynmp-auto,expdlist}
\usepackage[compat=1.1.0]{tikz-feynman}
\usepackage[normalem]{ulem}

\usepackage[T1]{fontenc}
\newcommand{\bea}{\begin{eqnarray}}
\newcommand{\eea}{\end{eqnarray}}
\newcommand{\be}{\begin{equation}}
\newcommand{\ee}{\end{equation}}
\newcommand{\bma}{\begin{pmatrix}}
\newcommand{\ema}{\end{pmatrix}}

\definecolor{nicered}{rgb}{0.5,0.,0.}
\definecolor{nicegreen}{rgb}{0.,0.5,0.}
\definecolor{niceblue}{rgb}{0.,0.,0.5}
\hypersetup{colorlinks,citecolor=nicegreen,linkcolor=nicered,urlcolor=niceblue}

\begin{document}

\title{Probing \boldmath{$\tau$} lepton dipole moments at multi-TeV muon colliders}

\author{Dario Buttazzo\orcidlink{0000-0002-4678-7975}}
\affiliation{Istituto Nazionale di Fisica Nucleare, Sezione di Pisa, I-56127 Pisa, Italy}
\author{Gabriele Levati\orcidlink{0009-0000-7693-2152}}
\affiliation{Albert Einstein Center for Fundamental Physics, Institute for
Theoretical Physics, University of Bern, Sidlerstrasse 5, 3012 Bern,
Switzerland}
\author{Yang Ma\orcidlink{0000-0002-9419-6598}}
\affiliation{Centre for Cosmology, Particle Physics and Phenomenology (CP3),
UCLouvain, 1348 Louvain-la-Neuve, Belgium}
\author{Fabio Maltoni\orcidlink{0000-0003-4890-0676}}
\affiliation{Centre for Cosmology, Particle Physics and Phenomenology (CP3),
UCLouvain, 1348 Louvain-la-Neuve, Belgium}
\affiliation{Dipartimento di Fisica e Astronomia, Università di Bologna and INFN,
Sezione di Bologna, via Irnerio 46, 40126 Bologna, Italy}
\author{Paride Paradisi\orcidlink{0000-0003-0556-7797}}
\affiliation{Dipartimento di Fisica e Astronomia `G. Galilei', Universit\`a di Padova, Italy}
\affiliation{Istituto Nazionale di Fisica Nucleare, Sezione di Padova, I--35131 Padova, Italy}
\author{ZeQiang Wang\orcidlink{0000-0002-1643-2727}}
\affiliation{Centre for Cosmology, Particle Physics and Phenomenology (CP3),
UCLouvain, 1348 Louvain-la-Neuve, Belgium}

\preprint{IRMP-CP3-26-09,COMETA-2026-42}

\begin{abstract}
\vspace*{1.2em}
We investigate the sensitivity of future multi-TeV muon colliders to the electromagnetic and weak dipole moments of the $\tau$ lepton within the dimension-six Standard Model Effective Field Theory. We consider $\tau^+\tau^-$ production through $\mu^+\mu^-$ annihilation and vector-boson scattering, electroweak-boson-associated $\tau$ production, and the radiative Higgs decay $H\to\tau^+\tau^-\gamma$. At high energies, the dipole-induced contributions quadratic in the Wilson coefficients become particularly important, with associated Higgs production, $\mu^+\mu^-\to\tau^+\tau^-H$, providing the strongest sensitivity through energy-enhanced contact interactions. The remaining channels offer complementary probes of the underlying dipole operators. For a $10$~TeV muon collider with an integrated luminosity of $10\,\mathrm{ab}^{-1}$, we obtain combined sensitivities at the nominal $95\%$ confidence level of $|\Delta a_\tau|<1.2\times10^{-5}$, $|d_\tau|<6.6\times10^{-20}\,e\,\mathrm{cm}$, $|\Delta a_\tau^Z|<1.8\times10^{-5}$, and $|d_\tau^Z|<1.0\times10^{-19}\,e\,\mathrm{cm}$. Compared with FCC-ee projections, the muon collider offers substantially stronger sensitivity to the electric dipole moments, while both facilities offer complementary probes of magnetic dipole moments. These results motivate further studies incorporating realistic detector effects and systematic uncertainties.
\end{abstract}

\maketitle
\clearpage
\tableofcontents
\section{Introduction}
\label{sec:int}

Lepton dipole moments provide some of the most stringent precision tests of the Standard Model (SM) and constitute powerful indirect probes of physics beyond the SM (BSM). The anomalous magnetic moments $a_\ell$ and electric dipole moments (EDMs) $d_\ell$ of the electron and muon have been measured with extraordinary precision, reaching $0.13$ parts per trillion for $a_e$~\cite{Fan:2022eto} and $124$ parts per billion for $a_\mu$~\cite{Muong-2:2023cdq,Muong-2:2025xyk}. The corresponding EDM searches set stringent bounds, $|d_e|<4.1\times10^{-30}\,e\,{\rm cm}$~\cite{Roussy:2022cmp} and $|d_\mu|<2\times10^{-19}\,e\,{\rm cm}$~\cite{Muong-2:2008ebm}. In the SM, charged-lepton EDMs are suppressed far below present experimental sensitivities; their observation would therefore provide evidence for new sources of CP violation. For anomalous magnetic moments, by contrast, the interpretation of a possible deviation depends critically on the precision of the SM prediction. In particular, recent lattice-QCD results have substantially changed the status of the muon anomaly: the latest SM evaluation, $a_\mu^{\rm SM}=116\,592\,033(62)\times10^{-11}$~\cite{Aliberti:2025beg,Aoyama:2012wk,Volkov:2019phy,Volkov:2024yzc,Aoyama:2024aly,Parker:2018vye,Morel:2020dww,Fan:2022eto,Czarnecki:2002nt,Gnendiger:2013pva,Ludtke:2024ase,Hoferichter:2025yih,RBC:2018dos,Giusti:2019xct,Borsanyi:2020mff,Lehner:2020crt,Wang:2022lkq,Aubin:2022hgm,Ce:2022kxy,ExtendedTwistedMass:2022jpw,RBC:2023pvn,Kuberski:2024bcj,Boccaletti:2024guq,Spiegel:2024dec,RBC:2024fic,Djukanovic:2024cmq,ExtendedTwistedMass:2024nyi,MILC:2024ryz,FermilabLatticeHPQCD:2024ppc,Keshavarzi:2019abf,DiLuzio:2024sps,Kurz:2014wya,Colangelo:2015ama,Masjuan:2017tvw,Colangelo:2017qdm,Colangelo:2017fiz,Hoferichter:2018dmo,Hoferichter:2018kwz,Eichmann:2019tjk,Bijnens:2019ghy,Leutgeb:2019gbz,Cappiello:2019hwh,Masjuan:2020jsf,Bijnens:2020xnl,Bijnens:2021jqo,Danilkin:2021icn,Stamen:2022uqh,Leutgeb:2022lqw,Hoferichter:2023tgp,Hoferichter:2024fsj,Estrada:2024cfy,Deineka:2024mzt,Eichmann:2024glq,Bijnens:2024jgh,Hoferichter:2024vbu,Hoferichter:2024bae,Holz:2024lom,Holz:2024diw,Cappiello:2025fyf,Colangelo:2014qya,Blum:2019ugy,Chao:2021tvp,Chao:2022xzg,Blum:2023vlm,Fodor:2024jyn}, 
is consistent with the experimental world average, $a_\mu^{\rm exp}=116\,592\,071.5(14.5)\times10^{-11}$~\cite{Aliberti:2025beg}.

The $\tau$ lepton presents a qualitatively different opportunity. Owing to its short lifetime, its dipole moments cannot be measured using the spin-precession techniques employed for the electron and muon, but must instead be inferred from production and decay processes involving $\tau$ leptons. 
At the same time, the larger mass of the $\tau$ relative to the electron and muon can enhance the size of BSM contributions to its dipole moments. For instance, naive dimensional analysis suggests
the scaling $a_\ell\propto m_\ell^2/\Lambda^2$, although the precise mass dependence is model dependent and can be altered by additional sources of chiral-symmetry breaking or interactions
preferentially coupled to third-generation
fermions~\cite{Giudice:2012ms,Crivellin:2021spu,Athron:2025ets}.
The SM prediction for the $\tau$ anomalous magnetic moment, $a_\tau^{\rm SM}=117\,720(4)\times10^{-8}$, is known much more precisely than can currently be tested experimentally~\cite{Eidelman:2007sb,Eidelman:2016aih,DiLuzio:2024sps,Hoferichter:2025fea,Wittig:2025azm}. Consequently, improved measurements of the $\tau$ magnetic and electric dipole moments can provide particularly clean tests of new physics.

The most stringent existing collider constraints on the $\tau$ anomalous magnetic moment come from measurements of photon-induced $\tau^+\tau^-$ production at the LHC. CMS finds $-0.0042<a_\tau<0.0062$ at $95\%$ confidence level (CL)~\cite{CMS:2024qjo}, while ATLAS obtains $a_\tau\in[-2.4,\,4.7]\times10^{-3}$ at the same CL~\cite{ATLAS:2025oiy}. For the electric dipole moment, Belle reports $d_\tau\in[-1.85,\,0.61]\times10^{-17}\,e\,{\rm cm}$ at $95\%$  CL~\cite{Belle:2021ybo}. Future measurements at Belle II and high-luminosity lepton colliders could improve these bounds substantially~\cite{Crivellin:2021spu,USBelleIIGroup:2022qro,Aihara:2024zds,Gogniat:2025eom,Hoferichter:2025zjp,Hoferichter:2025ijh}. In particular, the projected reach of FCC-ee for electromagnetic and weak $\tau$ dipole moments benefits from its large integrated luminosity, the possibility of scanning the $Z$ resonance, and the complementary sensitivity of several production and decay channels~\cite{Buttazzo:2026amk,Buttazzo:2026ulg}.

A natural question is whether future lepton colliders
operating at substantially higher energies can
complement the precision program at FCC-ee by
exploiting the energy dependence of the dipole
interactions. 
Multi-TeV muon colliders offer a particularly interesting possibility in this respect~\cite{Delahaye:2019omf,Long:2020wfp,Bartosik:2020xwr,Schulte:2021hgo,AlAli:2021let,Black:2022cth,MuonCollider:2022xlm,Aime:2022flm,MuonCollider:2022nsa,MuonCollider:2022ded,MuonCollider:2022glg,Accettura:2023ked,MuCoL:2024oxj,InternationalMuonCollider:2024jyv,InternationalMuonCollider:2025sys,MuCoL:2025quu}. At such energies, the effects of higher-dimensional interactions can grow relative to the SM contributions~\cite{Franceschini:2017xkh,Buttazzo:2020uzc}, opening up new opportunities to probe the $\tau$ dipole operators. In particular, while the interference of dipole amplitudes with the SM is often suppressed by the helicity flip associated with these operators, their quadratic contributions can become increasingly important. The resulting sensitivity is strongly process dependent: direct $\mu^+\mu^-\to\tau^+\tau^-$ annihilation, vector-boson scattering, and electroweak-boson-associated production probe different kinematic regions and combinations of dipole interactions.

Among these channels, $\tau$ production in association
with a Higgs or electroweak gauge boson provides
a distinctive high-energy probe.
Electroweak gauge invariance relates the anomalous $\tau^+\tau^-\gamma$ and $\tau^+\tau^-Z$ vertices to interactions involving the physical Higgs boson. As a result, processes such as $\mu^+\mu^-\to\tau^+\tau^-H$ receive contributions from dipole-induced contact interactions, which can be particularly effective at multi-TeV energies. The related channels $\mu^+\mu^-\to\tau^+\tau^-Z$ and $\mu^+\mu^-\to W^\pm\tau^\mp\nu_\tau$ provide complementary information, while the radiative Higgs decay $H\to\tau^+\tau^-\gamma$ probes the same underlying operators through a different interference pattern. Understanding the relative reach of these measurements, and their complementarity with FCC-ee, is a central goal of the present study.

In this work, we investigate the sensitivity of future multi-TeV muon colliders to the electromagnetic and weak dipole moments of the $\tau$ lepton within the dimension-six Standard Model Effective Field Theory (SMEFT). 
We study direct $\tau$-pair production, the inclusive $\tau^+\tau^-$ final states arising from vector-boson-scattering and related topologies, electroweak-boson-associated $\tau$ production, and the radiative Higgs decay $H\to\tau^+\tau^-\gamma$. We compare the sensitivity of the individual channels and their combination across several collider energies, and confront the resulting projections with those obtained for FCC-ee~\cite{Buttazzo:2026ulg}. Our analysis provides the technical details and additional results complementing Ref.~\cite{Buttazzo:2026amk}. The projected bounds are intended to characterize the statistical reach of these channels under specified selection and efficiency assumptions; a complete experimental assessment will require dedicated studies of detector response, beam-induced backgrounds, and systematic effects.

The paper is organized as follows. Section~\ref{sec:current_status} reviews the present theoretical and experimental status of the $\tau$ dipole moments. Section~\ref{sec:Theory} introduces their SMEFT description, including the matching to the physical dipole moments and the relevant consistency conditions. In Sec.~\ref{sec:pheno}, we describe the analysis framework and study the different production and decay channels before presenting their combined sensitivity and the comparison with FCC-ee. Finally, Sec.~\ref{sec:summary} summarizes our findings.

\section{Current status of the tau lepton dipole moments}
\label{sec:current_status}

The short lifetime of the $\tau$ prevents the use of the precision spin-precession techniques available for the electron and muon. Its dipole interactions must instead be constrained through kinematics and, when accessible, spin correlations of processes in which $\tau$ leptons are produced or decay. It is therefore useful to specify the dipole observables entering the interpretation of collider measurements, and to distinguish their electromagnetic and weak components.

For an on-shell charged fermion $f$ interacting with a photon of momentum $q=p'\!-\!p$, the electromagnetic vertex between external spinors can be expressed in terms of form factors as
\begin{align}
\Gamma^\mu_\gamma(q) = -ieQ_f\bigg[ &F_1(q^2)\gamma^\mu
 +\frac{i\sigma^{\mu\nu}q_\nu}{2m_f}F_2(q^2)
 +\frac{\sigma^{\mu\nu}q_\nu}{2m_f}\gamma_5 F_3(q^2)
 \nonumber\\[-1mm]
 &+\frac{q^2\gamma^\mu-q^\mu\slashed q}{m_f^2}\gamma_5F_A(q^2)\bigg],
 \label{eq:muc_em_form_factors}
\end{align}
where $Q_f$ is the electric charge in units of $e>0$, $\sigma^{\mu\nu}=i[\gamma^\mu,\gamma^\nu]/2$, and $F_A$ denotes the anapole form factor. In the conventions of Eq.~\eqref{eq:muc_em_form_factors}, the static magnetic and electric dipole moments are obtained from
\begin{equation}
 a_f\equiv\frac{g_f-2}{2}=F_2(0),\qquad
 d_f=\frac{eQ_f}{2m_f}F_3(0).
 \label{eq:muc_em_dipoles}
\end{equation}
The form factor $F_1$ fixes the electric charge, whereas the anapole term does not define a separately gauge-independent static observable for an elementary charged fermion. The magnetic and electric dipole structures differ in their transformation under CP: an EDM requires CP violation, while an anomalous magnetic moment does not.

An analogous decomposition applies to the $Zf\bar f$ vertex. In the normalization adopted here, its dipole part is
\begin{equation}
 \left.\Gamma_Z^\mu(q)\right|_{\rm dipole}
 =-ieQ_f\frac{\sigma^{\mu\nu}q_\nu}{2m_f}
 \left[iF_2^Z(q^2)+F_3^Z(q^2)\gamma_5\right],
 \label{eq:muc_z_form_factors}
\end{equation}
with the non-dipole vector, axial-vector and longitudinal terms omitted. The weak magnetic and electric dipole moments are conventionally specified at the $Z$ pole,
\begin{equation}
 a_f^Z=F_2^Z(M_Z^2),\qquad
 d_f^Z=\frac{eQ_f}{2m_f}F_3^Z(M_Z^2).
 \label{eq:muc_z_dipoles}
\end{equation}
Unlike the electromagnetic static moments, these quantities refer to a time-like momentum transfer. The weak magnetic form factor can consequently have an absorptive component even in the SM. The relations between these form factors and the gauge-invariant dimension-six operators considered in this work will be given in Sec.~\ref{sec:Theory}.

We separate possible new-physics contributions to the magnetic moments according to
\begin{equation}
 a_\tau=a_\tau^{\rm SM}+\Delta a_\tau,\qquad
 a_\tau^Z=a_\tau^{Z,\rm SM}+\Delta a_\tau^Z.
 \label{eq:muc_magnetic_split}
\end{equation}
The corresponding SM predictions are
\begin{equation}
 a_\tau^{\rm SM}=117\,720(4)\times10^{-8},\qquad
 a_\tau^{Z,\rm SM}=-(2.10+0.61\,i)\times10^{-6},
 \label{eq:muc_sm_predictions}
\end{equation}
where the quoted electromagnetic result follows Refs.~\cite{Eidelman:2007sb,Eidelman:2016aih,DiLuzio:2024sps,Hoferichter:2025fea,Wittig:2025azm}, while the weak result is taken from Refs.~\cite{Bernabeu:1994wh,Kosnik:2026cfq}. The imaginary part of $a_\tau^{Z,\rm SM}$ originates from absorptive electroweak contributions. The SM electric and weak electric dipole moments are much smaller than current experimental sensitivities~\cite{Hollik:1998wk,Yamaguchi:2020eub,Yamaguchi:2020dsy}; we therefore use $d_\tau$ and $d_\tau^Z$ to denote their new-physics contributions in the remainder of the paper.

Experimentally, bounds on $\tau$ dipole interactions have been obtained from several complementary environments. At LEP, radiative $Z$ decays and photon-induced production constrained the anomalous electromagnetic couplings~\cite{L3:1998gov,OPAL:1998dsa,DELPHI:2003nah}. Photon--photon collisions have also become a sensitive probe at the LHC, first in ultra-peripheral heavy-ion collisions~\cite{ATLAS:2022ryk,CMS:2022arf} and subsequently in proton--proton collisions~\cite{CMS:2024qjo,ATLAS:2025oiy}. Belle has independently constrained the electric dipole interaction using CP-sensitive information in $e^+e^-\to\tau^+\tau^-$~\cite{Belle:2021ybo}. Representative $95\%$ confidence-level intervals are collected in Table~\ref{tab:muc_status_limits}.

\begin{table}[t]
\centering
\renewcommand{\arraystretch}{1.17}
\begin{tabular}{lccc}
\hline\hline
 Measurement & Initial state & $a_\tau\;[10^{-3}]$ & $d_\tau\;[10^{-17}\,e\,{\rm cm}]$ \\
\hline
 DELPHI~\cite{DELPHI:2003nah} & $e^+e^-$ & $[-52,13]$ & $[-37,37]$ \\
 Belle~\cite{Belle:2021ybo} & $e^+e^-$ & --- & $[-1.85,0.61]$ \\
 ATLAS~\cite{ATLAS:2022ryk} & Pb+Pb & $[-57,24]$ & --- \\
 CMS~\cite{CMS:2024qjo} & $pp$ & $[-4.2,6.2]$ & $[-2.9,2.9]$ \\
 ATLAS~\cite{ATLAS:2025oiy} & $pp$ & $[-2.4,4.7]$ & $[-2.0,2.0]$ \\
\hline\hline
\end{tabular}
\caption{Selected experimental limits on the electromagnetic $\tau$ dipole moments at $95\%$ C.L., expressed in the conventions used here. The Belle entry refers to the constraint on $\operatorname{Re}(d_\tau)$; the corresponding interval on $\operatorname{Im}(d_\tau)$ is $[-1.03,0.23]\times10^{-17}\,e\,{\rm cm}$~\cite{Belle:2021ybo}. The table is not intended as an exhaustive survey of past measurements.}
\label{tab:muc_status_limits}
\end{table}

These measurements leave considerable room for improvement. The bounds on the magnetic moment still allow deviations much larger than the uncertainty of the SM prediction, and current EDM searches probe scales far above the tiny SM expectation. Moreover, experimental information on the weak dipole interactions is less direct and generally depends more strongly on the assumptions used in its extraction. This motivates treating the electromagnetic and weak couplings simultaneously rather than interpreting each measured process in isolation.

The high-statistics $e^+e^-$ program, including Belle II and FCC-ee, offers important opportunities to improve this situation~\cite{Crivellin:2021spu,USBelleIIGroup:2022qro,Aihara:2024zds,Gogniat:2025eom,Hoferichter:2025zjp,Hoferichter:2025ijh, Gogniat:2026zvf, Buttazzo:2026amk, Buttazzo:2026ulg}. The multi-TeV muon-collider program explored below addresses a complementary regime: instead of relying primarily on very large event samples or the $Z$ resonance lineshape scan, it can exploit the energy dependence of dipole-induced amplitudes and access production channels with additional electroweak bosons. The effective-field-theory framework needed for this comparison is introduced next.

\section{Tau dipole moments in the SMEFT}
\label{sec:Theory}

The electromagnetic and weak dipole interactions discussed in Sec.~\ref{sec:current_status} are not independent once the electroweak gauge symmetry is imposed. In the Standard Model Effective Field Theory (SMEFT), their leading contributions from heavy new physics are generated by two operators of dimension six, which couple the left- and right-handed charged leptons to the electroweak field strengths~\cite{Allwicher:2021jkr,Bresciani:2026acy}. Restricting the discussion to the dipole sector, we write
\begin{equation}
\label{eq:SMEFT}
\mathscr{L}\supset
\frac{\mathcal{C}_{\ell B}}{\Lambda^2}\,\bar\ell_L\sigma^{\mu\nu}e_R\varphi B_{\mu\nu}
+\frac{\mathcal{C}_{\ell W}}{\Lambda^2}\,\bar\ell_L\sigma^{\mu\nu}\tau_I e_R\varphi W^I_{\mu\nu}
+\mathrm{h.c.}\,,
\end{equation}
where $\Lambda$ denotes the scale suppressing the effective interactions. Here $\ell_L$ and $e_R$ are the lepton $SU(2)_L$ doublet and charged-lepton singlet, respectively, $\varphi$ is the Higgs doublet, and $B_{\mu\nu}$ and $W^I_{\mu\nu}$ are the $U(1)_Y$ and $SU(2)_L$ field-strength tensors. We use $\sigma^{\mu\nu}=i[\gamma^\mu,\gamma^\nu]/2$, with $\tau_I$ denoting the Pauli matrices. The coefficients $\mathcal C_{\ell B}$ and $\mathcal C_{\ell W}$ may be complex; their real and imaginary components lead to CP-even and CP-odd dipole interactions, respectively.

\begin{figure}[t]
\centering
\includegraphics[width=0.22\textwidth]{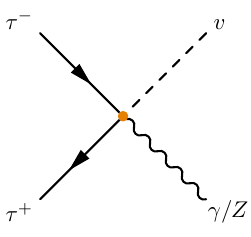}\hspace{0.045\textwidth}
\includegraphics[width=0.22\textwidth]{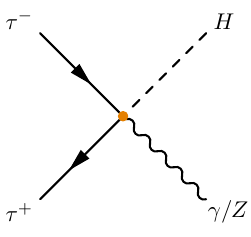}
\caption{Dipole-induced interactions after electroweak symmetry breaking. The left diagram shows the $\tau^+\tau^-V$ vertex proportional to the Higgs vacuum expectation value, while the right diagram contains the corresponding $\tau^+\tau^-HV$ contact interaction with a physical Higgs boson ($V=\gamma,Z$). The highlighted vertices represent insertions of the dimension-six dipole operators.}
\label{fig:FD_tthv}
\end{figure}

After electroweak symmetry breaking, $\varphi$ develops the vacuum expectation value $v=246\,\mathrm{GeV}$. In terms of the photon and $Z$ field strengths, the neutral dipole interactions take the form
\begin{equation}
\label{eq:SMEFT_az}
\mathscr{L}\supset \frac{v+H}{\sqrt{2}\Lambda^2}
\left(
\mathcal C_{\ell\gamma}\,\bar\ell\sigma^{\mu\nu}P_R\ell\,F_{\mu\nu}
+\mathcal C_{\ell Z}\,\bar\ell\sigma^{\mu\nu}P_R\ell\,Z_{\mu\nu}
\right)+\mathrm{h.c.}\,,
\end{equation}
with $P_R=(1+\gamma_5)/2$. The two coefficients appearing in Eq.~\eqref{eq:SMEFT_az} are linear combinations of the gauge-basis coefficients,
\begin{equation}
\label{eq:CgammaCz}
\begin{pmatrix}\mathcal C_{\ell\gamma}\\[1mm]\mathcal C_{\ell Z}\end{pmatrix}
=
\begin{pmatrix}c_W&-s_W\\-s_W&-c_W\end{pmatrix}
\begin{pmatrix}\mathcal C_{\ell B}\\[1mm]\mathcal C_{\ell W}\end{pmatrix},
\end{equation}
where $s_W$ and $c_W$ are the sine and cosine of the weak mixing angle. Thus, electromagnetic and weak dipoles probe different combinations of the same gauge-invariant interactions. More importantly for a high-energy muon collider, the common factor $v+H$ in Eq.~\eqref{eq:SMEFT_az} connects the ordinary $\tau^+\tau^-\gamma$ and $\tau^+\tau^-Z$ dipole vertices to contact interactions involving an additional physical Higgs boson. These vertices are illustrated in Fig.~\ref{fig:FD_tthv}. The Higgs-associated 
$\tau$ pair production process studied below therefore test the same operators as the dipole moments, through a different production topology.

Specializing to the $\tau$ lepton and matching onto the form factors defined in Sec.~\ref{sec:current_status}, the new-physics contributions to the four dipole moments are
\begin{equation}
\label{eq:adtau_final}
\begin{aligned}
\Delta a_\tau&=\frac{2\sqrt{2}\,m_\tau v}{e\Lambda^2}\,\mathrm{Re}\,\mathcal C_{\tau\gamma},
&\qquad d_\tau&=\frac{\sqrt{2}\,v}{\Lambda^2}\,\mathrm{Im}\,\mathcal C_{\tau\gamma},\\
\Delta a_\tau^Z&=\frac{2\sqrt{2}\,m_\tau v}{e\Lambda^2}\,\mathrm{Re}\,\mathcal C_{\tau Z},
&\qquad d_\tau^Z&=\frac{\sqrt{2}\,v}{\Lambda^2}\,\mathrm{Im}\,\mathcal C_{\tau Z}.
\end{aligned}
\end{equation}
The matching in Eq.~\eqref{eq:adtau_final} is at tree level and follows the conventions of the dipole form factors adopted in this work. It identifies the physical moments with the low-scale values of the relevant Wilson coefficients: $\mathcal C_{\tau\gamma}$ at $m_\tau$ and $\mathcal C_{\tau Z}$ at $M_Z$. 

At this order the induced dipole form factors are independent of the momentum transfer; corrections to this simple correspondence can arise from RG evolution and from operators beyond the dimension-six truncation. The numerical sensitivities presented in this work do not include RG evolution of the dipole coefficients. Using the leading-logarithmic expressions derived in Appendix~\ref{app:RGE}, we estimate that running from a reference scale $\Lambda=100$~TeV down to $m_\tau$ changes the electromagnetic dipole coefficient by about $15\%$, assuming a vanishing weak dipole coefficient at the high scale. This suggests that the neglected RG effects are moderate within the dipole-only approximation and do not alter the qualitative conclusions of our sensitivity study. A more general SMEFT analysis would also require RG mixing with other
dimension-six operators, such as semileptonic tensor operators.

A second issue is the range of validity of the effective description at high energies. The energy growth of amplitudes generated by the dipole operators is constrained by perturbative unitarity. In the normalization of Eq.~\eqref{eq:SMEFT}, the corresponding conditions are~\cite{Allwicher:2021jkr,Bresciani:2025toe,Bresciani:2026acy}
\begin{equation}
\label{eq:perturbative}
s\,\frac{|\mathcal C_{\tau B}|}{\Lambda^2}<4\pi,
\qquad
s\,\frac{|\mathcal C_{\tau W}|}{\Lambda^2}<4\pi\sqrt{\frac{2}{3}}\,.
\end{equation}
For fixed collision energy, these inequalities bound the allowed combinations of dipole coefficients in Eq.~\eqref{eq:adtau_final}. In particular, the corresponding bounds scale as $1/s$. 
It tests perturbative unitarity of the specified dipole amplitudes, but does not by itself guarantee convergence of the EFT expansion or bound Wilson coefficients of operators omitted from the benchmark. A broader SMEFT or ultraviolet interpretation also requires assumptions about the heavy-particle thresholds and relative sizes of higher-dimensional interactions.

The SMEFT interactions are implemented in \textsc{LEPTONdipole}~\cite{Ma:2026LEPTONdipole} using the Universal \textsc{FeynRules} Output (UFO)~\cite{Degrande:2011ua} model , which has been validated against \textsc{SMEFTsim}~\cite{Brivio:2020onw}. Numerical simulations are performed using \textsc{MadGraph5\_aMC@NLO}~\cite{Alwall:2014hca,Frederix:2018nkq} and cross-checked with \textsc{Whizard}~\cite{Kilian:2007gr}.

\section{Probing tau dipole moments at multi-TeV muon colliders}
\label{sec:pheno}

We now turn to the sensitivity of muon colliders to the $\tau$ dipole interactions. The channels considered below range from direct $\tau$-pair production to electroweak-boson-associated final states and radiative Higgs decays. Before discussing their different amplitudes and kinematic features, we specify the collider benchmarks and the common assumptions used to estimate their statistical reach.

\subsection{Analysis framework}
\label{sec:analysis_framework}

\paragraph*{Collider benchmarks.}
We consider four center-of-mass energies, $\sqrt{s}=3$, $6$, $10$, and $14$~TeV. For each energy we assume the integrated luminosity~\cite{Delahaye:2019omf,Accettura:2023ked}
\begin{equation}
\label{eq:muC_luminosity}
\mathcal L=10\left(\frac{\sqrt{s}}{10~\mathrm{TeV}}\right)^2\!\mathrm{ab}^{-1},
\end{equation}
corresponding to the benchmark values listed in Table~\ref{tab:muC_benchmarks}. This prescription, which scales the luminosity linearly with $s$, is used consistently when comparing the projected sensitivities at different energies; it is an input assumption rather than an experimentally established performance figure.
\begin{table}[htb]
\centering
\renewcommand{\arraystretch}{1.16}
\setlength{\tabcolsep}{13pt}
\begin{tabular}{c|cccc}
\hline
$\sqrt{s}$ [TeV] & 3 & 6 & 10 & 14 \\
\hline
$\mathcal L$ [ab$^{-1}$] & 0.9 & 3.6 & 10 & 19.6 \\
\hline
\end{tabular}
\caption{Muon-collider energies and integrated luminosities assumed in the numerical analysis, following Eq.~\eqref{eq:muC_luminosity}.}
\label{tab:muC_benchmarks}
\end{table}

\paragraph*{Event selection and reconstruction.}
A detector at a muon collider must cope with beam-induced background (BIB) from muon decays and secondary interactions in the machine. The shielding nozzles envisaged in detector studies reduce the forward angular acceptance. We model this limitation by requiring $10^\circ<\theta(\tau^\pm)<170^\circ$ for the reconstructed $\tau$ leptons. Visible final-state objects are subject to the isolation condition $\Delta R_{ij}>0.4$, with $\Delta R_{ij}=\sqrt{(\Delta\eta_{ij})^2+(\Delta\phi_{ij})^2}$. Unless stated otherwise, we impose $p_T>\sqrt{s}/10$ on the final-state objects to suppress soft contributions. Channel-specific selections, including invariant-mass requirements and boson-tagging assumptions, are given alongside the corresponding analyses.
In particular, the angular acceptance cut
$10^\circ<\theta<170^\circ$ is applied to 
all final-state particles, except for muons 
and neutrinos in VBS processes and for the 
final-state particles in VBF Higgs production.

The numerical study is performed at the $\tau$ production level: $\tau$ decays are not explicitly reconstructed. Instead, we assign a constant efficiency $\epsilon_\tau=0.8$ to each identified $\tau$. Thus the baseline $\tau$-pair efficiency is $\epsilon_\tau^2$, whereas additional channel-dependent selection or tagging efficiencies are applied where specified. Initial-state radiation affects the invariant-mass spectra, particularly in direct $\mu^+\mu^-\to\tau^+\tau^-$ production; its treatment and its consequences for the separation of production mechanisms are discussed in the relevant channel descriptions.

\paragraph*{Statistical procedure.}

We quantify deviations from the SM expectation using the counting-experiment significance estimator of Refs.~\cite{ParticleDataGroup:2020ssz,Cowan:2010js}, defined as
\begin{equation}
\label{eq:sen}
\mathcal S_i=\sqrt{2(S_i+B_i)\ln\!\left(1+\frac{S_i}{B_i}\right)-2S_i},
\end{equation}
where, for an analysis channel $i$,
\begin{equation}
\label{eq:muC_event_yields}
S_i=\epsilon_i\mathcal L\big[\sigma_i(\mathcal C_{\tau W},\mathcal C_{\tau B})-\sigma_i(0,0)\big],\qquad
B_i=\epsilon_i\mathcal L\,\sigma_i(0,0).
\end{equation}
Here $\sigma_i$ includes the kinematic selections for the channel and $\epsilon_i$ denotes its assumed net reconstruction and tagging efficiency, including the applicable powers of $\epsilon_\tau$. The quantity $S_i$ is a \emph{signed} change in the expected event count; interference can make it negative, provided $B_i+S_i>0$. Equation~\eqref{eq:sen} is the discovery-type Asimov counting-likelihood separation, evaluated here as a benchmark sensitivity measure. For an expected exclusion based on an SM Asimov dataset, the corresponding fixed-hypothesis Poisson likelihood distance would instead be $q_{\rm excl}=2\big[S_i-B_i\ln(1+S_i/B_i)\big]$. These agree at leading nonzero order for $|S_i|\ll B_i$, but not generally for larger deviations. 
We adopt the numerical prescription described above and use the reference thresholds $\mathcal S=2$ and $\mathcal S=3$ as \emph{nominal} sensitivity conventions, rather than as independently calibrated 95\% and 99\% confidence levels.

For statistically independent selections, the significances are combined in quadrature,
\begin{equation}
\label{eq:muC_significance_combined}
\mathcal S_{\mathrm{combined}}=\left(\sum_i\mathcal S_i^2\right)^{1/2}.
\end{equation}
This combination assumes statistically disjoint event categories and no correlated nuisance parameters. The present source does not provide the event-overlap information needed to validate this assumption for every pair of channels. Moreover, the contours in two coefficients and those with nearly quadratic dependence can require different critical values or explicit coverage calibration. 
We do not claim such a calibration for the projected sensitivities presented here.

\paragraph*{Interpretation of the projections.}
These estimates are intended to compare the intrinsic statistical sensitivity of the proposed channels under a common set of benchmark assumptions. They do not constitute detector-level projections: realistic $\tau$ decays and polarization reconstruction, detailed BIB mitigation, reducible backgrounds, channel-dependent systematic effects, and higher-order electroweak corrections are not modeled comprehensively here. The results should therefore be viewed as benchmarks for identifying promising measurements rather than as final experimental sensitivities. 
In particular, the ``95\% CL'' labels used in the tables and figures presented here correspond to the \emph{nominal} $\mathcal S=2$ criterion adopted in our analysis; they are not a claim of validated frequentist coverage for the one- or two-parameter regions.

\subsection{\boldmath{$\mu^+\mu^-\to\tau^+\tau^-$}}
\label{sec:muC_direct_tt}

In direct muon annihilation, the $\tau$ pair is produced through a neutral-current $s$-channel photon or $Z$ boson exchange. This process offers a particularly transparent test of the photon and $Z$ dipole interactions: both contribute to the same measured final state, while their relative impact is controlled by the structure of the electroweak couplings. At multi-TeV energies, the most important difference from measurements performed around the $Z$ resonance is that the chirality-flipping dipole amplitude contributes predominantly through its squared magnitude. We first establish this behavior analytically, before turning to the projected sensitivity in the following subsection.

\subsubsection{SMEFT description}
\label{sec:theory_mm2tt}

\begin{figure}[htb]
\includegraphics[width=0.25\textwidth]{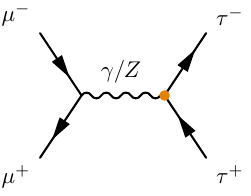}
\caption{Leading-order diagram for the $\mu^+\mu^- \to \tau^+\tau^-$ process. 
The orange vertex indicates sensitivity to the anomalous electromagnetic ($\mathcal{O}_{\tau \gamma}$) and weak ($\mathcal{O}_{\tau Z}$) dipole moments.
Removing the dipole insertion leaves the corresponding SM $s$-channel amplitude.} \label{fig:FD_mm2tt}
\end{figure}

At leading order, $\mu^+\mu^-\to\tau^+\tau^-$ proceeds through $\gamma$ and $Z$ exchange, as illustrated in Fig.~\ref{fig:FD_mm2tt}. The dipole operators introduced in Sec.~\ref{sec:Theory} modify the $\tau^+\tau^-\gamma$ and $\tau^+\tau^-Z$ vertices, respectively. Their contributions interfere with the SM amplitudes and also generate terms quadratic in the Wilson coefficients. No additional particles are required in the final state for these interactions to be probed.

The collision energy entering the hard scattering is reduced when photons are radiated from the incoming muons. For the annihilation process, the dominant initial-state radiation (ISR) is the QED contribution. In the structure-function approach employed in our analysis, the observable rate is obtained by convoluting the partonic cross section with the muon distributions,
\begin{equation}
\label{eq:muC_isr_explanation}
\hat s=\xi s,
\end{equation}
\begin{equation}
\sigma_{\mu^+\mu^-\to \tau^+\tau^-}(s) =\int_{\xi_0}^1 {\rm d}\xi \int_{\xi}^1 \frac{{\rm d}x}{x} f_{\mu^-/\mu^-}(x,Q^2) \, f_{\mu^+/\mu^+}\!\Big(\frac{\xi}{x},Q^2\Big) 
\hat{\sigma}_{\mu^+\mu^-\to \tau^+\tau^-}(\hat{s})\,,
\label{eq:sigma_muISR}
\end{equation}
where $f_{\mu^\pm/\mu^\pm}(x,Q^2)$ is the parton distribution function (PDF) that describes the probability of finding a $\mu^\pm$ with longitudinal momentum fraction $x$ of the beam $\mu^\pm$, and $\xi_0$ denotes the lower integration limit determined by the final-state phase space and any imposed selection. The leading-logarithmic QED treatment is available in event generators such as \textsc{MadGraph5\_aMC@NLO}~\cite{Alwall:2014hca,Frederix:2018nkq} and \textsc{Whizard}~\cite{Kilian:2007gr}. At a multi-TeV muon collider, broader electroweak radiation effects can also be important~\cite{Han:2020uid,Han:2021kes,Chen:2022msz}; their relevance depends on the channel and observable under consideration.

To display the dependence on the dipole interactions, it is useful to decompose the hard cross section according to the powers of the dimension-six coefficients,
\begin{equation}
   \hat{\sigma}_{\mu^+\mu^-\to\tau^+\tau^-} = \hat{\sigma}_{\mu\mu}^{\rm SM} + \hat{\sigma}_{\mu\mu}^{\rm Lin} + \hat{\sigma}_{\mu\mu}^{\rm Quad}\,. 
   \label{eq:mm2tata_xs}
\end{equation}
The first term is the SM contribution, the second denotes SM--dipole interference, and the third is the squared dimension-six amplitude. In the limit $\hat s\gg m_\tau^2$, the expressions used in our analysis read
\begin{subequations}
\label{eq:sigma_mm_all}
\begin{align}
    &\hat{\sigma}_{\mu\mu}^{\rm SM} = \frac{\hat{\beta}}{12 \pi} \Biggl[ \frac{e^4}{\hat{s}} + \frac{2e^2g_Z^2g_V^2\hat{s}\Delta}{\Delta^2\hat{s}^2+M_Z^2\Gamma_Z^2} + \frac{g_Z^4 \hat{s}\left(g_V^2+g_A^2\right)^2}{\Delta^2\hat{s}^2+M_Z^2\Gamma_Z^2} \Biggr], \label{eq:ee_sigma_sm} \\[10pt]
    &\hat{\sigma}_{\mu\mu}^{\rm Lin} = \frac{\hat{\beta} m_\tau v}{\sqrt{2}\pi \hat{s} \Lambda^2} \text{Re} \Biggl[ e^3{\mathcal C}_{\tau \gamma} +\frac{eg_Z^2g_V^2{\mathcal C}_{\tau \gamma}\Delta \hat{s}^2}{\Delta^2\hat{s}^2+M_Z^2\Gamma_Z^2} - \frac{g_Z g_V {\mathcal C}_{\tau Z} \hat{s}^2 (e^2\Delta+g_Z^2(g_V^2+g_A^2))}{\Delta^2\hat{s}^2+M_Z^2\Gamma_Z^2}\Biggr], \label{eq:ee_sigma_lin} \\[10pt]
    &\hat{\sigma}_{\mu\mu}^{\rm Quad} = \frac{\hat{\beta} v^2}{12 \pi \Lambda^4} \Biggl[ e^2 {|\mathcal C}_{\tau\gamma}|^2 +\frac{g_Z^2 |{\mathcal C}_{\tau Z}|^2 \hat{s}^2(g_V^2+g_A^2)}{\Delta^2\hat{s}^2+M_Z^2\Gamma_Z^2} - \frac{2 e g_Z g_V \text{Re}({\mathcal C}_{\tau\gamma}^*{\mathcal C}_{\tau Z}) \Delta \hat{s}^2}{\Delta^2\hat{s}^2+M_Z^2\Gamma_Z^2} \Biggr], \label{eq:mmtt_sigma_quad}
\end{align}
\end{subequations}
Here $g_V=-1/4+s_W^2$ and $g_A=1/4$ are the vector and axial couplings appearing in the $Z$-exchange amplitude, $g_Z=e/(s_Wc_W)$. We use $\hat\beta=\sqrt{1-4m_\tau^2/\hat s}$ and $\Delta=1-M_Z^2/\hat s$; the $Z$ mass and width are denoted by $M_Z$ and $\Gamma_Z$. When $\hat s\gg M_Z^2$, one has $\hat\beta\simeq\Delta\simeq1$, while the finite-width contribution becomes negligible away from the resonance.

The hierarchy among these terms has a simple helicity interpretation. Standard Model gauge interactions preserve chirality in the high-energy limit, whereas the dipole operators flip it. Their interference therefore requires a mass insertion and is suppressed relative to the squared-dipole contribution in the relevant multi-TeV kinematics. In the expressions above, the linear term carries the characteristic factor $m_\tau v/\hat s$~\cite{Buttazzo:2026ulg}. Although the numerical balance also depends on the coefficient values, this suppression explains why the high-energy rate is particularly sensitive to the terms quadratic in $\mathcal C_{\tau\gamma}$ and $\mathcal C_{\tau Z}$.

An additional feature is visible in Eq.~\eqref{eq:mmtt_sigma_quad}. The interference between the photon- and $Z$-mediated dipole amplitudes is proportional to the small vector coupling $g_V\simeq-0.02$. Consequently, the leading quadratic dependence is approximately diagonal in the $(\mathcal C_{\tau\gamma},\mathcal C_{\tau Z})$ basis. In the phenomenological analysis this translates into a weak correlation between the electromagnetic and weak directions, unlike the correlations that can arise in measurements dominated by interference with the SM.

Our cross sections retain the SM contribution, its interference with the dimension-six $\tau$ dipole amplitudes, and the squares of those amplitudes. This is a \emph{dipole-only quadratic dimension-six benchmark}, not a complete SMEFT expansion through $\mathcal O(\Lambda^{-4})$. At that order, interference between the SM and dimension-eight amplitudes also contributes in general. While the helicity suppression described above applies to the chirality-flipping dipole structures, it does not establish a corresponding suppression for all possible dimension-eight operators, including chirality-preserving interactions. The projected sensitivities therefore constrain the two dipole coefficients \emph{conditional on the assumption} that additional effective interactions have negligible impact on the selected observables. No dimension-eight contribution has been calculated or included in the present numerical analysis.

\subsubsection{Phenomenological analysis}
\label{sec:Pheno_mm2tt}
\label{sec:Pheno_tt}

The direct annihilation topology is most readily identified at large tau-pair invariant mass. As illustrated by the distribution in Fig.~\ref{fig:muC_dmtt}, events from $\mu^+\mu^-\to\tau^+\tau^-$ accumulate near $m(\tau^+\tau^-)\simeq\sqrt{s}$, with initial-state photon radiation extending the spectrum toward lower masses. In contrast, the vector-boson-scattering contributions predominantly populate the lower invariant-mass region. We therefore select
\begin{equation}
\label{eq:direct_tt_mass_cut}
 m(\tau^+\tau^-)>0.8\sqrt{s},
\end{equation}
which defines a high-mass sample dominated by the $s$-channel annihilation process. The detector acceptance and tau-identification assumptions are those specified in Sec.~\ref{sec:analysis_framework}. The complementary lower-mass region, where different production mechanisms must be considered together, will be addressed separately in the VBS analysis.

\begin{figure}[htb]
\includegraphics[width=0.42\textwidth]{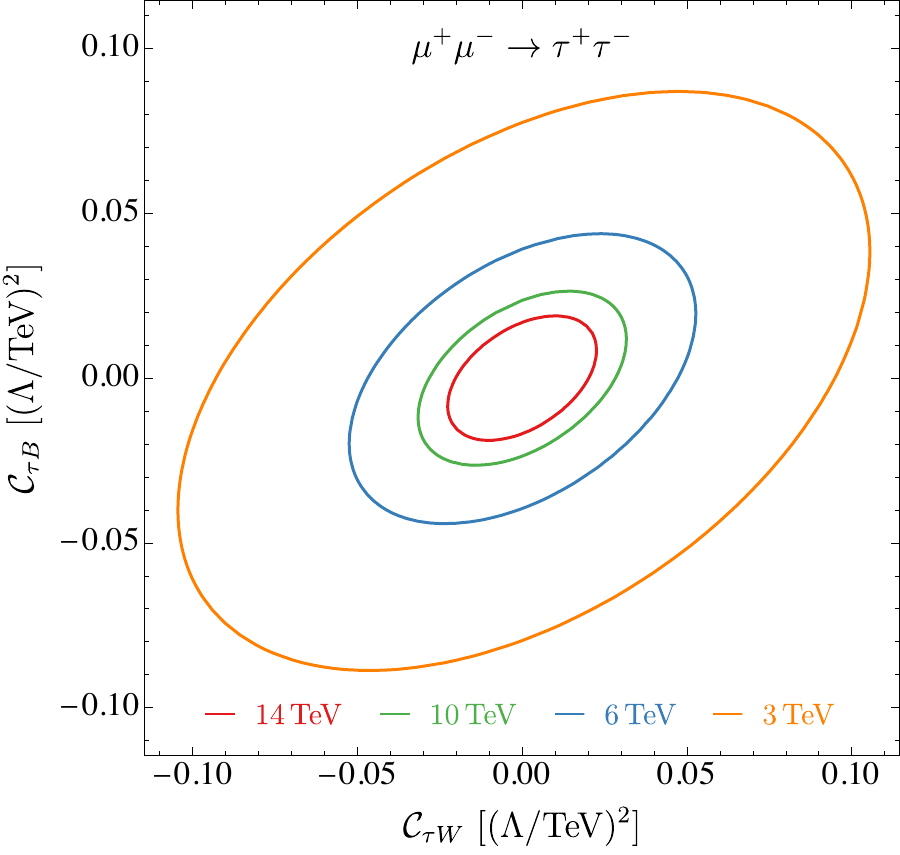}
\includegraphics[width=0.425\textwidth]{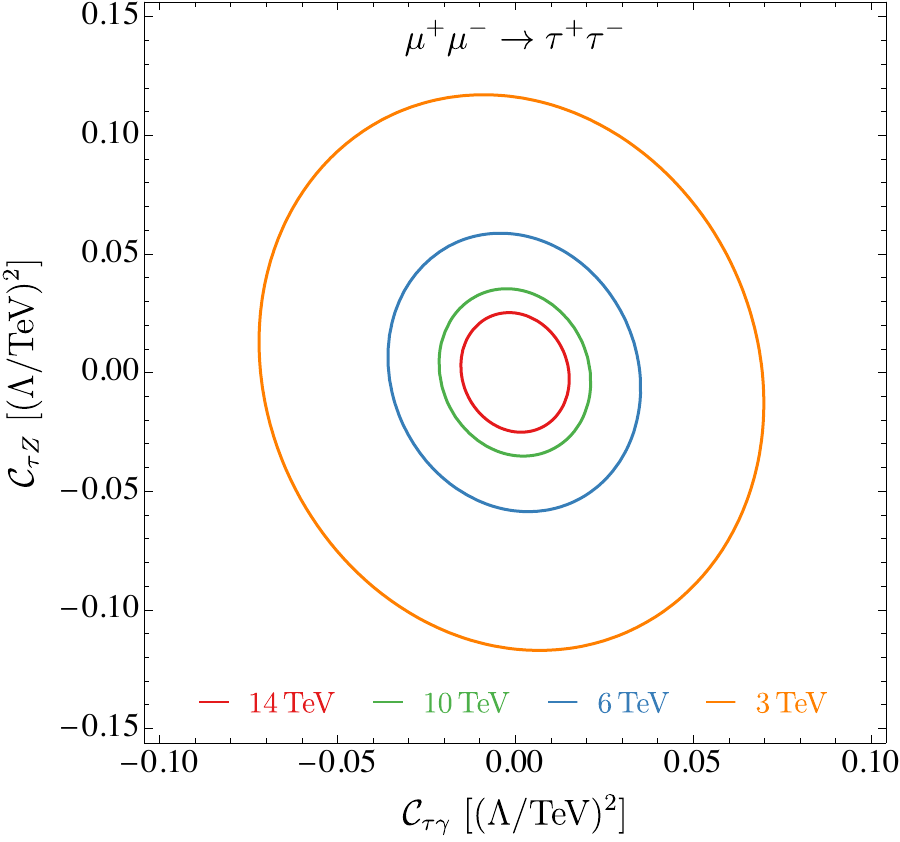}
\caption{Expected sensitivity reach on the $\tau$ dipole Wilson coefficients at the 95\% CL in the (left) ($\mathcal{C}_{\tau W}, \mathcal{C}_{\tau B}$) plane and (right) in the ($\mathcal{C}_{\tau \gamma}, \mathcal{C}_{\tau Z}$) plane, derived from the $\mu^+\mu^-\to\tau^+\tau^-$ process at a multi-TeV muon collider for center-of-mass energies $\sqrt{s}=3,\,6,\,10$, and 14 TeV.} \label{fig:muC_contour_mm2tt}
\end{figure}

For the four benchmark collider energies, the projected 95\% CL contours are displayed in Fig.~\ref{fig:muC_contour_mm2tt}. The left panel uses the gauge-eigenstate coefficients $(\mathcal C_{\tau W},\mathcal C_{\tau B})$, while the right panel shows the mass-eigenstate basis $(\mathcal C_{\tau\gamma},\mathcal C_{\tau Z})$. The red, green, blue and orange curves correspond, respectively, to $\sqrt{s}=14$, 10, 6 and 3~TeV. Expressing the result in both bases makes it possible to distinguish correlations induced by the electroweak rotation from those intrinsic to the scattering amplitude.

The corresponding bounds on the electromagnetic and weak magnetic and electric dipole moments are collected in Table~\ref{tab:constraints_mmtt}. 
The two coefficients are marginalized in the $(\mathcal C_{\tau\gamma},\mathcal C_{\tau Z})$ plane, while the CP-even and CP-odd limits are extracted
assuming purely real and purely imaginary coefficients, respectively.

These are projected statistical sensitivities under the common assumptions of Sec.~\ref{sec:analysis_framework}, not detector-level confidence intervals.

\begin{table}[htb]
\centering
\renewcommand{\arraystretch}{1.18}
\setlength{\tabcolsep}{5pt}
\begin{tabular}{c|cccc}
\hline
$\sqrt{s}$ [TeV]
& 3
& 6
& 10
& 14
\\
\hline
$|\Delta a_\tau|$                    & $2.9\times10^{-4}$  & $1.5\times10^{-4}$  & $8.8\times10^{-5}$  & $6.3\times10^{-5}$  \\
$|\Delta a^Z_\tau|$                  & $4.7\times10^{-4}$  & $2.4\times10^{-4}$  & $1.4\times10^{-4}$  & $1.0\times10^{-4}$  \\
$|d_\tau|$ [$e\,{\rm cm}$]           & $1.6\times10^{-18}$ & $8.1\times10^{-19}$ & $4.8\times10^{-19}$ & $3.5\times10^{-19}$ \\
$|d^Z_\tau|$ [$e\,{\rm cm}$]         & $2.7\times10^{-18}$ & $1.3\times10^{-18}$ & $8.0\times10^{-19}$ & $5.7\times10^{-19}$ \\
\hline
\end{tabular}
\caption{
Projected $95\%$ CL constraints on the $\tau$ dipole moments from
$\mu^+\mu^-\to\tau^+\tau^-$ at high-energy muon colliders. The
constraints on $\Delta a^{(Z)}_\tau$ are derived assuming a purely real
$\mathcal{C}_{\tau V}$, while those on $d^{(Z)}_\tau$ assume a purely
imaginary $\mathcal{C}_{\tau V}$.}
\label{tab:constraints_mmtt}
\end{table}

The main features of the contours follow directly from the high-energy behavior derived in Sec.~\ref{sec:theory_mm2tt}. The interference between the SM and a dipole insertion is suppressed by the chirality flip; the characteristic factor $m_\tau v/s$ is approximately $5\times10^{-5}$ at 3~TeV and $4\times10^{-6}$ at 10~TeV. For the coefficient values relevant to these projections, the quadratic contribution thus controls the excess event rate. In addition, the term mixing $\mathcal C_{\tau\gamma}$ and $\mathcal C_{\tau Z}$ is proportional to $g_V\simeq-0.02$. The two dipole directions are therefore only weakly correlated in the mass-eigenstate basis, explaining why the marginalized bounds are close to those obtained when one operator at a time is turned on.

The improvement with energy reflects both the behavior of the hard cross sections and the luminosity benchmark. At fixed Wilson coefficients the SM $2\to2$ rate falls approximately as $1/s$, whereas the leading squared-dipole contribution approaches an energy-independent value. With the assumed $\mathcal L\propto s$, the background yield is consequently approximately constant and the dipole-induced event excess grows proportionally to $s$. In the regime $S\ll B$, the significance in Eq.~\eqref{eq:sen} reduces to $\mathcal S\simeq S/\sqrt{B}$. For illustration, taking $\mathcal C_{\tau Z}=0$, this gives
\begin{equation}
\mathcal{S}(\mathcal{C}_{\tau\gamma}) \propto \frac{s \,\mathcal{C}_{\tau\gamma}^2}{\Lambda^4},
\end{equation}
so that a fixed-significance bound on $|\mathcal C_{\tau\gamma}|/\Lambda^2$ scales as the inverse collider energy $1/\sqrt{s}$. The numerical limits in Table~\ref{tab:constraints_mmtt} closely follow this trend. The same scaling applies to the dipole moments after using the matching relations of Sec.~\ref{sec:Theory}; it is conditional on the adopted luminosity prescription and the statistical approximation, and should not be interpreted as a universal scaling for all collider scenarios.

\subsection{Vector-boson scattering: $\mathrm{VBS}\to\tau^+\tau^-$}
\label{sec:muC_vbs}

Unlike direct muon annihilation, the multi-particle final states that contain a $\tau^+\tau^-$ pair can be populated by the scattering of electroweak gauge bosons radiated from the incoming muons. At multi-TeV energies, such emissions provide access to vector-boson-scattering configurations~\cite{Buttazzo:2018qqp,Costantini:2020stv,Han:2020uid,Han:2021kes,BuarqueFranzosi:2021wrv,Ruiz:2021tdt,Accettura:2023ked,InternationalMuonCollider:2025sys,Frixione:2025guf,Dahlen:2025udl,Han:2026edr,Garosi:2023bvq}. We consider two experimentally distinct classes of four-particle final states,
\begin{equation}
\mu^+\mu^-\to\tau^+\tau^-\mu^+\mu^-,\qquad
\mu^+\mu^-\to\tau^+\tau^-\nu_\ell\bar\nu_\ell,
\quad \ell=e,\mu,\tau.
\label{eq:muC_vbs_channels}
\end{equation}
They receive dipole contributions from the interactions introduced in Sec.~\ref{sec:Theory}, but are not restricted to pure boson--boson scattering diagrams. This distinction will be relevant when defining the event sample used for the sensitivity estimates.

\subsubsection{SMEFT description}
\label{sec:theory_VBS}

\begin{figure}[htb]
\includegraphics[width=0.22\textwidth]{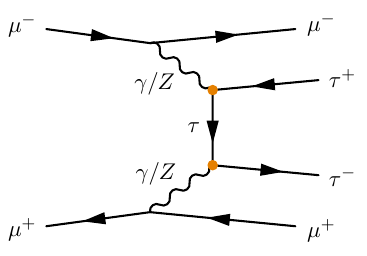}
\includegraphics[width=0.22\textwidth]{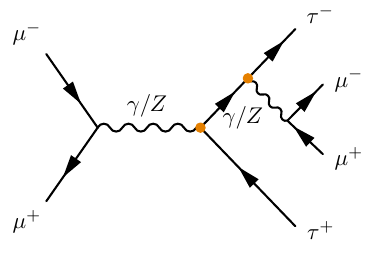}
\includegraphics[width=0.22\textwidth]{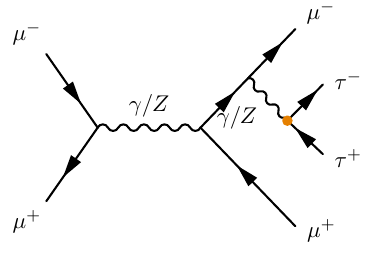}\\
\includegraphics[width=0.22\textwidth]{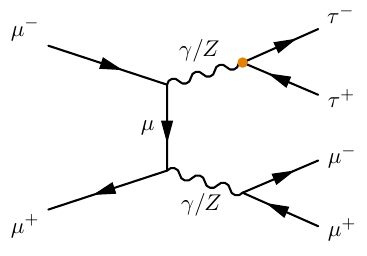}
\includegraphics[width=0.22\textwidth]{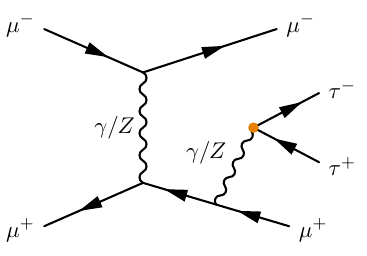}
\includegraphics[width=0.22\textwidth]{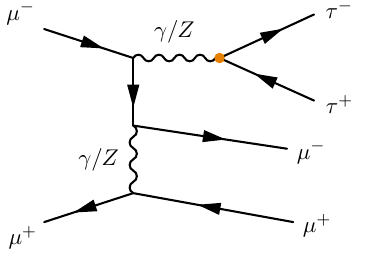}
\caption{Leading-order Feynman diagrams for $\mu^+\mu^-\to\tau^+\tau^-\mu^+\mu^-$. The orange nodes represent vertices sensitive to both anomalous electromagnetic ($\mathcal{O}_{\tau \gamma}$) and weak ($\mathcal{O}_{\tau Z}$) dipole moments. For the BSM diagrams that possess a SM counterpart, the removal of these colored vertex modifications recovers the corresponding tree-level SM interactions.}\label{fig:FD_mm2tt2mu}
\end{figure}

For the final state with two additional muons, the representative leading-order diagrams are collected in Fig.~\ref{fig:FD_mm2tt2mu}. The dominant scattering topology involves neutral-boson scattering, with $\gamma\gamma$, $\gamma Z$, and $ZZ$ initial states contributing to the production of the $\tau$ pair. The dimension-six dipole interactions modify the $\tau^+\tau^-\gamma$ and $\tau^+\tau^-Z$ vertices entering these amplitudes. The outgoing muons preferentially follow the beam directions and may remain outside the central detector acceptance. Other contributions to the same four-lepton final state are present, including configurations in which an intermediate on-shell $Z$ decays to $\tau^+\tau^-$. The latter can be reduced by excluding the $Z$-resonance region in the tau-pair invariant mass.

\begin{figure}[htb]
\includegraphics[width=0.22\textwidth]{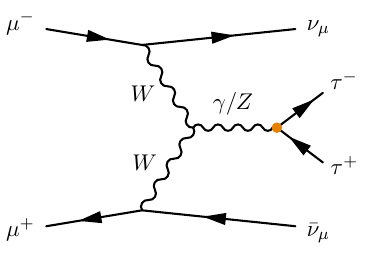}
\includegraphics[width=0.22\textwidth]{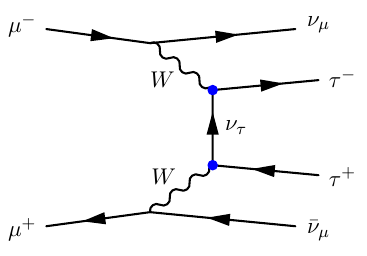}
\includegraphics[width=0.22\textwidth]{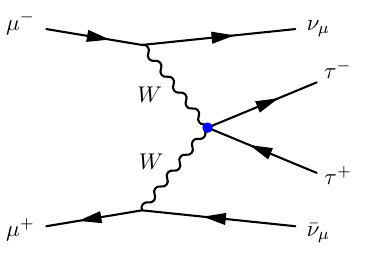}
\includegraphics[width=0.22\textwidth]{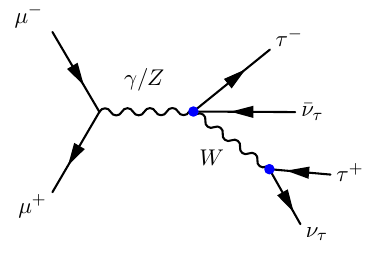}
\includegraphics[width=0.22\textwidth]{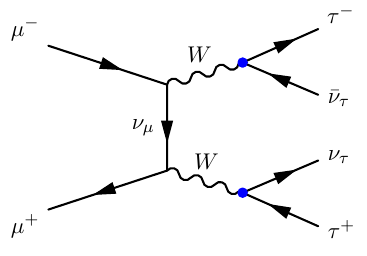}
\includegraphics[width=0.22\textwidth]{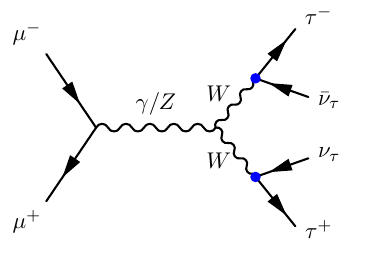}
\includegraphics[width=0.22\textwidth]{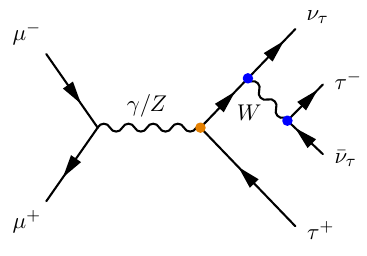}
\includegraphics[width=0.22\textwidth]{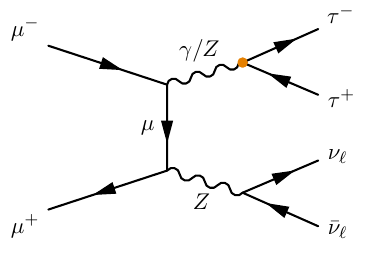}
\includegraphics[width=0.22\textwidth]{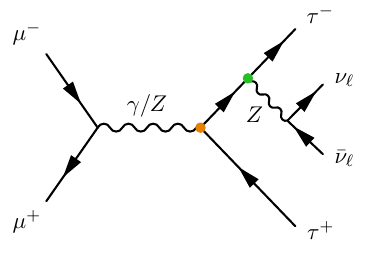}
\includegraphics[width=0.22\textwidth]{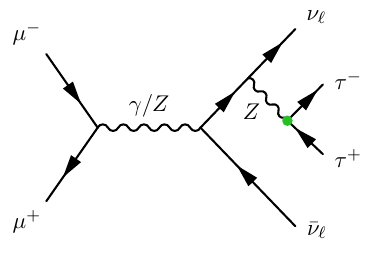}
\includegraphics[width=0.22\textwidth]{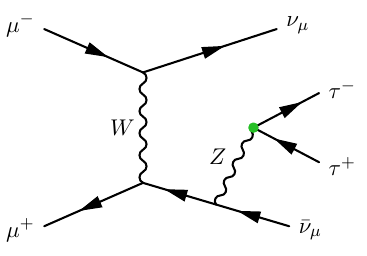}
\includegraphics[width=0.22\textwidth]{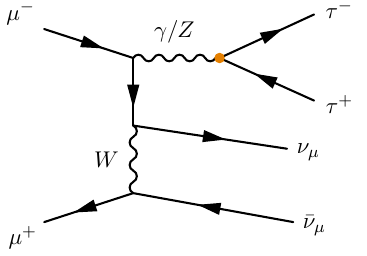}
\caption{Leading-order Feynman diagrams for $\mu^+\mu^-\to\tau^+\tau^-\nu_\ell\bar{\nu}_\ell$. The orange nodes represent vertices sensitive to both anomalous electromagnetic ($\mathcal{O}_{\tau \gamma}$) and weak ($\mathcal{O}_{\tau Z}$) dipole moments; blue nodes denote vertices modified exclusively by the $\mathcal{O}_{\tau W}$ operator; and the green node signifies a vertex specifically modified by the $\mathcal{O}_{\tau Z}$ operator. For the BSM diagrams that possess a SM counterpart, the removal of these colored vertex modifications recovers the corresponding tree-level SM interactions.}\label{fig:FD_mm2tt2nu}
\end{figure}

The neutrino-associated final state involves a broader set of electroweak topologies, as shown in Fig.~\ref{fig:FD_mm2tt2nu}. In particular, $W^+W^-$ scattering with $\nu_\mu\bar\nu_\mu$ spectators provides an important contribution at low $m(\tau^+\tau^-)$. We sum inclusively over $\nu_e$, $\nu_\mu$, and $\nu_\tau$, since these species cannot be separated from their missing-momentum signatures event by event. In addition to the dipole insertions in the neutral-current $\tau$ vertices, the diagrams contain interactions induced specifically by the $SU(2)_L$ dipole operator $\mathcal O_{\tau W}$, reflecting its charged-current components. The non-scattering contributions cannot be neglected throughout the whole phase space: removing $Z\to\tau^+\tau^-$ with an invariant-mass selection does not also eliminate resonant $Z\to\nu_\ell\bar\nu_\ell$ or $W\to\tau\nu_\tau$ configurations. These become increasingly relevant in the higher-$m(\tau^+\tau^-)$ part of the neutrino-associated spectrum.

\begin{figure}[htb]
\includegraphics[width=0.6\textwidth]{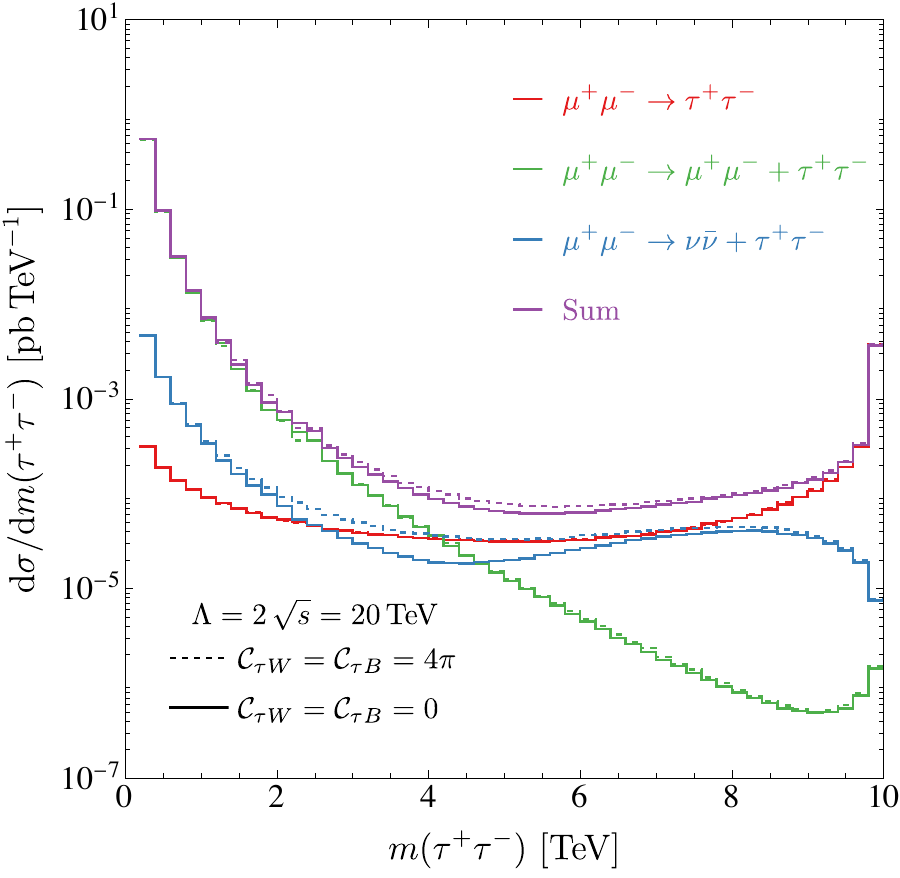}
\caption{Invariant mass distribution ${\rm d}\sigma/{\rm d}m(\tau^+\tau^-)$ for $\tau^+\tau^-$ pair production at a $\sqrt{s}=10$ TeV muon collider. Solid lines represent the SM predictions, while dashed lines denote the BSM benchmark with Wilson coefficients $\mathcal{C}_{\tau W} = \mathcal{C}_{\tau B} = 4\pi$ and a new physics scale $\Lambda = 2\sqrt{s} = 20$ TeV. A kinematic veto of $m(\tau^+\tau^-) > 200$ GeV is applied to suppress the resonant $Z \to \tau^+\tau^-$ contribution. To ensure detector acceptance and clear particle identification, we further require the scattering angles of the $\tau$ leptons to satisfy $10^\circ< \theta(\tau^\pm) < 170^\circ$ with respect to the beam axis, and a minimum angular separation of $\Delta R(\tau^+\tau^-) > 0.4$. The purple line represents the total integrated signal from muon annihilation and the VBS processes.} \label{fig:muC_dmtt}
\end{figure}

The different kinematics can be seen directly from the invariant-mass distributions at $\sqrt{s}=10$~TeV in Fig.~\ref{fig:muC_dmtt}. 
The curves use the benchmark $\mathcal C_{\tau W}=\mathcal C_{\tau B}=4\pi$, $\Lambda=20$~TeV, with the kinematic selections specified in the figure caption. 
The direct annihilation process is concentrated near $m(\tau^+\tau^-)\simeq\sqrt{s}$, apart from the radiative tail generated by initial-state photon emission. The neutral-current four-lepton channel instead has a large contribution close to the lower invariant-mass cut and extends across the spectrum. A much smaller high-mass feature arises from configurations with $\mu^+\mu^-\to\tau^+\tau^-\gamma$ and subsequent $\gamma\to\mu^+\mu^-$. For the neutrino final state, $W^+W^-$ scattering is especially important below approximately $2$~TeV, whereas the additional resonant gauge-boson-decay topologies increase the rate at larger tau-pair invariant masses.

These features explain why the direct and multi-particle production mechanisms probe complementary kinematic regions. They also make a strict event-by-event separation of scattering and non-scattering topologies impractical within a common low-mass tau-pair sample. In the following phenomenological analysis, we therefore use the term \emph{VBS sample} as a shorthand for the inclusive set of contributing production mechanisms, including the direct $\mu^+\mu^-\to\tau^+\tau^-$ channel in the selected region, rather than as a claim that every accepted event originates from a pure vector-boson-scattering diagram. The detailed kinematic selections and optimized sensitivity bounds are presented separately in Sec.~\ref{sec:Pheno_VBS}.

\subsubsection{Phenomenological analysis}
\label{sec:Pheno_VBS}

The invariant-mass spectrum discussed in Sec.~\ref{sec:theory_VBS} motivates a selection complementary to the high-mass annihilation sample of Sec.~\ref{sec:Pheno_tt}. At low $m(\tau^+\tau^-)$, the neutral-current four-body final state benefits from substantial photon-initiated contributions. The large event rate is potentially attractive for probing the electromagnetic dipole interaction, particularly where SM--BSM interference is relevant. In practice, however, the low-mass region is affected by sizable QCD-induced dijet backgrounds~\cite{Han:2021kes}, complicating the identification of hadronic $\tau$ decays. We therefore assess the sensitivity after imposing a lower threshold on the reconstructed $\tau$-pair invariant mass.

The sample includes the contributions from direct $\mu^+\mu^-\to\tau^+\tau^-$ production and from the $\tau^+\tau^-\mu^+\mu^-$ and $\tau^+\tau^-\nu_\ell\bar\nu_\ell$ final states. As explained above, we retain the designation ``VBS sample'' for this inclusive selection, without attributing every event to a genuine vector-boson-scattering diagram. In addition to the baseline acceptance and isolation requirements of Sec.~\ref{sec:analysis_framework}, we impose
\begin{equation}
\label{eq:muC_VBS_selection}
|\eta(\mu^\pm)|>2.5,\qquad
m_{\mathcal F}^{\rm min}<m(\tau^+\tau^-)<0.8\sqrt{s},\qquad
p_T(\tau^\pm)>\frac{m_{\mathcal F}^{\rm min}}{10}\,.
\end{equation}
The forward-muon requirement is applied to the charged-muon final state to select configurations in which the outgoing muons escape the central detector acceptance; it also suppresses contributions with centrally produced muon pairs. The upper invariant-mass boundary removes the region reserved for direct annihilation, while the adjustable lower boundary regulates the experimentally challenging low-mass domain. We parameterize the latter as $x=m_{\mathcal F}^{\rm min}/\sqrt{s}$ and optimize $x$ for each dipole observable.

\begin{figure}[htb]
\includegraphics[width=0.44\textwidth]{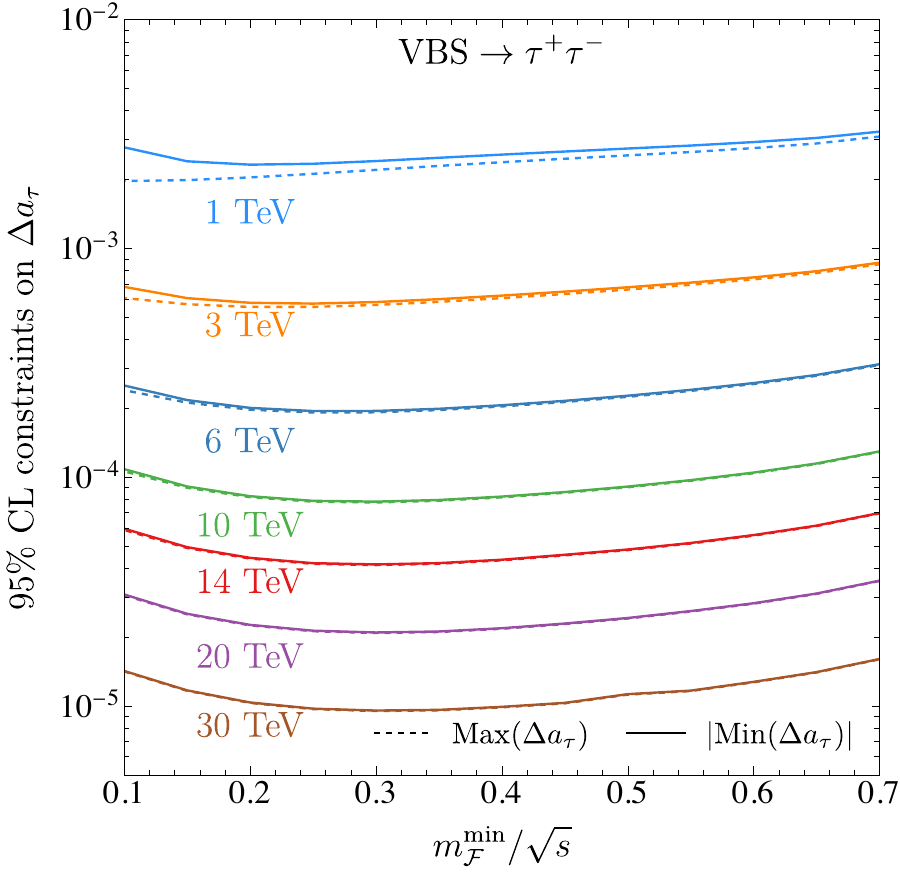}
\includegraphics[width=0.44\textwidth]{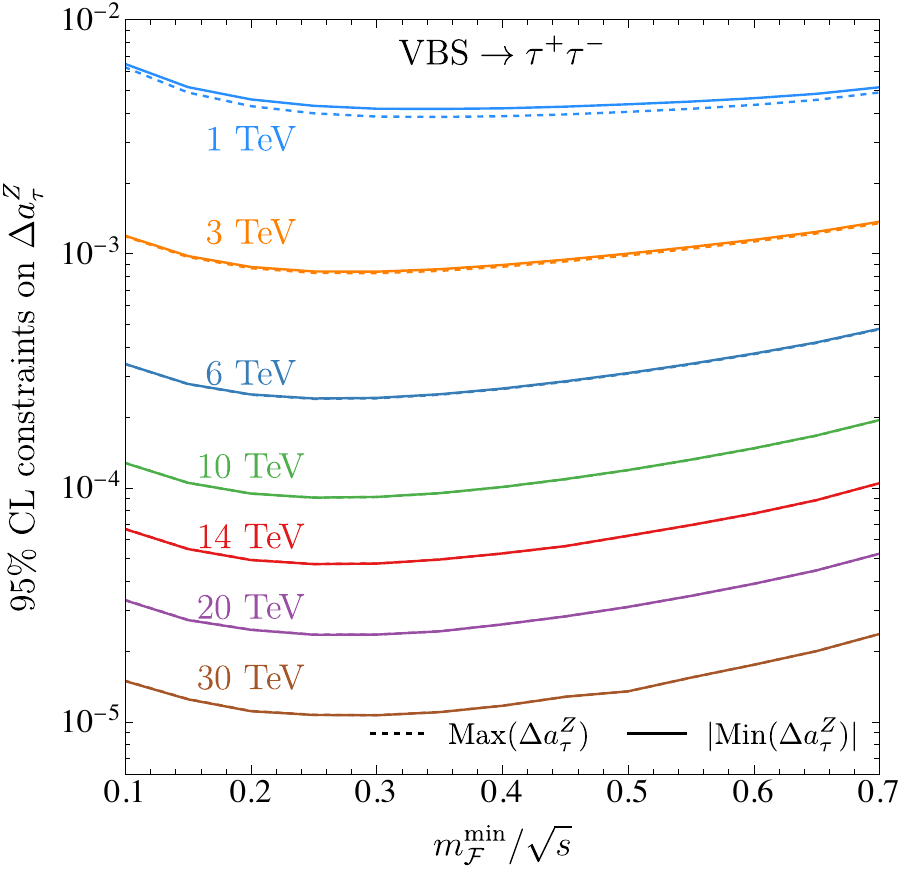}
\includegraphics[width=0.44\textwidth]{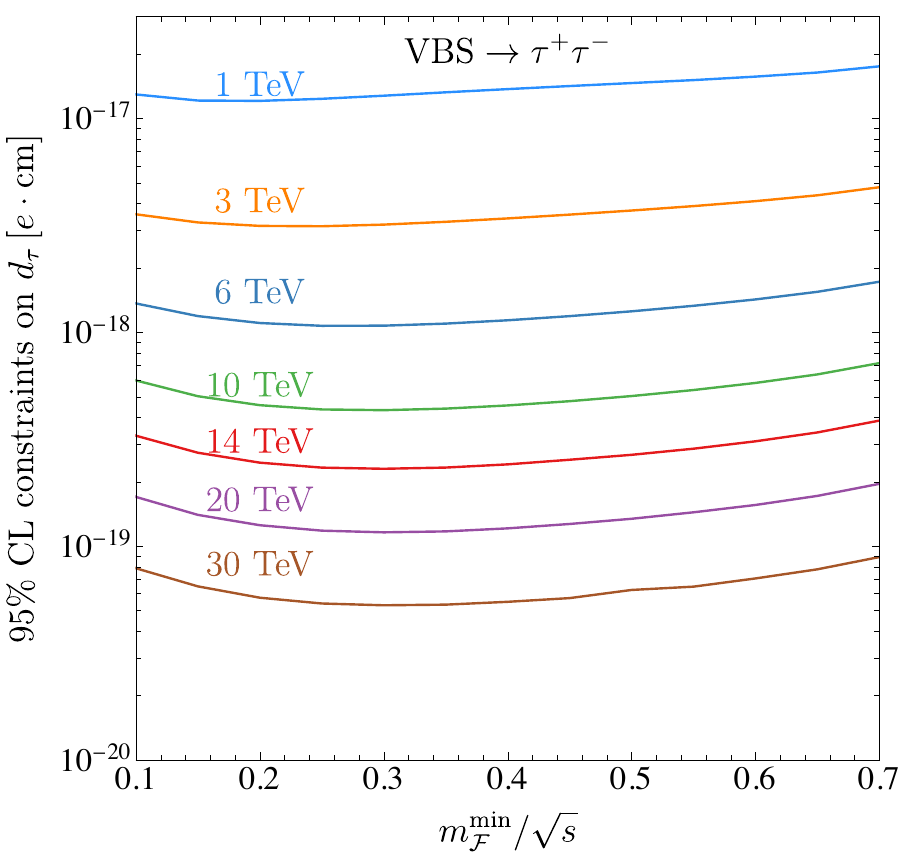}
\includegraphics[width=0.44\textwidth]{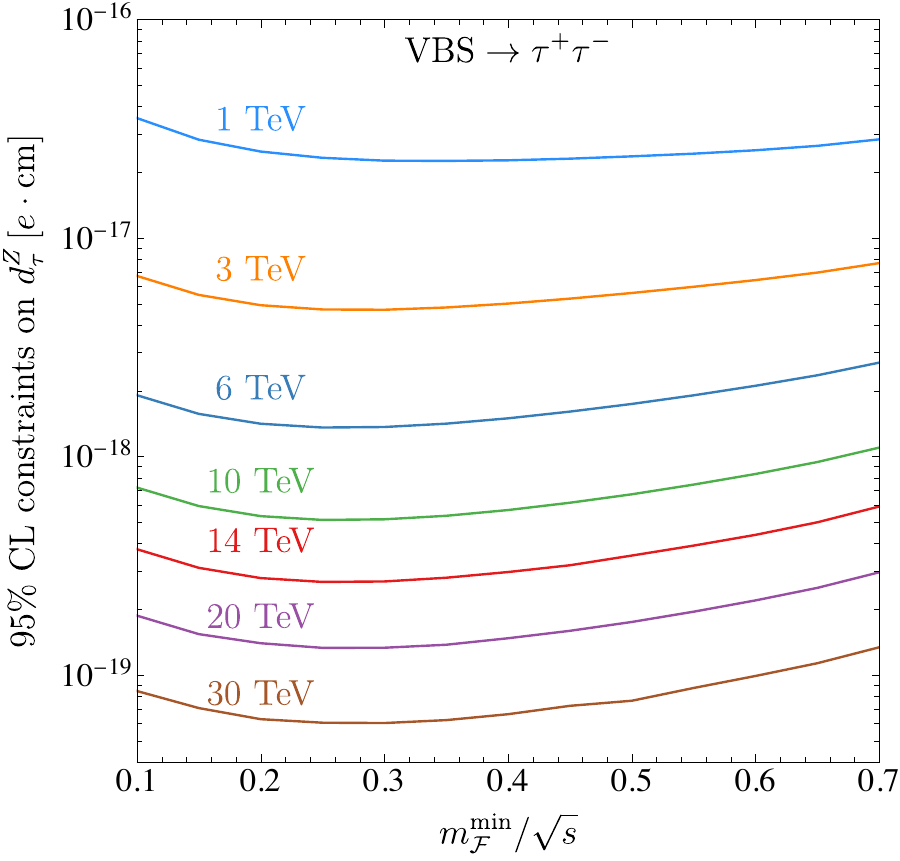}
\caption{Projected 95\% CL bounds on $\Delta a^{(Z)}_\tau$ and $d^{(Z)}_\tau$ from the VBS $\to \tau^+\tau^-$ process at a multi-TeV muon collider, under different $m_{\mathcal{F}}^{\rm min} < m(\tau^+\tau^-) < 0.8 \sqrt{s}$ and $p_T(\tau^\pm)>m_{\mathcal{F}}^{\rm min}/10$ cuts at various energies. For $\Delta a^{(Z)}_\tau$, the bounds are derived assuming purely real Wilson coefficients $\mathcal{C}_{\tau V}$; solid (dashed) curves correspond to bounds from the negative (positive) side. The bounds on $d^{(Z)}_\tau$ are obtained assuming purely imaginary $\mathcal{C}_{\tau V}$.} \label{fig:muC_MttCut_VBS}
\end{figure}

Figure~\ref{fig:muC_MttCut_VBS} shows the projected 95\% CL sensitivities as functions of the lower mass threshold, using the luminosity prescription of Sec.~\ref{sec:analysis_framework}. The scan covers $\sqrt{s}=1,3,6,10,14,20$, and $30$~TeV; the four main benchmark energies are used for the tabulated results. For the magnetic moments, real Wilson coefficients are assumed, and the solid and dashed curves distinguish the bounds obtained on the negative and positive sides of the respective parameters. 
The separation between the positive and negative magnetic-moment limits becomes less pronounced as the energy increases, consistent with an increasingly important contribution quadratic in the dipole coefficients. For the electric moments, the analysis probes purely imaginary Wilson coefficients.

The optimized threshold is not universal across the four observables. Table~\ref{tab:x_opt} lists the values of $x$ selected at each benchmark energy. For the electromagnetic magnetic and electric moments, the preferred threshold moves from $x=0.25$ at 3--6~TeV to $x=0.30$ at 10--14~TeV; for the weak moments the preferred thresholds are generally $x=0.25$, with $x=0.30$ for $|\Delta a_\tau^Z|$ at 3~TeV. The corresponding one-parameter 95\% CL sensitivities are collected in Table~\ref{tab:constraints_VBFtt}.

\begin{table}[htb]
\centering
\renewcommand{\arraystretch}{1.18}
\setlength{\tabcolsep}{5pt}
\begin{tabular}{c|cccc}
\hline
$\sqrt{s}$ [TeV]
& $|\Delta a_\tau|$
& $|d_\tau|$
& $|\Delta a^Z_\tau|$
& $|d^Z_\tau|$
\\
\hline
$3$  & $0.25$ & $0.25$ & $0.30$ & $0.25$ \\
$6$  & $0.25$ & $0.25$ & $0.25$ & $0.25$ \\
$10$ & $0.30$ & $0.30$ & $0.25$ & $0.25$ \\
$14$ & $0.30$ & $0.30$ & $0.25$ & $0.25$ \\
\hline
\end{tabular}
\caption{Optimal cut on $x = m(\tau^+\tau^-)/\sqrt{s}$ for the VBF $\to\tau^+\tau^-$ process.}
\label{tab:x_opt}
\end{table}

\begin{table}[htb]
\centering
\renewcommand{\arraystretch}{1.18}
\setlength{\tabcolsep}{5pt}
\begin{tabular}{c|cccc}
\hline
$\sqrt{s}$ [TeV]
& 3
& 6
& 10
& 14
\\
\hline
$|\Delta a_\tau|$                    & $5.7\times10^{-4}$  & $1.9\times10^{-4}$  & $7.8\times10^{-5}$  & $4.2\times10^{-5}$  \\
$|\Delta a^Z_\tau|$                  & $8.4\times10^{-4}$  & $2.4\times10^{-4}$  & $9.1\times10^{-5}$  & $4.7\times10^{-5}$  \\
$|d_\tau|$ [$e\,{\rm cm}$]           & $3.1\times10^{-18}$ & $1.1\times10^{-18}$ & $4.3\times10^{-19}$ & $2.3\times10^{-19}$ \\
$|d^Z_\tau|$ [$e\,{\rm cm}$]         & $4.7\times10^{-18}$ & $1.4\times10^{-18}$ & $5.2\times10^{-19}$ & $2.7\times10^{-19}$ \\
\hline
\end{tabular}
\caption{Same as Table~\ref{tab:constraints_mmtt}, but for VBS $\to\tau^+\tau^-$ processes at high-energy muon colliders.}
\label{tab:constraints_VBFtt}
\end{table}

The numerical results indicate a marked improvement with increasing collider energy. For example, the projected reach on $|\Delta a_\tau|$ changes from $5.7\times10^{-4}$ at 3~TeV to $4.2\times10^{-5}$ at 14~TeV, while that on $|d_\tau|$ improves from $3.1\times10^{-18}$ to $2.3\times10^{-19}\,e\,{\rm cm}$. The gains reflect the combined effects of higher integrated luminosity, changes in the electroweak production mechanisms, and the growing relative importance of dipole-induced contributions in the selected phase space. They should not be interpreted as arising solely from the pure VBS topology, since the inclusive sample also contains non-VBS and resonant contributions.

\begin{figure}[htb]
\includegraphics[width=0.42\textwidth]{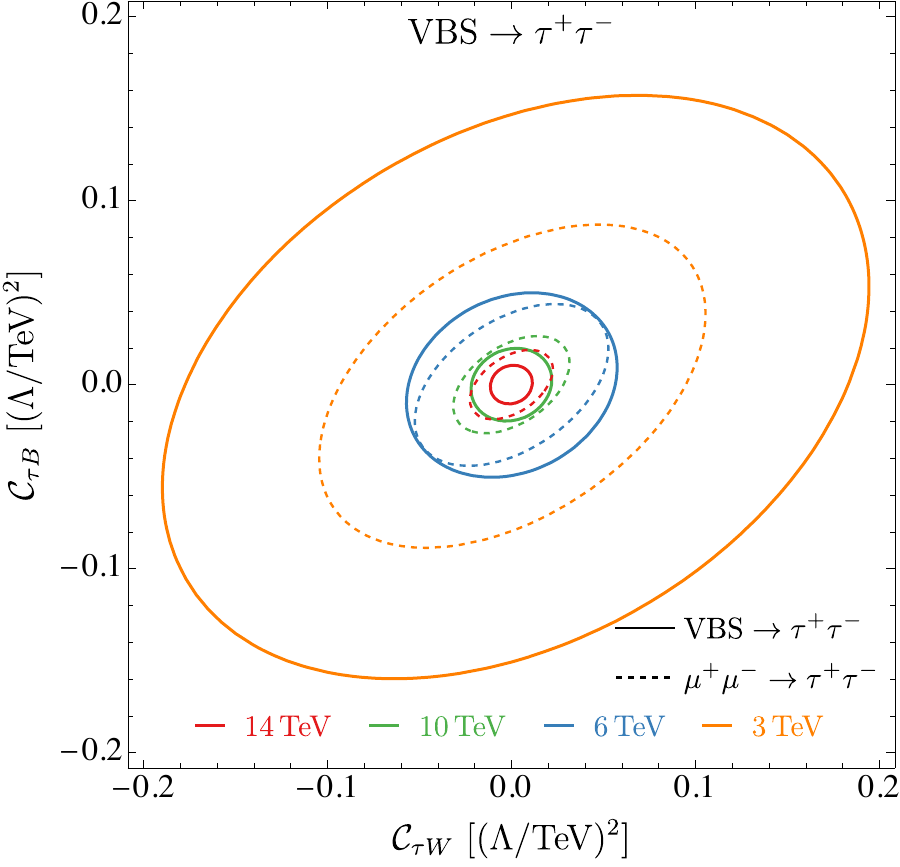}
\includegraphics[width=0.42\textwidth]{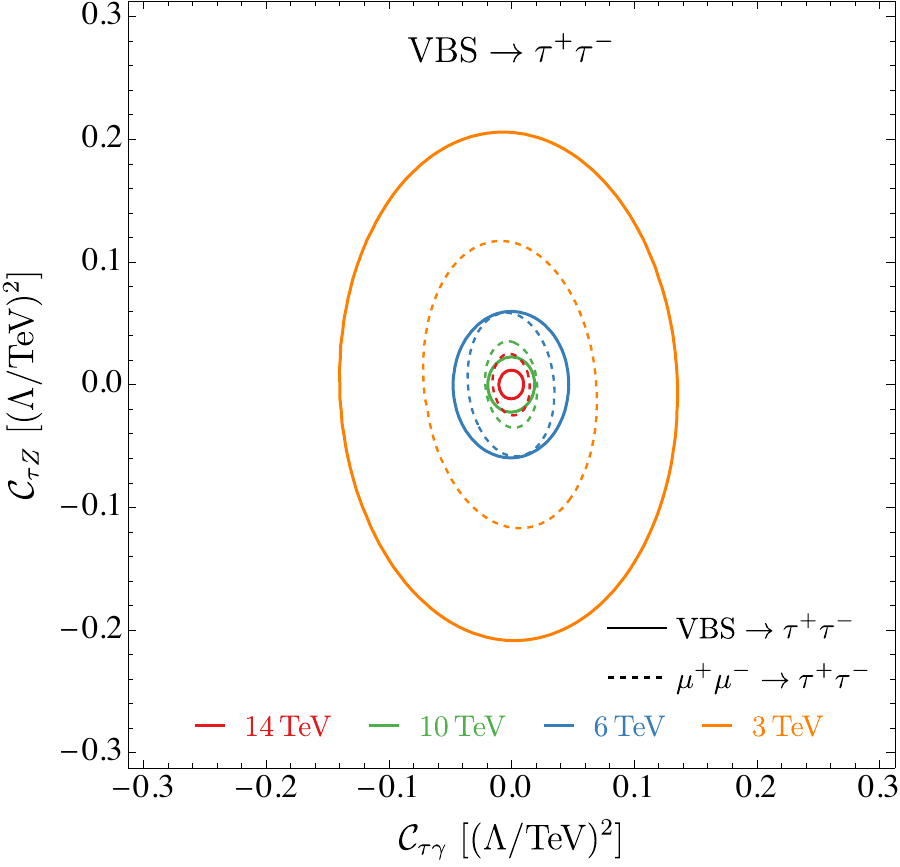}
\caption{Expected sensitivity reach on the $\tau$ dipole Wilson coefficients at the 95\% CL in the (left) ($\mathcal{C}_{\tau W}, \mathcal{C}_{\tau B}$) plane and (right) in the ($\mathcal{C}_{\tau \gamma}, \mathcal{C}_{\tau Z}$) plane, derived from the $\text{VBS}\to\tau^+\tau^-$ process at a multi-TeV muon collider for center-of-mass energies $\sqrt{s}=3,\,6,\,10$, and 14 TeV. In each plot, the solid contours represent the VBS sensitivities, while the dashed contours show the $\mu^+\mu^-\to\tau^+\tau^-$ benchmark for comparison.} \label{fig:muC_contour_VBF2t}
\end{figure}

To display the interplay of the two gauge-invariant operators, Fig.~\ref{fig:muC_contour_VBF2t} presents the 95\% CL contours in both the $(\mathcal C_{\tau W},\mathcal C_{\tau B})$ and $(\mathcal C_{\tau\gamma},\mathcal C_{\tau Z})$ planes, with the lower-mass threshold optimized for $\Delta a_\tau$. The direct-annihilation contours are overlaid in dashed contours for comparison. At 3 and 6~TeV, the direct channel generally provides a stronger constraint, whereas the inclusive VBS sample becomes competitive and eventually more sensitive at energies around and above 10~TeV. This crossover is also apparent in the one-parameter bounds: at 10~TeV the VBS reach on $|\Delta a_\tau|$ is $7.8\times10^{-5}$, compared with $8.8\times10^{-5}$ from direct annihilation. The comparison illustrates why it is useful to retain the two kinematically distinct samples in the combined analysis, whose global sensitivity is discussed later in this section.

\subsection{Electroweak boson-associated tau production}
\label{sec:EW_associated}
\subsubsection{SMEFT description}
\label{sec:theory_mm2ttB}

At multi-TeV muon colliders, tau-pair production in association with an electroweak boson offers an additional way of accessing the dimension-six dipole interactions. Of particular interest is $\mu^+\mu^-\to\tau^+\tau^-H$: electroweak symmetry breaking relates the dipole coupling of a tau pair to a photon or $Z$ boson to an interaction involving the same fields and a physical Higgs boson. The latter produces a contact contribution to Higgs-associated tau production that has no counterpart in the renormalizable SM. This feature distinguishes the channel from direct $\tau^+\tau^-$ production and is ultimately responsible for its high-energy sensitivity.

\begin{figure}[htb]
\includegraphics[width=0.25\textwidth]{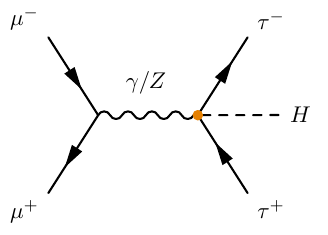}
\includegraphics[width=0.25\textwidth]{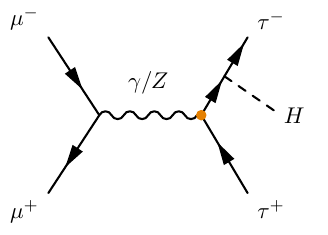}
\includegraphics[width=0.25\textwidth]{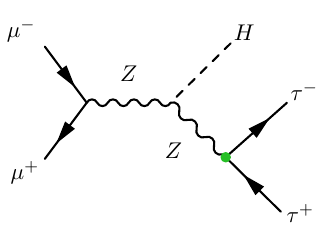}
\caption{Same as Fig.~\ref{fig:FD_mm2tt} but for $\mu^+\mu^- \to \tau^+\tau^- H$. The orange nodes denote vertices sensitive to both the anomalous electromagnetic ($\mathcal{O}_{\tau \gamma}$) and weak ($\mathcal{O}_{\tau Z}$) dipole moments, while the green node represents a vertex modified exclusively by $\mathcal{O}_{\tau Z}$. The first diagram exists only in the BSM scenario; for the second and third diagrams, the removal of the colored vertices recovers the pure SM contribution.
} \label{fig:FD_mm2tth}
\end{figure}

The representative leading-order topologies are displayed in Fig.~\ref{fig:FD_mm2tth}. The first diagram contains the dipole-induced $\tau^+\tau^-HV$ vertex ($V=\gamma,Z$). The other two diagrams are present already in the SM: in one, the Higgs is emitted from a final-state tau line; in the other, it is produced in association with a $Z$ boson which subsequently decays to $\tau^+\tau^-$. Dipole insertions can also modify the latter topologies. It is useful to distinguish their dynamical suppressions before assessing the relative size of these contributions.

Radiation of a Higgs boson from a tau line is controlled by the small Yukawa coupling $y_\tau=\sqrt{2}m_\tau/v\simeq0.01$. The Higgsstrahlung topology, instead, is concentrated near the $Z\to\tau^+\tau^-$ resonance and can be reduced by requiring the tau-pair invariant mass to lie above the resonance region. The dipole modifications of these two topologies inherit the corresponding Yukawa or resonant restrictions. By contrast, the four-point contact contribution is neither proportional to $y_\tau$ nor tied to the on-shell $Z$ region.

The interference pattern also reflects the different helicity structures. The contact amplitude and the SM tau-Yukawa radiation amplitude can interfere without an additional $m_\tau/E_\tau$ helicity penalty, although this interference remains proportional to the small Yukawa coupling. Interference of the contact amplitude with the SM $Z$-decay topology is chirality suppressed. Consequently, after suppressing the resonant background, the dominant dipole contribution at high energy is the square of the contact amplitude, rather than its interference with the SM.

This behavior can be made explicit in the limit $\sqrt{s}\gg M_Z\gg m_\tau$. Retaining the leading contact contribution quadratic in the dipole coefficients, the differential cross section takes the form

\begin{align}
\frac{{\rm d}\sigma_{\tau\tau H}}{{\rm d}x_1 dx_2}
\simeq &\frac{1}{192\pi^3}\frac{s}{\Lambda^4}
(1-x_1-x_2+2x_1x_2)
\Biggl[ |{\mathcal C}_{\tau \gamma}|^2 e^2 \nonumber \\
&- 2 e g_Z g_V {\rm Re}({\mathcal C}^*_{\tau Z} {\mathcal C}_{\tau \gamma}) \!+\! |{\mathcal C}_{\tau Z}|^2 g_Z^2 (g_A^2 \!+\! g_V^2)
\Biggr],
\label{eq:mm2tth}
\end{align}
Here $Q$ is the total incoming four-momentum and $k_i$ ($i=1,2$) are the tau four-momenta; the Dalitz variables are defined by $x_i=2Q\cdot k_i/Q^2$. Equation~\eqref{eq:mm2tth} exhibits both the quadratic dependence on the electromagnetic and weak dipole coefficients and the characteristic $s/\Lambda^4$ enhancement typical of dimension-6 effective interactions~\cite{Franceschini:2017xkh,Buttazzo:2020uzc,Buttazzo:2020ibd}. As in direct pair production, the mixed $\gamma$--$Z$ contribution is proportional to the small vector coupling $g_V$, so the corresponding interference between the two dipole directions is limited.

The origin of the energy enhancement is especially transparent when this process is compared with $\mu^+\mu^-\to\tau^+\tau^-$. In the direct channel, the Higgs field in the dimension-six operator is replaced by its vacuum expectation value, and the leading quadratic contribution approaches a constant proportional to $v^2/\Lambda^4$. In Higgs-associated production, the physical Higgs is emitted directly through the four-point interaction; unlike the SM tau-radiation topology, the contact diagram does not contain an additional off-shell tau propagator. The increase of the three-body phase space with the collision energy therefore survives in the cross section, yielding the $s/\Lambda^4$ behavior in Eq.~\eqref{eq:mm2tth}. With the resonant $Z$ contribution removed, the SM tau-radiation rate is instead Yukawa suppressed and decreases at high energy. These different asymptotic behaviors motivate the scaling comparison in Table~\ref{tab:scaling}; the entries for the other associated channels will be discussed below.

\begin{table}[htb]
    \centering
    {
    \renewcommand{\arraystretch}{1.3}
    \setlength{\tabcolsep}{10pt}
    \begin{tabular}{cccc}
    \hline
    Process               &   $\sigma^{\rm SM}_{\mu\mu}$   &   $\sigma^{\rm Lin}_{\mu\mu}$   &   $\sigma^{\rm Quad}_{\mu\mu}$   \\
    \hline
    $\mu^+\mu^-\to\tau^+\tau^-$      &  $\frac{1}{s}$ & $\frac{1}{s}\frac{m_\tau v}{\Lambda^2}$ & $\frac{v^2}{\Lambda^4}$\\
    $\mu^+\mu^-\to\tau^+\tau^- H$    &  $\frac{m_\tau^2}{v^2}\frac{1}{s}$ & $\frac{m_\tau}{v}\frac{1}{\Lambda^2}$ & $\frac{s}{\Lambda^4}$\\
    $\mu^+\mu^-\to\tau^+\tau^- Z, \tau \nu_\tau W$    &  $\frac{1}{s}$ & $\frac{m_\tau}{v}\frac{1}{\Lambda^2}$  & $\frac{s}{\Lambda^4}$\\
    \hline
    \end{tabular}
    }
    \caption{Asymptotic energy scaling behavior of the SM, linear, and quadratic cross-section components for direct and EW boson associated $\tau$ flavor production processes. Note that the resonant $W/Z$-decay contributions in the $2 \to 3$ gauge boson channels have been removed by imposing a dedicated invariant mass veto away from the gauge boson mass poles.}
    \label{tab:scaling}
\end{table}

The same gauge-invariant dipole operators also contribute to the production of a tau pair with a $Z$ boson and to charged-current associated production,
\begin{equation}
 \mu^+\mu^-\to\tau^+\tau^-Z,\qquad
 \mu^+\mu^-\to W^+\tau^-\bar\nu_\tau\quad\text{or}\quad W^-\tau^+\nu_\tau.
\end{equation}
Their relation to the Higgs channel is most readily understood before considering the individual diagrams. The $SU(2)_L$ structure of the operators in Sec.~\ref{sec:Theory} generates interactions with the electroweak Goldstone modes in addition to those involving the physical Higgs. At energies large compared with $M_W$ and $M_Z$, the Goldstone-boson equivalence theorem relates amplitudes involving these modes to amplitudes with longitudinally polarized $W$ and $Z$ bosons, up to corrections suppressed by the gauge-boson masses divided by the characteristic energy. Thus the energy growth of the leading quadratic dipole contribution found for $\tau^+\tau^-H$ production also occurs in the longitudinal components of the two gauge-boson-associated channels. As summarized in Table~\ref{tab:scaling}, the corresponding high-energy behavior is $\sigma^{\rm Quad}\propto s/\Lambda^4$ once resonant contributions are excluded.

The two channels nevertheless probe different combinations of the SMEFT parameters. The neutral-current process $\mu^+\mu^-\to\tau^+\tau^-Z$ receives contributions involving both $\mathcal C_{\tau B}$ and $\mathcal C_{\tau W}$, equivalently the electromagnetic and weak dipole combinations introduced in Sec.~\ref{sec:Theory}. By contrast, the charged-current dipole interaction relevant to $\mu^+\mu^-\to W^\pm\tau^\mp\nu_\tau$ originates from $\mathcal C_{\tau W}$ alone. This distinction makes the charged-current channel complementary to $\tau^+\tau^-H$ and $\tau^+\tau^-Z$ when constraining the two gauge-invariant operators. In addition, the $\tau^+\tau^-Z$ final state has a direct experimental connection to Higgs-associated production: a hadronically decaying $Z$ can mimic a Higgs candidate and hence provides an irreducible physics background once misidentification is taken into account.

\begin{figure}[htb]
\includegraphics[width=0.24\textwidth]{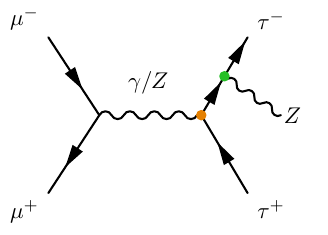}
\includegraphics[width=0.24\textwidth]{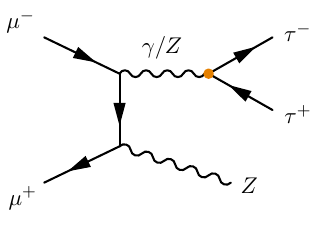}
\caption{Same as Fig.~\ref{fig:FD_mm2tth} but for $\mu^+\mu^- \to \tau^+\tau^- Z$.
} 
\label{fig:FD_mm2ttz}
\end{figure}

\begin{figure}[htb]
\includegraphics[width=0.19\textwidth]{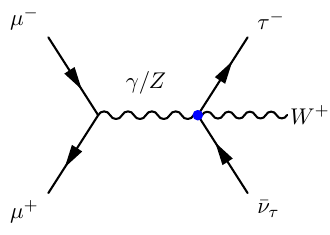}
\includegraphics[width=0.19\textwidth]{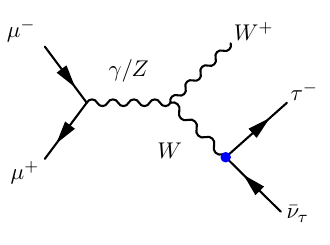}
\includegraphics[width=0.19\textwidth]{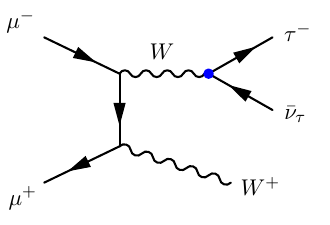}
\includegraphics[width=0.19\textwidth]{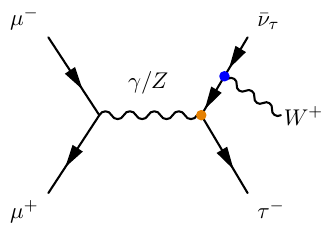}
\includegraphics[width=0.19\textwidth]{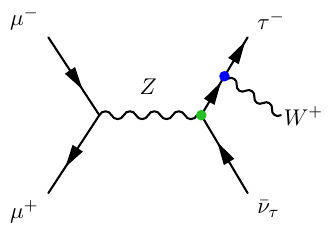}
\caption{Same as Fig.~\ref{fig:FD_mm2tth} but for $\mu^+\mu^- \to W^\pm \tau^\mp \nu_\tau$, where the blue nodes represent vertices modified exclusively by the $\mathcal{O}_{\tau W}$ operator. The leftmost diagram represents a purely BSM contribution; for the remaining diagrams, the removal of the colored vertex modifications recovers the corresponding SM interactions.
} \label{fig:FD_mm2wtv}
\end{figure}

The representative leading-order diagrams are shown in Figs.~\ref{fig:FD_mm2ttz} and \ref{fig:FD_mm2wtv}. The highlighted insertions in the neutral-current diagrams modify the tau dipole interactions, while the blue insertions in the charged-current diagrams correspond specifically to $\mathcal O_{\tau W}$. Both processes also contain substantial SM contributions that do not share the high-energy behavior of the dipole-induced contact terms. In particular, the $\tau^+\tau^-Z$ sample receives a resonant contribution from $Z\to\tau^+\tau^-$, which can be reduced by selecting events with tau-pair invariant masses away from the $Z$ pole. The analogous charged-current contamination involves $W\to\tau\nu_\tau$ and is more difficult to separate because the neutrino is not observed.

In principle, the charged-current topology can be constrained by reconstructing the missing neutrino momentum from the visible tau and $W$ kinematics. Such a reconstruction is sensitive to the detector resolution and to additional unobserved longitudinal momentum from initial-state radiation, so its practical effectiveness is less straightforward than an invariant-mass veto on a fully visible tau pair. 
The numerical analysis therefore adopts the benchmark requirements $m(\tau^+\tau^-)>\sqrt{s}/10$ for the neutral-current channel and $m(\tau\nu_\tau)>\sqrt{s}/10$ for the charged-current channel.
These requirements select the high-energy regions relevant for the dipole signal; their experimental implementation, especially for the neutrino-associated final state, warrants a dedicated detector-level study.

\begin{figure}
\includegraphics[width=0.24\textwidth]{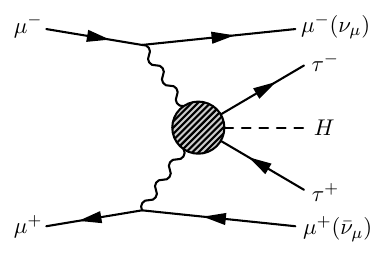}
\includegraphics[width=0.24\textwidth]{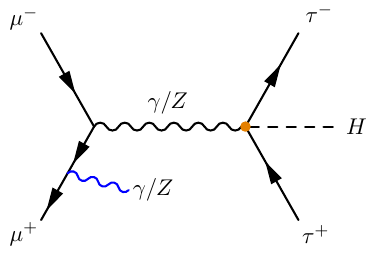}
\includegraphics[width=0.24\textwidth]{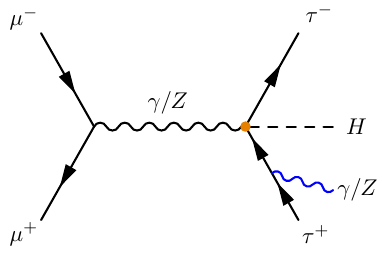}
\caption{Representative Feynman diagrams for the $2\to5$ multi-body final states $\mu^+\mu^-\to \tau^+\tau^- H +\mu^+\mu^-$ ($\nu_\ell \bar{\nu}_\ell$). The leftmost diagram illustrates the VBS topology. The second and third diagrams depict the BSM contact configurations, where an additional photon or $Z$ boson (highlighted in blue) is emitted from either the initial-state muon or the final-state $\tau$-lepton lines, sequentially splitting into a pair of muons or neutrinos.}
\label{fig:FD_VBS2tth_assoc}
\end{figure}

Multi-body production provides a further class of diagrams with the same underlying interactions. For example, the Higgs-associated final state can be accompanied by a muon pair or by a neutrino pair,
\begin{equation}
 \mu^+\mu^-\to\tau^+\tau^-H\,\mu^+\mu^-,\qquad
 \mu^+\mu^-\to\tau^+\tau^-H\,\nu_\ell\bar\nu_\ell.
\end{equation}
Representative $2\to5$ topologies are displayed in Fig.~\ref{fig:FD_VBS2tth_assoc}. They include genuine vector-boson-scattering configurations as well as contact-production diagrams supplemented by the radiation of a photon or $Z$ boson that subsequently produces the additional fermion pair.
These radiative contributions probe the same dipole
Wilson coefficients as the $2\to3$ signal, without
introducing sensitivity to additional independent
operators.

In the setup of the present study, these extra final states have much smaller dipole-sensitive rates than the leading associated-production channels. We consequently do not include them in the numerical sensitivity projections. 
The phenomenological analysis in the following subsection focuses on the three $2\to3$ associated-production channels.

\subsubsection{Phenomenological analysis}
\label{sec:Pheno_mm2tth}

We begin with the Higgs-associated channel, $\mu^+\mu^-\to\tau^+\tau^-H$, for which the dipole-induced four-point interaction identified above provides the dominant high-energy BSM contribution. The most useful events are those in which the tau pair and the Higgs carry large momenta and are produced in the annihilation regime. Our kinematic selection therefore supplements the baseline angular acceptance, object isolation and ISR treatment described in Sec.~\ref{sec:analysis_framework} with
\begin{equation}
\label{eq:mmtth_selection}
 m(\tau^+\tau^-H)>0.8\sqrt{s},\qquad
 m(\tau^+\tau^-)>\frac{\sqrt{s}}{10},\qquad
 p_T(\tau^\pm,H)>\frac{\sqrt{s}}{10}.
\end{equation}
The first requirement retains events with a three-body invariant mass close to the nominal collision energy, thereby reducing contamination from lower-energy configurations after ISR. The lower bound on $m(\tau^+\tau^-)$ removes much of the resonant $Z\to\tau^+\tau^-$ contribution to Higgsstrahlung, while the transverse-momentum thresholds preferentially select the hard kinematics characteristic of the contact interaction. 
These conditions define the kinematic selections adopted for this channel;
they should be understood as production-level selections rather than a detector-optimized reconstruction strategy.

We identify the Higgs through $H\to b\bar b$, taking $\mathrm{BR}(H\to b\bar b)=58.24\%$~\cite{LHCHiggsCrossSectionWorkingGroup:2016ypw}. The assumed single-jet $b$-tagging efficiency is $\epsilon_b=0.8$. With a requirement of at least one tagged $b$ jet from the decay, the corresponding Higgs-tagging efficiency is
\begin{equation}
\epsilon_{H\to b\bar b}^{\rm tag}=1-(1-\epsilon_b)^2=0.96.
\end{equation}
The signal yield must additionally include the Higgs branching fraction and the two-tau reconstruction efficiency $\epsilon_\tau^2$ defined in Sec.~\ref{sec:analysis_framework}. The most relevant physics background in the adopted setup is $\mu^+\mu^-\to\tau^+\tau^-Z$, followed by $Z\to b\bar b$ with $\mathrm{BR}(Z\to b\bar b)=15.12\%$. 
We assume a benchmark misidentification probability of $\epsilon_{H/Z}=15\%$ for a $Z$ candidate misidentified as a Higgs boson.
This contribution is treated separately from the irreducible SM $\tau^+\tau^-H$ rate. More realistic event-by-event separation of the Higgs and $Z$ peaks and reducible backgrounds requires a detector-level study and is not assumed here.

After the selections, the signal from the dimension-six operators is driven predominantly by the terms quadratic in their Wilson coefficients. 
This is the phenomenological consequence of the contact amplitude and its energy growth derived in Sec.~\ref{sec:theory_mm2ttB}: the SM contribution is suppressed by the small tau Yukawa coupling once the resonant $ZH$ topology is reduced. In the background-dominated limit of the counting analysis, with $S\ll B$, the significance scales schematically as
\begin{equation}
\label{eq:significance_tth_scaling}
 \mathcal S\simeq\frac{S}{\sqrt B}\propto
 \frac{s^2}{\Lambda^4}\left|\mathcal C_{\tau V}\right|^2,
\end{equation}
where the estimate uses $\sigma_{\rm Quad}\propto s/\Lambda^4$, $\sigma_{\rm SM}\propto1/s$ and the luminosity prescription $\mathcal L\propto s$. The notation $\mathcal C_{\tau V}$ here represents the relevant combination of electromagnetic and weak dipole coefficients, not a single independent operator. Thus, under these assumptions, the bound on a Wilson coefficient divided by $\Lambda^2$ improves approximately as $1/s$, faster than in direct tau-pair production. This scaling refers to the benchmark counting analysis and should not be extrapolated without qualification to a systematics-dominated regime.

\begin{figure}[htb]
\centering
\includegraphics[width=0.46\textwidth]{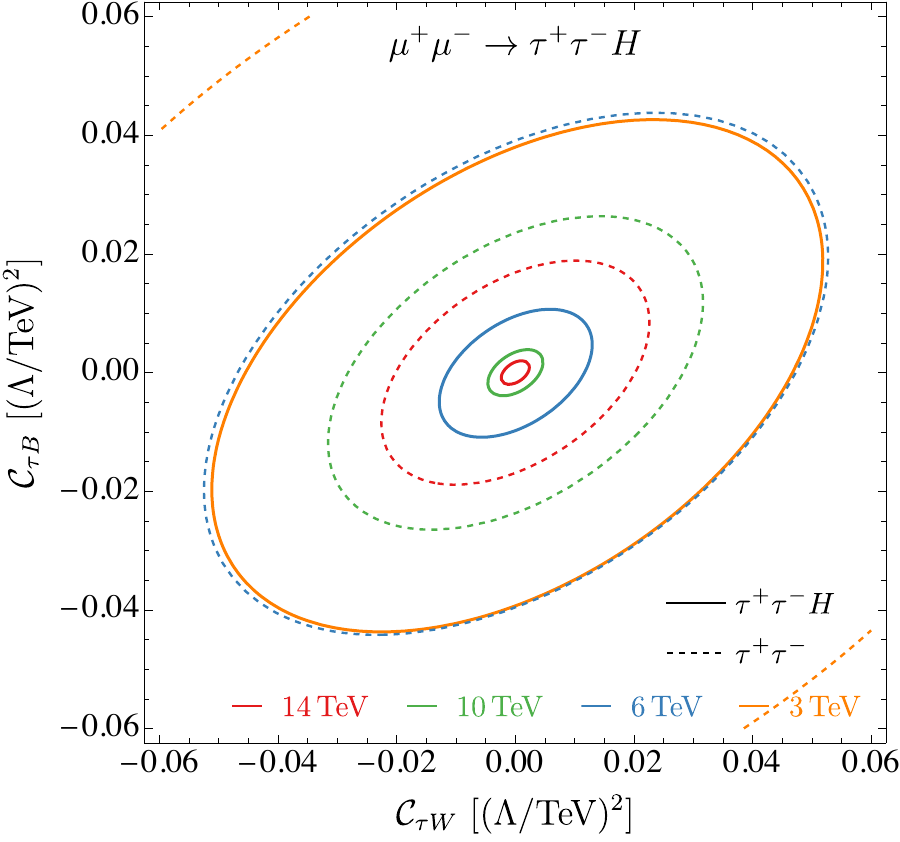}
\includegraphics[width=0.46\textwidth]{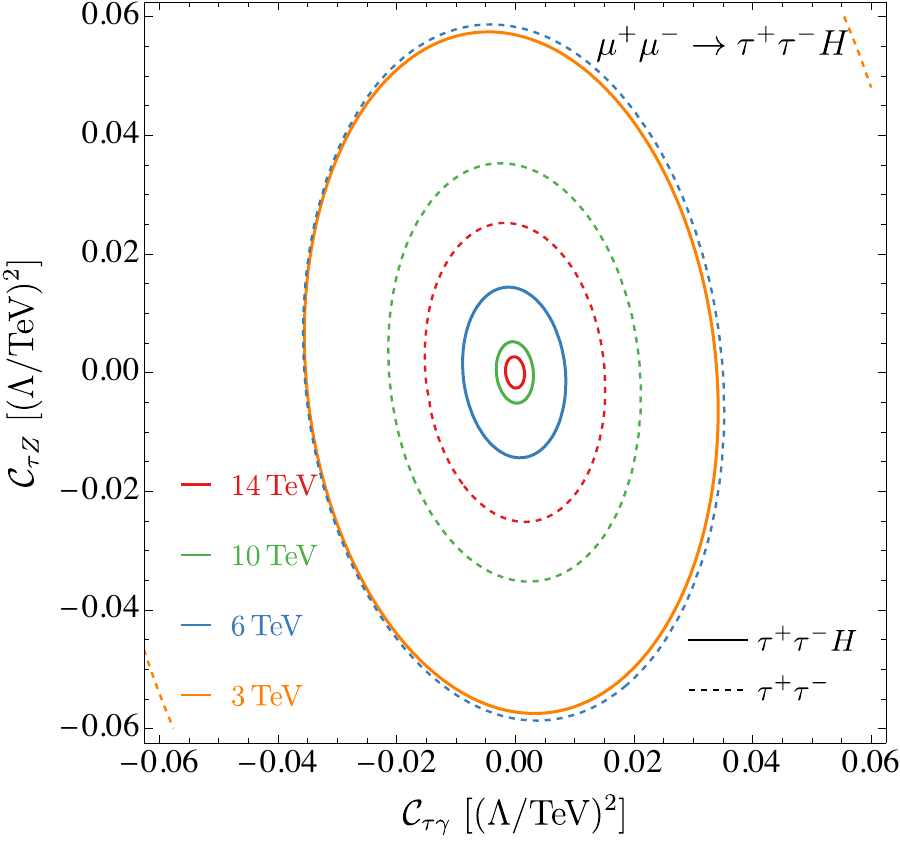}
\caption{Projected 95\% CL sensitivity to the tau dipole Wilson coefficients from $\mu^+\mu^-\to\tau^+\tau^-H$ at $\sqrt{s}=3,\,6,\,10$, and 14~TeV. The left panel shows the $(\mathcal C_{\tau W},\mathcal C_{\tau B})$ basis and the right panel the $(\mathcal C_{\tau\gamma},\mathcal C_{\tau Z})$ basis. Solid contours represent Higgs-associated production, while dashed contours display the direct $\mu^+\mu^-\to\tau^+\tau^-$ benchmark.}
\label{fig:muC_contour_mmtth}
\end{figure}

The resulting 95\% CL contours are shown in Fig.~\ref{fig:muC_contour_mmtth}. The two parameter bases emphasize complementary features: the $(\mathcal C_{\tau W},\mathcal C_{\tau B})$ plane displays the gauge-invariant operator combinations, whereas in the $(\mathcal C_{\tau\gamma},\mathcal C_{\tau Z})$ basis the contours are nearly aligned with the coordinate axes. As in $\mu^+\mu^-\to\tau^+\tau^-$, the mixed photon--$Z$ term is suppressed by the small vector coupling of the tau to the $Z$ boson. The solid Higgs-associated contours tighten markedly as the energy increases and become substantially stronger than the dashed $\mu^+\mu^-\to\tau^+\tau^-$ reference. 
The resulting contours become substantially tighter with increasing collider energy and provide significantly stronger constraints than direct tau-pair production. The corresponding projected sensitivities are summarized in Table~\ref{tab:constraints_mmtth}.

\begin{table}[htb]
\centering
\renewcommand{\arraystretch}{1.18}
\setlength{\tabcolsep}{5pt}
\begin{tabular}{c|cccc}
\hline
$\sqrt{s}$ [TeV]
& 3
& 6
& 10
& 14
\\
\hline
$|\Delta a_\tau|$                    & $1.4\times10^{-4}$  & $3.4\times10^{-5}$  & $1.2\times10^{-5}$  & $6.3\times10^{-6}$  \\
$|\Delta a^Z_\tau|$                  & $2.2\times10^{-4}$  & $5.5\times10^{-5}$  & $2.0\times10^{-5}$  & $1.0\times10^{-5}$  \\
$|d_\tau|$ [$e\,{\rm cm}$]           & $7.5\times10^{-19}$ & $1.9\times10^{-19}$ & $6.8\times10^{-20}$ & $3.5\times10^{-20}$ \\
$|d^Z_\tau|$ [$e\,{\rm cm}$]         & $1.2\times10^{-18}$ & $3.1\times10^{-19}$ & $1.1\times10^{-19}$ & $5.7\times10^{-20}$ \\
\hline
\end{tabular}
\caption{Projected 95\% CL bounds on the tau electromagnetic and weak dipole moments from $\mu^+\mu^-\to\tau^+\tau^-H$ at the four muon-collider benchmarks, using the common analysis assumptions of Sec.~\ref{sec:analysis_framework}.}
\label{tab:constraints_mmtth}
\end{table}

At $\sqrt{s}=10$~TeV the projected one-dimensional bounds from this channel are $|\Delta a_\tau|<1.2\times10^{-5}$ and $|\Delta a_\tau^Z|<2.0\times10^{-5}$, together with $|d_\tau|<6.8\times10^{-20}\,e\,\mathrm{cm}$ and $|d_\tau^Z|<1.1\times10^{-19}\,e\,\mathrm{cm}$. The corresponding 14-TeV constraints improve to $6.3\times10^{-6}$, $1.0\times10^{-5}$, $3.5\times10^{-20}\,e\,\mathrm{cm}$ and $5.7\times10^{-20}\,e\,\mathrm{cm}$, respectively. These are statistical projections within the benchmark assumptions; their interpretation as bounds on the physical moments follows the matching conventions of Sec.~\ref{sec:Theory}. The comparatively small SM rate, coupled to the energy-enhanced dipole signal, explains why this channel is expected to be the leading individual probe among the electroweak-boson-associated modes at multi-TeV muon colliders.

\paragraph*{$\mu^+\mu^-\to\tau^+\tau^-Z$.}
We next consider associated production of a tau pair with a $Z$ boson. As discussed in Sec.~\ref{sec:theory_mm2ttB}, the high-energy signal receives dipole-induced contributions that grow with the partonic collision energy, while the Standard Model rate is larger than in the Higgs-associated final state. This makes the $Z$ channel a useful, though generally less sensitive, probe of the same gauge-invariant operators. It also constitutes a relevant background to the Higgs analysis when the two bosons are reconstructed through similar hadronic final states.

For the present projections we include $Z$ decays to hadrons and to charged electrons or muons, using the branching fractions $\mathrm{BR}(Z\to\mathrm{hadrons})=69.91\%$ and $\mathrm{BR}(Z\to e^+e^-,\mu^+\mu^-)=6.73\%$~\cite{ParticleDataGroup:2024cfk}. The tau-pair selection follows that used for $\mu^+\mu^-\to\tau^+\tau^-H$ in this subsection, replacing the Higgs by the reconstructed $Z$ boson:
\begin{equation}
\label{eq:mmttz_selection}
 m(\tau^+\tau^-Z)>0.8\sqrt{s},\qquad
 m(\tau^+\tau^-)>\frac{\sqrt{s}}{10},\qquad
 p_T(\tau^\pm,Z)>\frac{\sqrt{s}}{10}.
\end{equation}
The high invariant-mass requirement selects the annihilation-dominated regime after ISR. The cut on $m(\tau^+\tau^-)$ suppresses the resonant $Z\to\tau^+\tau^-$ configurations, while the transverse-momentum thresholds preferentially retain hard associated production. All common angular-acceptance and tau-reconstruction assumptions are those of Sec.~\ref{sec:analysis_framework}. 
The quoted sensitivity includes the relevant $Z$-boson decay branching fractions, while no additional channel-specific detector response model is included.

\begin{figure}[htb]
\centering
\includegraphics[width=0.46\textwidth]{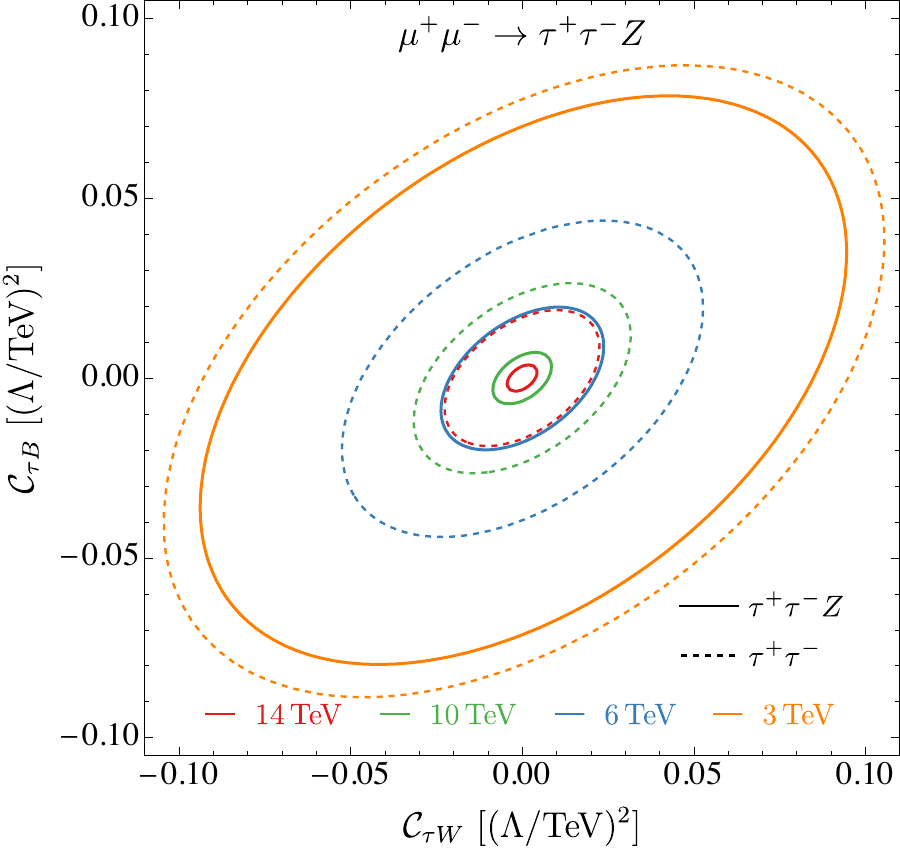}
\includegraphics[width=0.46\textwidth]{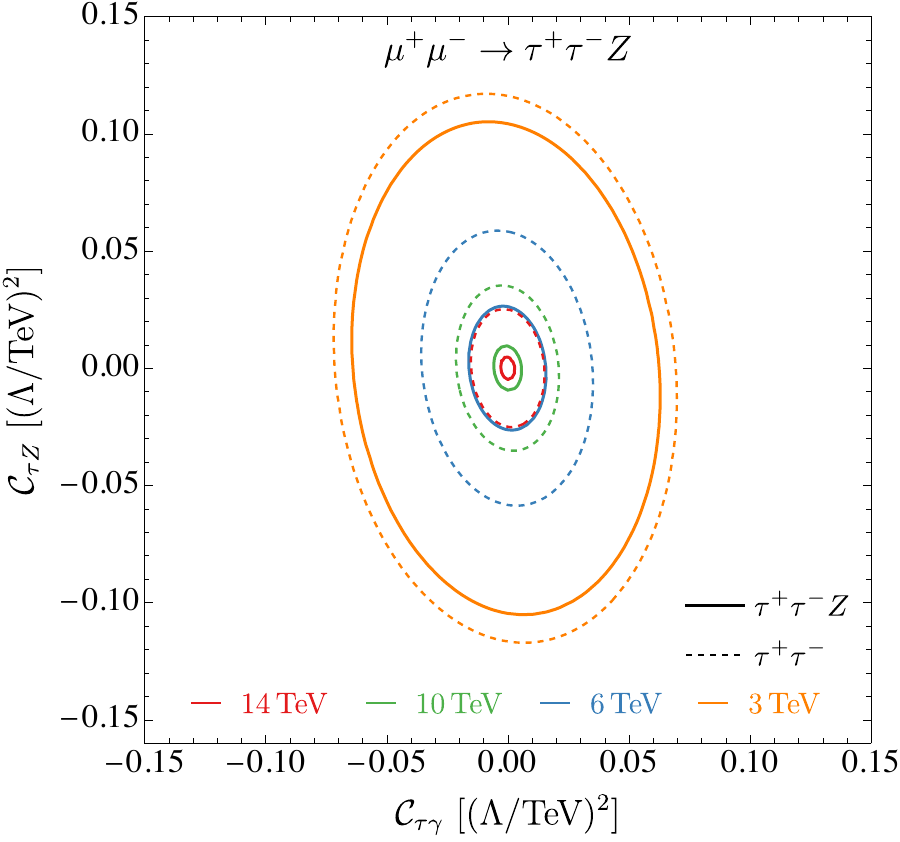}
\caption{Projected 95\% CL sensitivities from $\mu^+\mu^-\to\tau^+\tau^-Z$ at $\sqrt{s}=3,\,6,\,10$, and 14~TeV. The left and right panels show the $(\mathcal C_{\tau W},\mathcal C_{\tau B})$ and $(\mathcal C_{\tau\gamma},\mathcal C_{\tau Z})$ planes, respectively. Solid curves correspond to the $Z$-associated channel; dashed curves give the direct tau-pair benchmark. }
\label{fig:muC_contour_mmttz}
\end{figure}

The resulting confidence regions in the two Wilson-coefficient bases are displayed in Fig.~\ref{fig:muC_contour_mmttz}. In the physical $(\mathcal C_{\tau\gamma},\mathcal C_{\tau Z})$ plane, the two dipole directions remain weakly correlated, consistently with the high-energy contact-interaction structure discussed above. The sensitivity becomes stronger at larger $\sqrt{s}$ because the quadratic signal grows with energy whereas the SM background decreases. With the adopted luminosity scaling, this gives an approximate $1/s$ improvement in the bounds on the dipole coefficients divided by $\Lambda^2$, subject to the event-selection and background assumptions. The contours are less constraining than the Higgs-associated ones, reflecting the larger SM $\tau^+\tau^-Z$ contribution.

\begin{table}[htb]
\centering
\renewcommand{\arraystretch}{1.18}
\setlength{\tabcolsep}{5pt}
\begin{tabular}{c|cccc}
\hline
$\sqrt{s}$ [TeV]
& 3
& 6
& 10
& 14
\\
\hline
$|\Delta a_\tau|$                    & $2.6\times10^{-4}$  & $6.6\times10^{-5}$  & $2.4\times10^{-5}$  & $1.2\times10^{-5}$  \\
$|\Delta a^Z_\tau|$                  & $4.2\times10^{-4}$  & $1.1\times10^{-4}$  & $3.8\times10^{-5}$  & $2.0\times10^{-5}$  \\
$|d_\tau|$ [$e\,{\rm cm}$]           & $1.4\times10^{-18}$ & $3.6\times10^{-19}$ & $1.3\times10^{-19}$ & $6.7\times10^{-20}$ \\
$|d^Z_\tau|$ [$e\,{\rm cm}$]         & $2.4\times10^{-18}$ & $6.0\times10^{-19}$ & $2.2\times10^{-19}$ & $1.1\times10^{-19}$ \\
\hline
\end{tabular}
\caption{Projected 95\% CL bounds on the tau electromagnetic and weak dipole moments from $\mu^+\mu^-\to\tau^+\tau^-Z$ at the four muon-collider benchmark energies. Magnetic and electric dipole constraints are evaluated with real and imaginary Wilson coefficients, respectively, following the common statistical assumptions of Sec.~\ref{sec:analysis_framework}.}
\label{tab:constraints_mmttz}
\end{table}

Table~\ref{tab:constraints_mmttz} summarizes the projected 95\% CL sensitivities at the four benchmark energies. At $\sqrt{s}=10$~TeV, for instance, the one-dimensional bounds are
\begin{align}
\label{eq:mmttz_bounds_10tev}
|\Delta a_\tau|&<2.4\times10^{-5},\qquad\qquad\qquad
|d_\tau|<1.3\times10^{-19}\,e\,\mathrm{cm},\nonumber\\
|\Delta a_\tau^Z|&<3.8\times10^{-5}, \qquad\qquad\qquad
 |d_\tau^Z|<2.2\times10^{-19}\,e\,\mathrm{cm}.
\end{align}
These sensitivities are weaker than their $\tau^+\tau^-H$ counterparts but improve substantially with collider energy. The different signal and background compositions also provide a complementary check of the dipole interpretation. The projections remain statistical and are based on the production-level reconstruction assumptions described above; in particular, they are not a substitute for a dedicated study of hadronic $Z$ identification, reducible backgrounds or correlations between associated channels.

\paragraph*{$\mu^+\mu^-\to W^\pm\tau^\mp\nu_\tau$.}
The charged-current final state provides a third, experimentally distinct test of the dipole interactions. Unlike the neutral associated-production channels, its dipole contribution is generated directly by the gauge-invariant operator $\mathcal O_{\tau W}$, as explained in Sec.~\ref{sec:theory_mm2ttB}. It therefore probes a particular direction in the $(\mathcal C_{\tau W},\mathcal C_{\tau B})$ parameter space, while the transformation to the electromagnetic and weak basis maps this direction into a correlated combination of $\mathcal C_{\tau\gamma}$ and $\mathcal C_{\tau Z}$. This different operator dependence makes the channel useful in a combined analysis, even though its individual sensitivity is generally weaker than that of Higgs-associated production.

For the event selection we retain the hadronic $W$ decay mode, with $\mathrm{BR}(W\to\mathrm{hadrons})=67.41\%$~\cite{ParticleDataGroup:2024cfk}. The experimental treatment differs from that of $\tau^+\tau^-H$ and $\tau^+\tau^-Z$ because the neutrino is not observed. In particular, the full invariant mass $m(W\tau\nu_\tau)$ cannot be reconstructed straightforwardly, and the requirement $m(W\tau\nu_\tau)>0.8\sqrt{s}$ adopted for visible three-body final states has no direct experimental analogue. 
In the simulated event sample, however, the selected events remain concentrated near the full center-of-mass energy, even after including ISR.
We therefore retain the charged-current sample without imposing this global invariant-mass cut. 
In addition to the common acceptance and isolation criteria of Sec.~\ref{sec:analysis_framework}, we apply the following kinematic selections:
\begin{equation}
\label{eq:wtv_selection}
 m(W\tau)>\frac{\sqrt{s}}{10},\qquad
 m(\tau\nu_\tau)>\frac{\sqrt{s}}{10},\qquad
 p_T(\tau,W)>\frac{\sqrt{s}}{10}.
\end{equation}
The cut on $m(\tau\nu_\tau)$ reduces the resonant $W\to\tau\nu_\tau$ background, while the requirements on $m(W\tau)$ and the transverse momenta favor the hard configurations relevant to the dipole signal. The neutrino-dependent invariant mass is thus part of the benchmark kinematic selection, rather than a claim that this observable can be reconstructed without ambiguity in a realistic detector.

An approximate neutrino momentum may be inferred from momentum balance using the reconstructed visible objects, provided longitudinal momentum losses from ISR are negligible. At a muon collider this approximation must be treated with care, and the additional missing momentum from the subsequent decay of the tau introduces further complications. 
Accordingly, the results presented below are based on the production-level treatment described above and should not be interpreted as a validated 
detector-level reconstruction of the neutrino or the $W\tau\nu_\tau$ final state.

\begin{figure}[htb]
\centering
\includegraphics[width=0.46\textwidth]{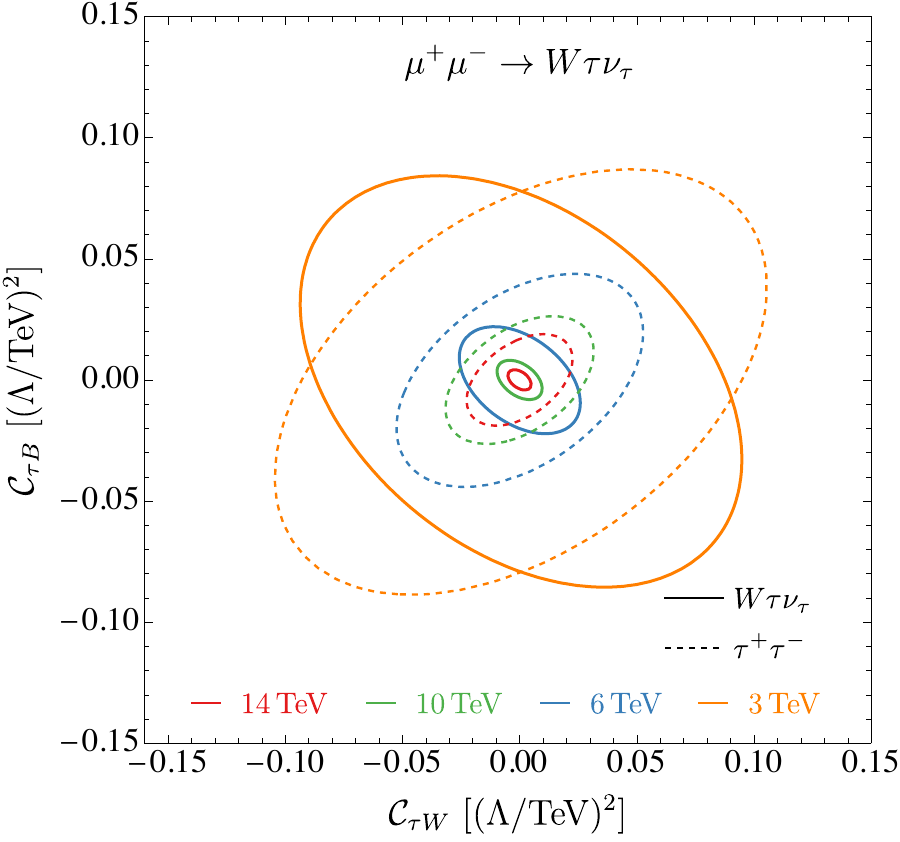}
\includegraphics[width=0.46\textwidth]{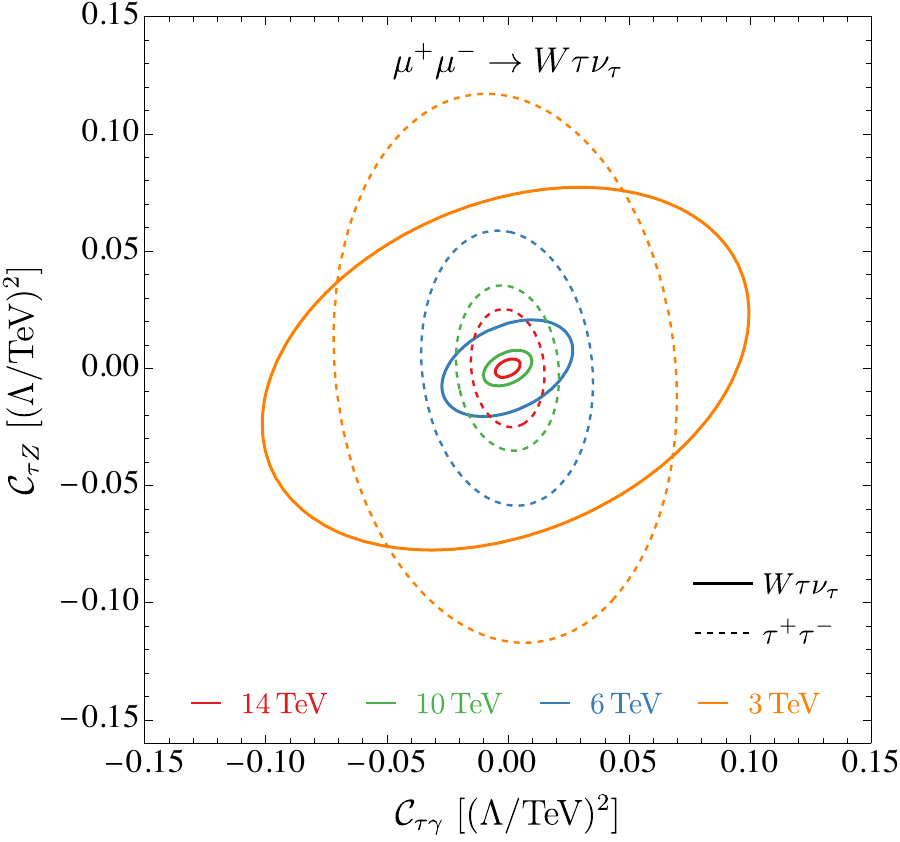}
\caption{Projected 95\% CL sensitivity to the tau dipole Wilson coefficients from the charged-current channel $\mu^+\mu^-\to W^\pm\tau^\mp\nu_\tau$ at $\sqrt{s}=3,\,6,\,10,$ and $14$~TeV. The left and right panels show the $(\mathcal C_{\tau W},\mathcal C_{\tau B})$ and $(\mathcal C_{\tau\gamma},\mathcal C_{\tau Z})$ planes. Solid contours correspond to charged-current associated production and dashed contours to the direct tau-pair benchmark. }
\label{fig:muC_contour_mmwtv}
\end{figure}

As for the other associated-production modes, the selected high-energy signal is predominantly quadratic in the dipole coefficients. The charged-current amplitude, however, contains structures associated specifically with $\mathcal O_{\tau W}$ in addition to the neutral-current dipole combinations. Consequently, its quadratic contribution generates a different pattern of coefficient dependence. This is visible in Fig.~\ref{fig:muC_contour_mmwtv}: in contrast to the nearly axis-aligned contours of the $\tau^+\tau^-H$ and $\tau^+\tau^-Z$ channels in the physical basis, the charged-current confidence regions are tilted. In the gauge basis they likewise exhibit a different orientation, making explicit the complementary information supplied by charged-current production. The solid contours improve with increasing collider energy and can 
be compared directly with the dashed $\mu^+\mu^-\to\tau^+\tau^-$ benchmarks shown in the same figure.

Table~\ref{tab:constraints_mmtvtw} summarizes the projected 95\% CL bounds at the four reference energies. The numerical results and the corresponding Wilson-coefficient contours are presented below.
At $\sqrt{s}=10$~TeV, the bounds read $|\Delta a_\tau|<4.1\times10^{-5}$, $|\Delta a_\tau^Z|<3.1\times10^{-5}$, $|d_\tau|<2.3\times10^{-19}\,e\,\mathrm{cm}$ and $|d_\tau^Z|<1.7\times10^{-19}\,e\,\mathrm{cm}$. By $14$~TeV they tighten to $2.1\times10^{-5}$, $1.6\times10^{-5}$, $1.2\times10^{-19}\,e\,\mathrm{cm}$ and $9.0\times10^{-20}\,e\,\mathrm{cm}$, respectively. As expected for the energy-enhanced quadratic contributions under the luminosity prescription used here, these projections improve rapidly as the collider energy increases.

\begin{table}[htb]
\centering
\renewcommand{\arraystretch}{1.18}
\setlength{\tabcolsep}{5pt}
\begin{tabular}{c|cccc}
\hline
$\sqrt{s}$ [TeV]
& 3
& 6
& 10
& 14
\\
\hline
$|\Delta a_\tau|$                    & $4.1\times10^{-4}$  & $1.1\times10^{-4}$  & $4.1\times10^{-5}$  & $2.1\times10^{-5}$  \\
$|\Delta a^Z_\tau|$                  & $3.1\times10^{-4}$  & $8.3\times10^{-5}$  & $3.1\times10^{-5}$  & $1.6\times10^{-5}$  \\
$|d_\tau|$ [$e\,{\rm cm}$]           & $2.3\times10^{-18}$ & $6.1\times10^{-19}$ & $2.3\times10^{-19}$ & $1.2\times10^{-19}$ \\
$|d^Z_\tau|$ [$e\,{\rm cm}$]         & $1.8\times10^{-18}$ & $4.7\times10^{-19}$ & $1.7\times10^{-19}$ & $9.0\times10^{-20}$ \\
\hline
\end{tabular}
\caption{Projected 95\% CL bounds on the electromagnetic and weak tau dipole moments from the $\mu^+\mu^-\to W^\pm\tau^\mp\nu_\tau$ channel at the four collider benchmark energies, using the common statistical assumptions of Sec.~\ref{sec:analysis_framework}.}
\label{tab:constraints_mmtvtw}
\end{table}

The relative importance of the three electroweak-associated channels can be understood by considering both their Standard Model rates and their operator dependence. The Higgs channel benefits from the smallest background and sets the strongest individual limits on the electromagnetic dipole moments. The $\tau^+\tau^-Z$ process has a similar contour geometry but a larger SM background, whereas $W\tau\nu_\tau$ probes a different combination of the underlying SMEFT parameters. At $10$~TeV, for example, the charged-current constraint on $|\Delta a_\tau^Z|$ is slightly stronger than the corresponding $\tau^+\tau^-Z$ result, while remaining weaker than the Higgs-associated limit. Thus, the main advantage of the charged-current measurement is not only its standalone reach but also its ability to help distinguish the two gauge-invariant dipole operators when multiple channels are considered together. The combined reach of all channels will be examined in the global comparison in Sec.~\ref{sec:global_muc}.

\subsection{Radiative Higgs decay}
\label{sec:Pheno_VBS2h}

The decay $H\to\tau^+\tau^-\gamma$ probes the electromagnetic $\tau$ dipole interaction in a setting distinct from high-energy $\tau$-pair production. Its sensitivity follows from the gauge-invariant origin of the dipole operator: after electroweak symmetry breaking, the same interaction that corrects the $\tau^+\tau^-\gamma$ vertex also generates a local $H\tau^+\tau^-\gamma$ coupling, as discussed in Sec.~\ref{sec:Theory}. A muon collider supplies Higgs bosons predominantly through electroweak vector-boson fusion, so the radiative decay can be examined in final states accompanied by forward muons or missing energy. Unlike the high-energy production channels discussed above, the interference contribution to this decay does not need to be suppressed relative to the quadratic dimension-six term.

\subsubsection{SMEFT description}
\label{sec:muc_h2ttg_smeft}

\begin{figure}[htb]
\centering
\includegraphics[width=0.25\textwidth]{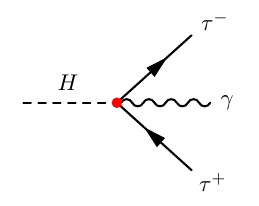}
\includegraphics[width=0.25\textwidth]{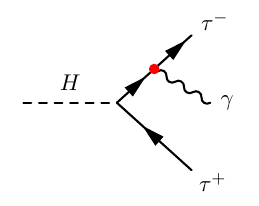}
\includegraphics[width=0.25\textwidth]{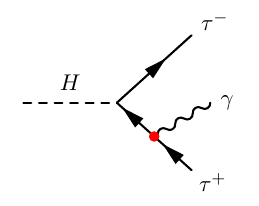}
\caption{Representative tree-level contributions to $H\to\tau^+\tau^-\gamma$. The first topology contains the dimension-six $H\tau\tau\gamma$ contact interaction, while photon emission from the $\tau$ legs receives both SM and dipole-induced contributions. The marked vertices indicate insertions of the electromagnetic dipole operator $\mathcal O_{\tau\gamma}$.}
\label{fig:muc_h2tta_diagrams}
\end{figure}

At tree level, the relevant dipole amplitude contains a contact contribution from $H\tau^+\tau^-\gamma$, together with photon emission from the external $\tau$ legs in the presence of a dipole insertion. These structures are displayed in Fig.~\ref{fig:muc_h2tta_diagrams}. The SM amplitude instead arises from photon radiation accompanying the Higgs Yukawa coupling to $\tau$ leptons. Consequently, the interference between the SM and contact amplitudes is proportional to the $\tau$ Yukawa coupling but is not subject to the additional high-energy suppression characteristic of the interference between vector-current and dipole amplitudes in $\mu^+\mu^-\to\tau^+\tau^-$. The radiative decay is thus particularly sensitive to the real part of the electromagnetic Wilson coefficient.

For illustration, the new-physics shift in the radiative partial width can be organized into terms linear and quadratic in the dipole coefficient. In the leading-order approximation adopted in Ref.~\cite{Buttazzo:2026ulg}, it reads
\begin{equation}
\Gamma^{\rm NP}(H\to\tau^+\tau^-\gamma)
=\frac{e\,y_\tau M_H^3}{64\pi^3}\,
  \frac{\operatorname{Re}\mathcal C_{\tau\gamma}}{\Lambda^2}
 +\frac{M_H^5}{768\pi^3}\,
  \frac{|\mathcal C_{\tau\gamma}|^2}{\Lambda^4}\,.
\label{eq:muc_higgs_decay_width}
\end{equation}
Here $y_\tau$ denotes the $\tau$ Yukawa coupling in the conventions of that calculation. The first term represents the SM--dipole interference, whereas the second originates from the squared dipole amplitude. Because the CP-odd electric-dipole interaction is associated with 
$\operatorname{Im}\mathcal C_{\tau\gamma}$, it does not contribute linearly to the CP-even decay rate. Therefore, the decay-rate sensitivity to $d_\tau$ begins at quadratic order, in contrast to the leading linear dependence on $\Delta a_\tau \propto \operatorname{Re}\mathcal C_{\tau\gamma}$. 
Equation~\eqref{eq:muc_higgs_decay_width} describes the
dipole dependence at the decay level, whereas the numerical
projections presented below are based on the full production
processes, including the relevant kinematic selections.

\begin{figure}[htb]
\includegraphics[width=0.32\textwidth]{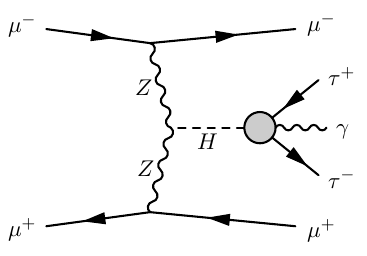}
\includegraphics[width=0.32\textwidth]{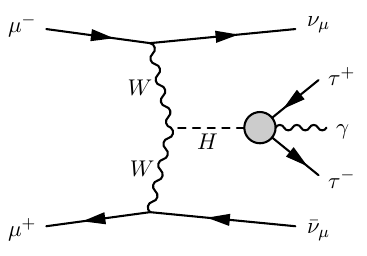}
\caption{Representative diagrams for the processes $\mu^+\mu^- \to \mu^+\mu^-\tau^+\tau^-\gamma$ and $\mu^+\mu^- \to \nu_\mu\bar{\nu}_\mu\tau^+\tau^-\gamma$. The large gray circle marks the $H \to \tau^+\tau^-\gamma$ decay sub-process.
}\label{fig:FD_VBS2tth}
\end{figure}

The collider environment is incorporated through the two vector-boson-fusion production modes, $\mu^+\mu^-\to H\nu_\mu\bar\nu_\mu$ and $\mu^+\mu^-\to H\mu^+\mu^-$. At $\sqrt{s}=10$~TeV, the baseline estimates used in the present study give a $WW$-fusion cross section of approximately $1$~pb and a $ZZ$-fusion rate about one order of magnitude lower. For $\mathcal L=10$~ab$^{-1}$, the $WW$-fusion rate alone corresponds to roughly $10^7$ produced Higgs bosons before decay selection. To retain the non-resonant contributions to the same observable final states, we simulate $\mu^+\mu^-\to\tau^+\tau^-\gamma\nu_\ell\bar\nu_\ell$ and $\mu^+\mu^-\to\tau^+\tau^-\gamma\mu^+\mu^-$ rather than factorizing Higgs production and decay. In Fig.~\ref{fig:FD_VBS2tth}, the Higgs-mediated subset is indicated by the radiative decay subdiagram; other diagrams include photon radiation in continuum $\tau$-pair production. Their relative contributions, and the corresponding event selections, are examined next.

\subsubsection{Phenomenological analysis}
\label{sec:muc_h2ttg_pheno}

The radiative Higgs decay is studied through the complete $\tau^+\tau^-\gamma$ final states introduced above, rather than by treating the Higgs production rate and its branching fraction as separate observables. This choice includes the continuum amplitudes leading to the same final states, which must be suppressed before the dipole-induced decay can be isolated. Of particular importance is the neutral-current process $\mu^+\mu^-\to\mu^+\mu^-\tau^+\tau^-\gamma$: photon emission from charged external legs produces a sizable Standard Model contribution, with enhanced rates in soft and collinear regions. 
The charged-current final state with missing energy also contributes to the selected sample, with a relatively larger contribution from Higgs-mediated diagrams.

In addition to the common polar-angle and isolation requirements of Sec.~\ref{sec:analysis_framework}, we impose
\begin{equation}
\begin{aligned}
 p_T(\tau^\pm,\gamma) &> 50~\mathrm{GeV},\\
 \bigl|m(\tau^+\tau^-\gamma)-M_H\bigr| &< 10~\mathrm{GeV},\\
 \bigl|m(\tau^+\tau^-)-M_Z\bigr| &> 10~\mathrm{GeV}.
\end{aligned}
\label{eq:muc_h2ttg_selection}
\end{equation}
The transverse-momentum requirement removes the soft-photon region and rejects low-momentum visible particles. The window around the Higgs mass selects configurations compatible with $H\to\tau^+\tau^-\gamma$, while the last requirement suppresses resonant $Z\to\tau^+\tau^-$ production accompanied by photon radiation. In the simulation these selections substantially reduce the neutral-current continuum, so that the charged-current $WW$-initiated contribution becomes the dominant residual background. 
As throughout this study, kinematic selections are applied
at the generated-particle level; assessing the corresponding
reconstructed-mass resolution requires a dedicated treatment
of $\tau$ decays and detector response.

The relative sensitivity to electromagnetic and weak dipole operators is then controlled by two effects. First, the contact interaction responsible for the radiative Higgs decay depends on $\mathcal C_{\tau\gamma}$, making this process primarily a probe of the electromagnetic dipole. Second, vetoing the $Z$ pole reduces contributions associated with the weak dipole interaction in non-resonant production. The residual reach along the $\mathcal C_{\tau Z}$ direction is therefore substantially weaker. Moreover, the SM--dipole interference in the decay rate is linear in $\operatorname{Re}\mathcal C_{\tau\gamma}$, whereas the dependence on $\operatorname{Im}\mathcal C_{\tau\gamma}$ begins at quadratic order for the CP-even event-rate observable. This difference leads to a stronger reach in $\Delta a_\tau$ than in $d_\tau$ under the assumptions used here.

\begin{figure}[htb]
\centering
\includegraphics[width=0.42\textwidth]{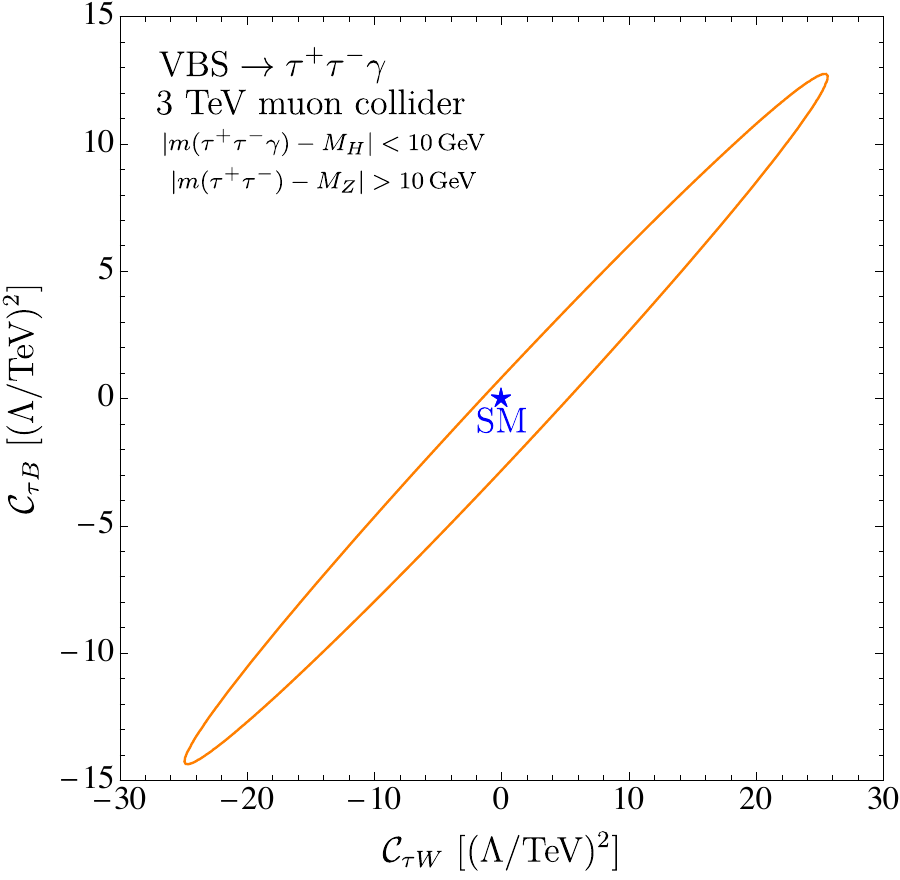}
\includegraphics[width=0.41\textwidth]{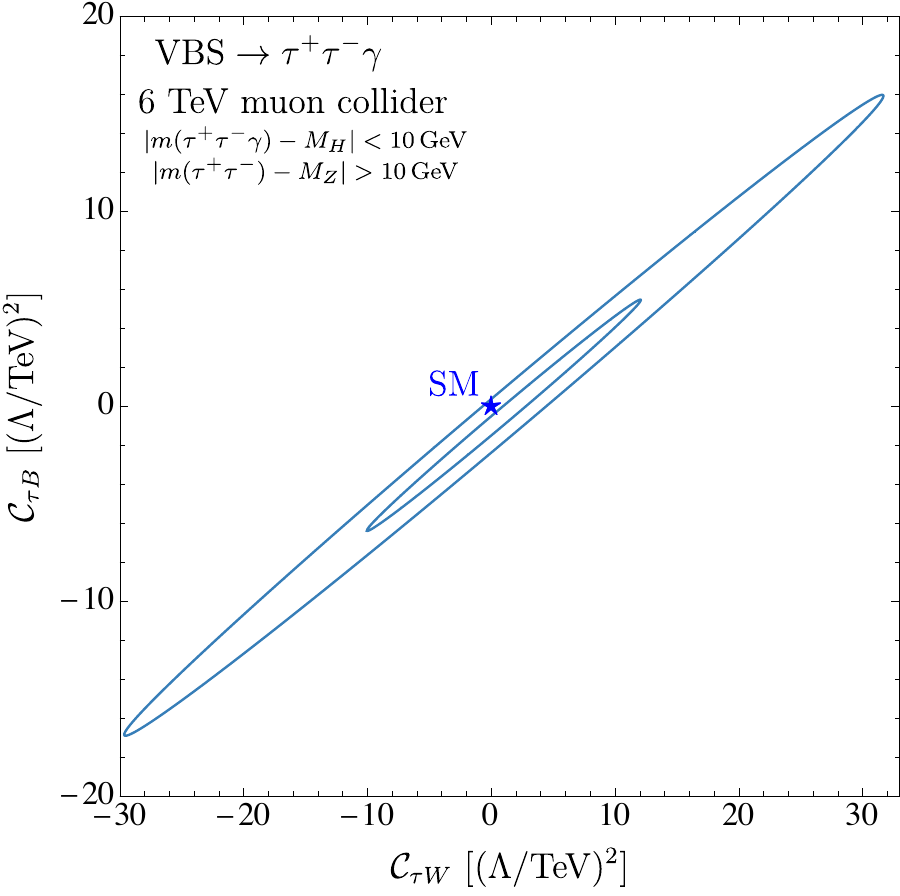}
\includegraphics[width=0.42\textwidth]{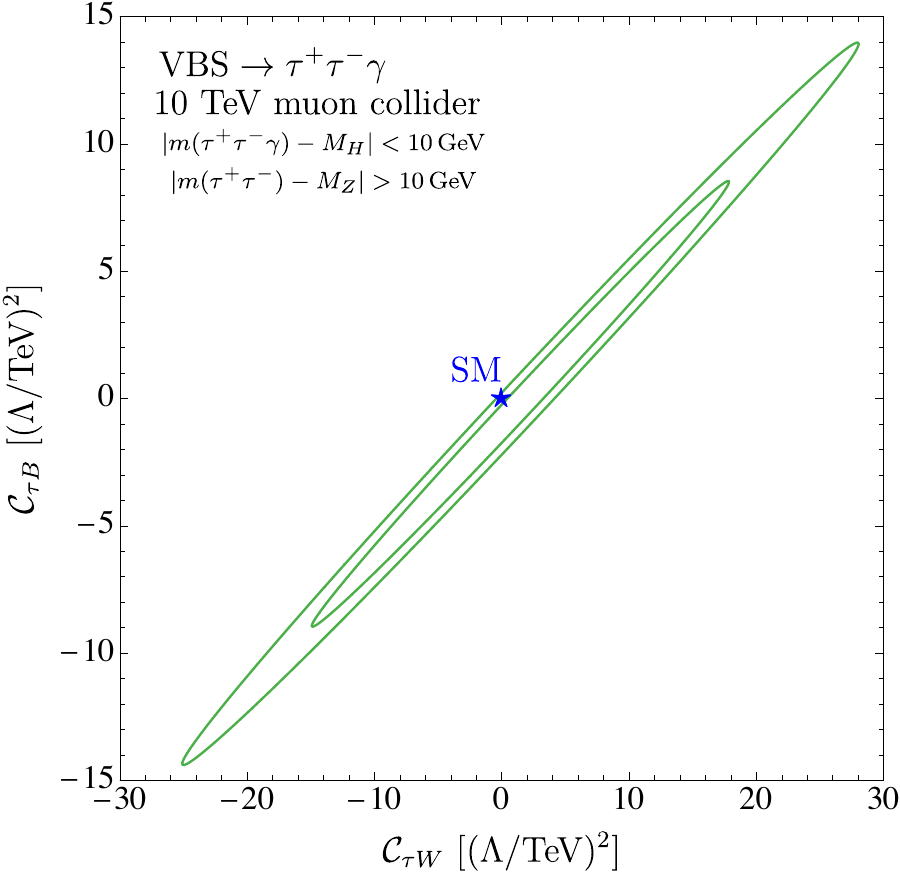}
\includegraphics[width=0.42\textwidth]{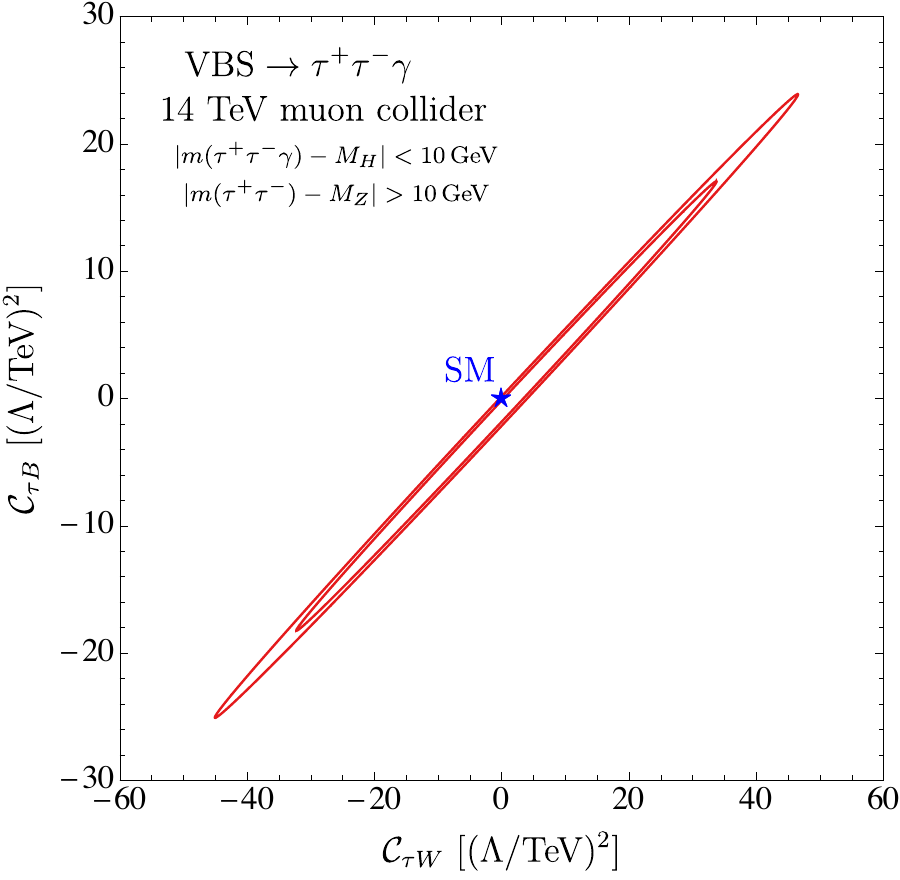}
\caption{Projected 95\% CL sensitivity contours for the $\tau$ dipole Wilson coefficients in the $(\mathcal{C}_{\tau W}, \mathcal{C}_{\tau B})$ plane from $H\to\tau^+\tau^-\gamma$ decay processes at a multi-TeV muon collider. The orange, blue, green, and red contours illustrate the sensitivities derived at center-of-mass energies $\sqrt{s}=3,\,6,\,10$, and 14 TeV. The SM prediction is indicated by the blue star at the origin.} \label{fig:contours_VBFh2tta_1}
\end{figure}

\begin{figure}[htb]
\centering
\includegraphics[width=0.42\textwidth]{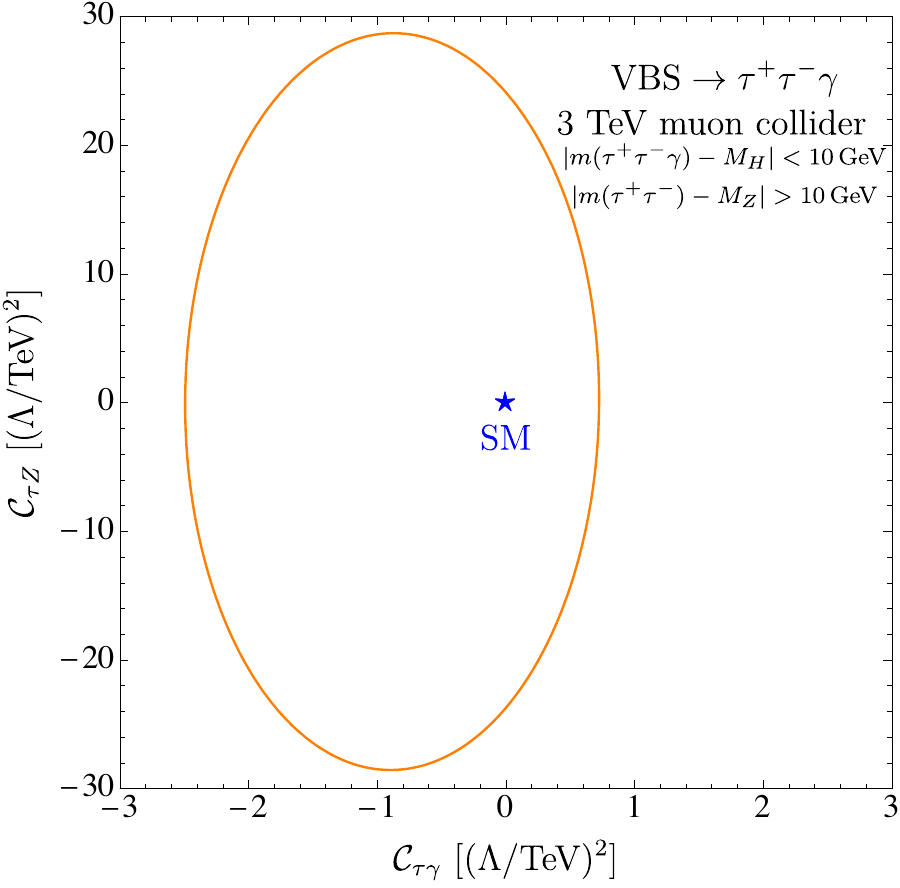}
\includegraphics[width=0.42\textwidth]{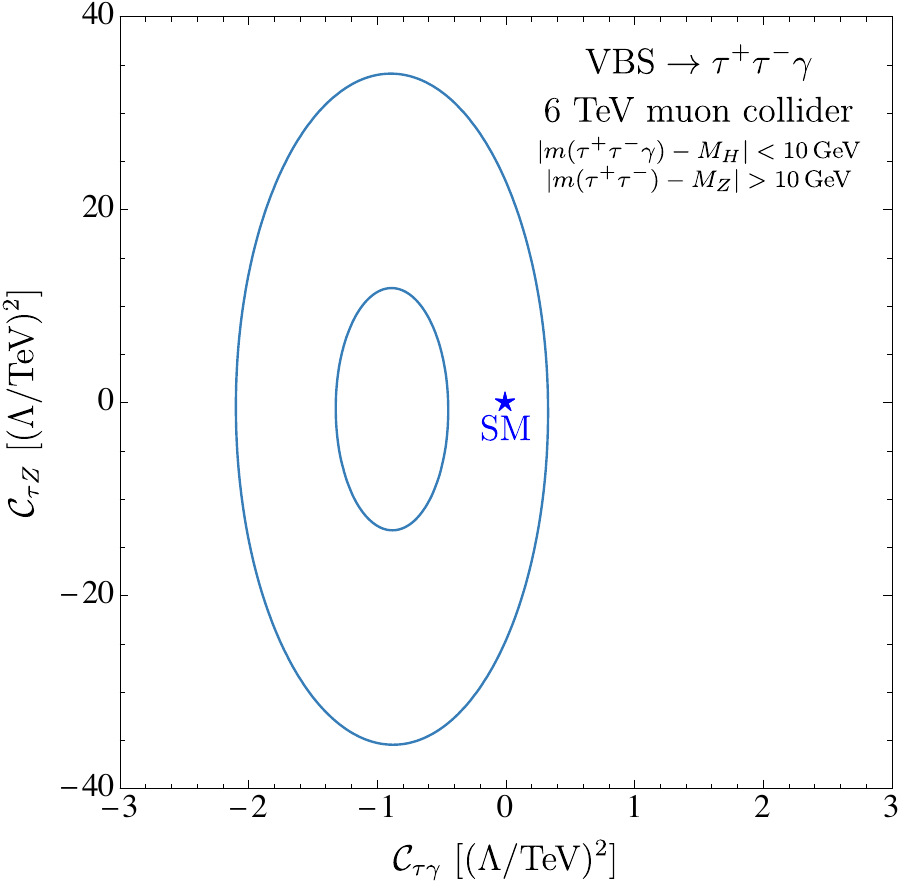}
\includegraphics[width=0.42\textwidth]{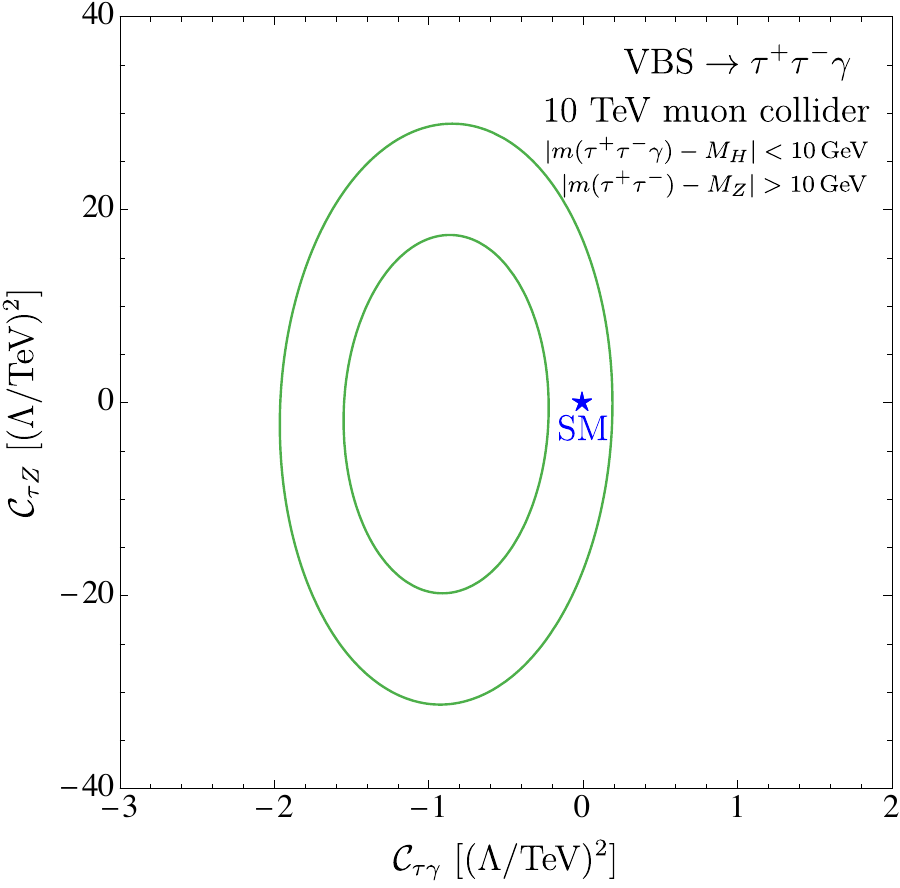}
\includegraphics[width=0.42\textwidth]{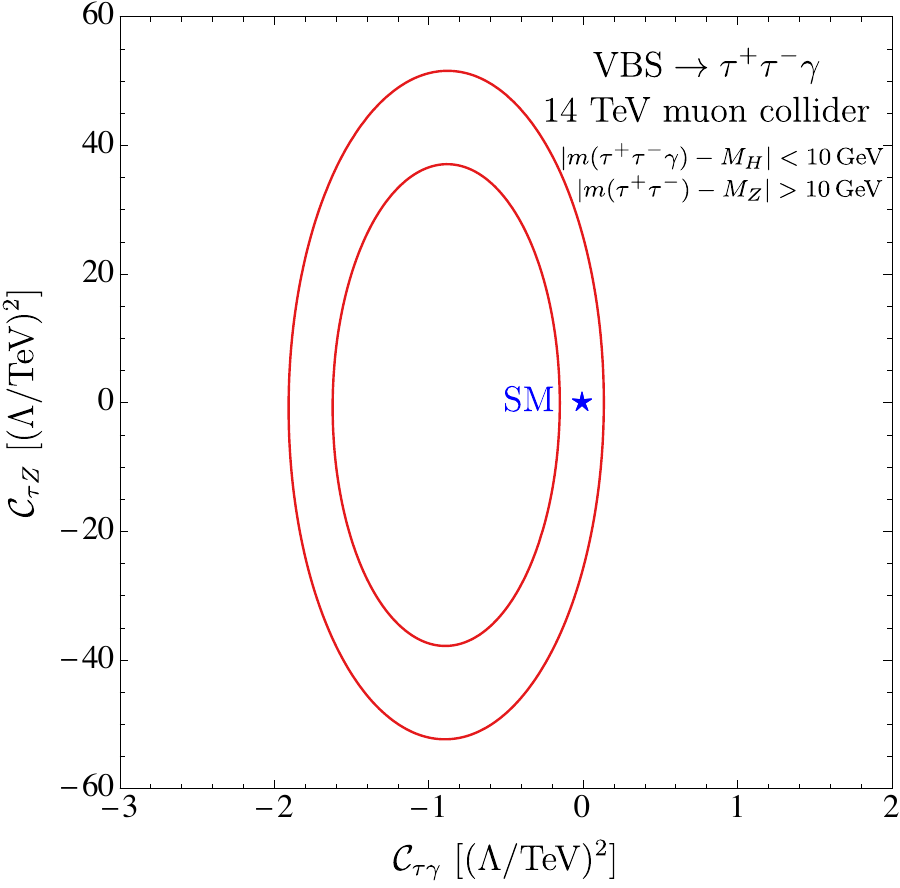}
\caption{Projected 95\% CL sensitivity contours for the same collider energies as Fig.~\ref{fig:contours_VBFh2tta_1}, shown in the $(\mathcal{C}_{\tau \gamma}, \mathcal{C}_{\tau Z})$ plane.} \label{fig:contours_VBFh2tta_2}
\end{figure}

The resulting projected 95\% CL contours are shown in Figs.~\ref{fig:contours_VBFh2tta_1} and \ref{fig:contours_VBFh2tta_2}, for the four collider benchmarks $\sqrt{s}=3$, $6$, $10$, and $14$~TeV. The first figure uses the gauge-invariant basis $(\mathcal C_{\tau W},\mathcal C_{\tau B})$; the second rotates to the physical basis $(\mathcal C_{\tau\gamma},\mathcal C_{\tau Z})$. The comparison makes the much stronger restriction of the electromagnetic combination apparent, while the weak direction remains comparatively poorly determined. The contours are not generally centered on the SM point, reflecting the interference term in the CP-even amplitude.

An unusual feature of the contours at $\sqrt{s}\geq 6$~TeV is the appearance of separated boundary branches, producing a ring-like or ``donut'' structure in parts of the parameter space. This behavior results from the competition between terms linear and quadratic in the Wilson coefficients: different nonzero parameter choices can yield similar shifts in the selected rate. The allowed gap between the boundaries becomes narrower for the higher-energy benchmarks shown. These degeneracies are specific to an event-rate measurement and should not be interpreted as independent solutions for a dipole moment without stating which coefficients are held fixed.

\begin{table}[htb]
\centering
\renewcommand{\arraystretch}{1.18}
\setlength{\tabcolsep}{5pt}
\begin{tabular}{c|cccc}
\hline
$\sqrt{s}$ [TeV]
& 3
& 6
& 10
& 14
\\
\hline
Max$(\Delta a_\tau)$                 & $3.0\times10^{-3}$   & $1.3\times10^{-3}$   & $7.8\times10^{-4}$   & $5.5\times10^{-4}$   \\
Min$(\Delta a_\tau)$                 & $-1.0\times10^{-2}$  & $-1.8\times10^{-3}$  & $-9.1\times10^{-4}$  & $-6.1\times10^{-4}$  \\
Max$(\Delta a^Z_\tau)$               & $9.7\times10^{-2}$   & $9.2\times10^{-2}$   & $6.8\times10^{-2}$   & $1.0\times10^{-1}$   \\
Min$(\Delta a^Z_\tau)$               & $-9.5\times10^{-2}$  & $-9.9\times10^{-2}$  & $-7.0\times10^{-2}$  & $-1.0\times10^{-1}$  \\
$|d_\tau|$ [$e\,{\rm cm}$]           & $3.1\times10^{-17}$  & $1.9\times10^{-17}$  & $1.4\times10^{-17}$  & $1.2\times10^{-17}$  \\
$|d^Z_\tau|$ [$e\,{\rm cm}$]         & $5.4\times10^{-16}$  & $5.4\times10^{-16}$  & $3.9\times10^{-16}$  & $5.9\times10^{-16}$  \\
\hline
\end{tabular}
\caption{
Projected $95\%$ CL constraints on the $\tau$ dipole moments
$\Delta a^{(Z)}_\tau$ and $|d^{(Z)}_\tau|$ from the
$H\to\tau^+\tau^-\gamma$ decay at high-energy muon colliders. Bounds on
the magnetic dipole moments assume real Wilson coefficients, while
those on the electric dipole moments assume purely imaginary
coefficients. For $\Delta a_\tau$ and $|d_\tau|$, the $Z$ dipole
operator is switched off ($\mathcal{C}_{\tau Z}=0$), whereas for
$\Delta a^Z_\tau$ and $|d^Z_\tau|$ the photon dipole coupling is set to
zero ($\mathcal{C}_{\tau\gamma}=0$).}
\label{tab:constraints_HiggsDecay}
\end{table}

For this reason, Table~\ref{tab:constraints_HiggsDecay} quotes one-operator-at-a-time limits. The electromagnetic moments are determined with $\mathcal C_{\tau Z}=0$ and the weak moments with $\mathcal C_{\tau\gamma}=0$; real and purely imaginary coefficients are treated separately for magnetic and electric dipoles. In contrast to channels dominated by the squared dipole amplitude, the interference here yields asymmetric upper and lower bounds for $\Delta a_\tau$. At $10$~TeV the tabulated interval is
\begin{equation}
-9.1\times10^{-4}<\Delta a_\tau<7.8\times10^{-4},
\qquad |d_\tau|<1.4\times10^{-17}\,e\,\mathrm{cm},
\label{eq:muc_h2ttg_10tev}
\end{equation}
with significantly weaker corresponding constraints on the weak dipole moments. Increasing the collider energy improves the electromagnetic reach through the larger vector-boson-fusion Higgs yield and the luminosity benchmarks, although the weak-dipole bounds need not improve monotonically because the selected rate has little sensitivity to that direction.

The radiative Higgs decay channel is less constraining than direct $\tau$-pair production or electroweak-boson-associated production in the multi-TeV scenarios studied here. Its value is instead in the qualitatively different interference pattern and the Higgs dipole contact vertex, which together provide a complementary check of the interpretation of the stronger production-channel limits. The comparison among all channels will be presented in the following subsection.

\clearpage
\subsection{Combined sensitivity and global picture}
\label{sec:global_muc}

\begin{figure}[htb]
\includegraphics[width=0.45\textwidth]{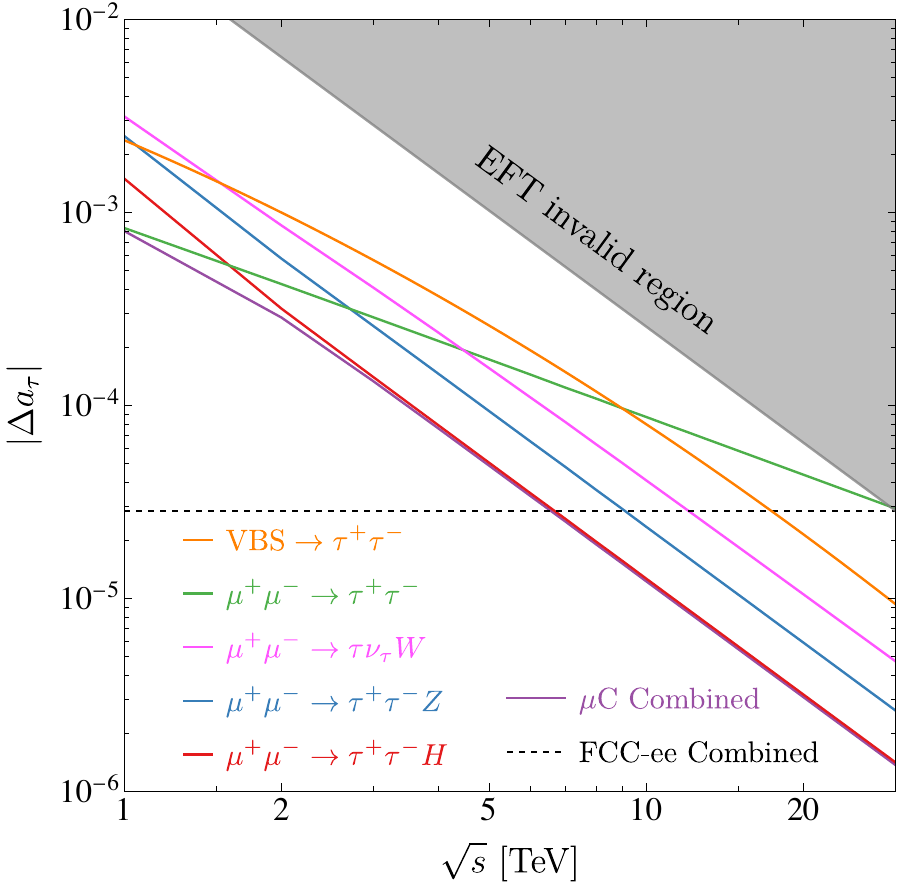}
\includegraphics[width=0.45\textwidth]{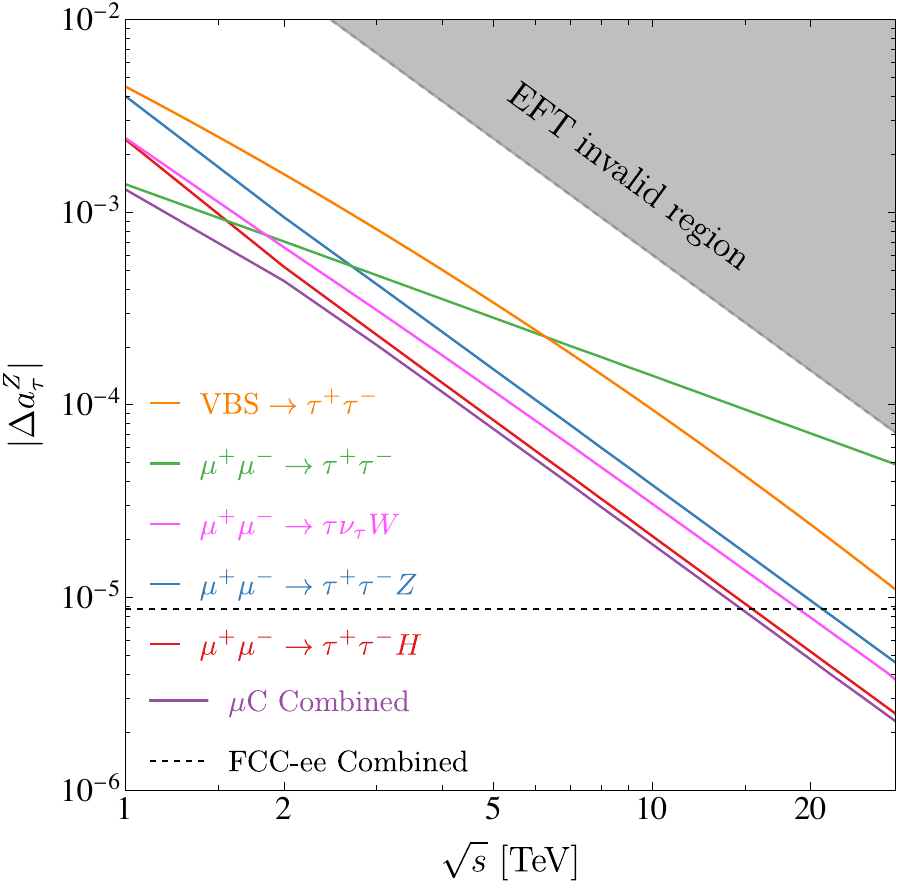}\\
\includegraphics[width=0.45\textwidth]{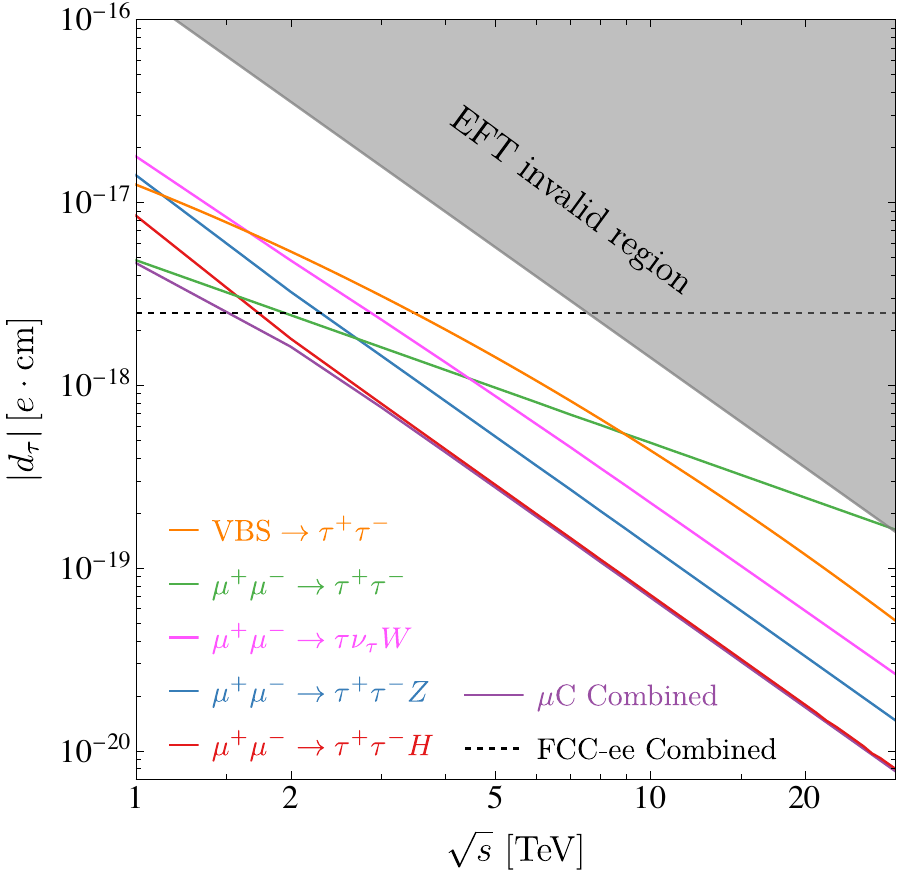}
\includegraphics[width=0.45\textwidth]{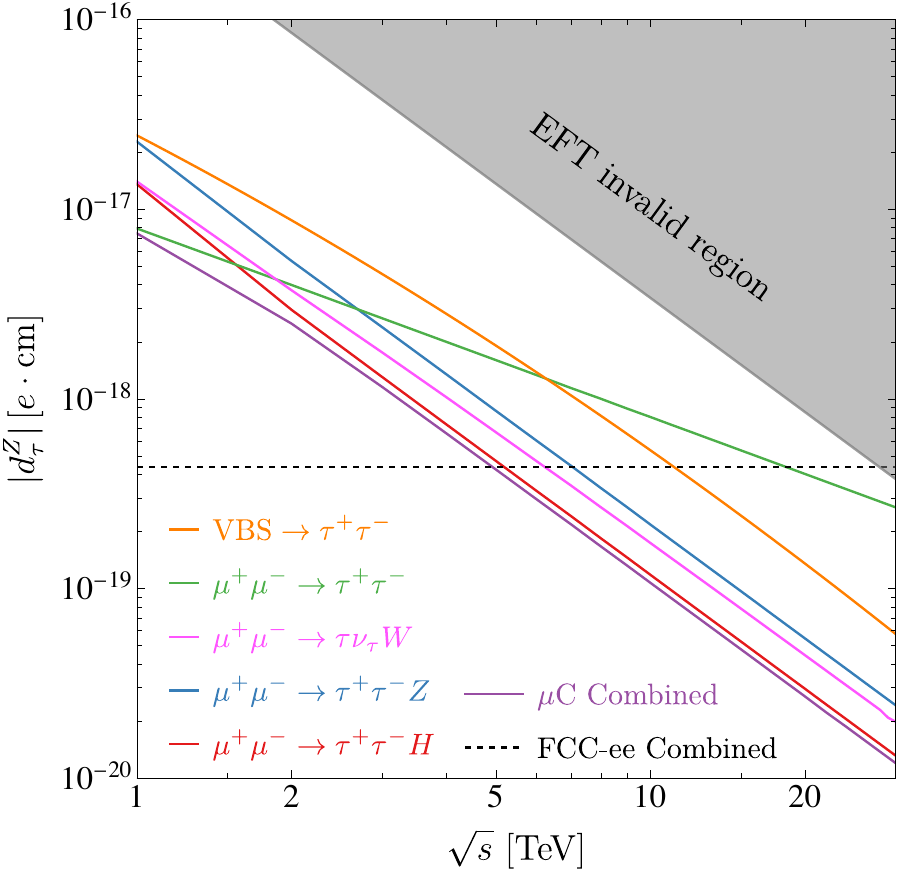}
\caption{Projected 95\% CL bounds on $\Delta a_\tau$ (upper left), $\Delta a^Z_\tau$ (upper right), $d_\tau$ (lower left), and $d^Z_\tau$ (lower right) at a multi-TeV muon collider, derived from different processes as functions of $\sqrt{s}$. The bounds on $\Delta a^{(Z)}_\tau$ are obtained assuming purely real $\mathcal{C}_{\tau V}$, while those for $d^{(Z)}_\tau$ assume purely imaginary $\mathcal{C}_{\tau V}$. The dashed black line represents the combined sensitivity from all stages of the FCC-ee~\cite{Buttazzo:2026ulg} for comparison, and the gray shaded region indicates the parameter space excluded by perturbative unitarity violation.} \label{fig:muC_constraint}
\end{figure}

The results of the preceding subsections show that the relative reach of the different channels is not fixed, but changes substantially with the collision energy. To compare their impact on a common footing, Fig.~\ref{fig:muC_constraint} collects the projected 95\% CL sensitivities to the electromagnetic and weak dipole moments over the range $\sqrt{s}=1$--$30$~TeV. The individual channels are shown alongside their statistical combination. The FCC-ee projections of Ref.~\cite{Buttazzo:2026ulg} provide a reference at lower energies, while the shaded regions indicate the perturbative-unitarity criterion of Eq.~\eqref{eq:perturbative}. The continuous energy curves extend beyond the four benchmark operating points of Sec.~\ref{sec:pheno} and should be read within the luminosity and analysis assumptions specified there.

Two characteristic trends emerge. For direct $\mu^+\mu^-\to\tau^+\tau^-$ annihilation, the high-energy quadratic dipole contribution is a constant, whereas the SM cross section decreases approximately as $1/s$. With the assumed luminosity $\mathcal L\propto s$, this leads to a projected bound on a dipole coefficient scaling approximately as $s^{-1/2}$. By contrast, the contact-interaction contribution to electroweak-boson-associated production grows as $s/\Lambda^4$. Under the corresponding background and efficiency assumptions, the limits on the dipole coefficients then improve approximately as $s^{-1}$. These powers describe the observed high-energy behavior of the projections; they are not universal scaling laws independent of selection or background composition.

In the associated-production class, $\mu^+\mu^-\to\tau^+\tau^-H$ gives the best individual reach. Its advantage stems from the energy-enhanced dipole contact interaction combined with the comparatively small selected SM background. The $\tau^+\tau^-Z$ and $W^\pm\tau^\mp\nu_\tau$ channels exhibit related energy enhancement but weaker individual bounds, while probing complementary combinations of Wilson coefficients. The inclusive sample denoted VBS also becomes increasingly competitive. In particular, its high-energy behavior receives an important contribution from topologies involving associated $Z$ production followed by invisible decay, and is therefore not attributable exclusively to genuine vector-boson scattering. This explains why its projected limits approach the $1/s$ behavior of the associated-production channels and can overtake direct annihilation around $\sqrt{s}\sim10$~TeV.

The channel hierarchy is consequently energy dependent: $\mu^+\mu^-\to\tau^+\tau^-$ provides the leading sensitivity near the lower end of the energy range considered, whereas Higgs-associated 
$\tau$ pair production becomes dominant above roughly $2$~TeV in the adopted setup. At the upper end of the scan, perturbative unitarity must be assessed for each inferred bound. In particular, portions of the $\mu^+\mu^-\to\tau^+\tau^-$ reach enter the shaded EFT-inconsistent region; sensitivity curves entering that region should not be interpreted as reliable EFT exclusions without additional model information.

The combined bounds at the four benchmark energies are given in Table~\ref{tab:constraints_muC}. They are derived by combining the independent channels with the statistical prescription of Sec.~\ref{sec:pheno}, assuming purely real dipole coefficients for the magnetic observables and purely imaginary ones for the electric observables. These are projections within the stated event-selection and efficiency model, rather than detector-level limits. At $\sqrt{s}=10$~TeV, the combination yields
\begin{equation}
\begin{aligned}
 |\Delta a_\tau|&<1.2\times10^{-5}, &\qquad |d_\tau|&<6.6\times10^{-20}\,e\,{\rm cm},
 \\
  |\Delta a_\tau^Z|&<1.8\times10^{-5}, & |d_\tau^Z|&<1.0\times10^{-19}\,e\,{\rm cm}.
\end{aligned}
\label{eq:muc_combined_10tev}
\end{equation}
The steady improvement across the benchmark energies mainly follows the growth of the associated-production sensitivity. A comparison of the combined constraints with FCC-ee, including the correlations in the two Wilson-coefficient bases, is presented next.

\begin{table}[htb]
\centering
\renewcommand{\arraystretch}{1.18}
\setlength{\tabcolsep}{5pt}
\begin{tabular}{c|cccc}
\hline
$\sqrt{s}$ [TeV]
& 3
& 6
& 10
& 14
\\
\hline
$|\Delta a_\tau|$                    & $1.3\times10^{-4}$  & $3.2\times10^{-5}$  & $1.2\times10^{-5}$  & $5.9\times10^{-6}$  \\
$|\Delta a^Z_\tau|$                  & $2.0\times10^{-4}$  & $5.0\times10^{-5}$  & $1.8\times10^{-5}$  & $9.3\times10^{-6}$  \\
$|d_\tau|$ [$e\,{\rm cm}$]           & $7.2\times10^{-19}$ & $1.8\times10^{-19}$ & $6.6\times10^{-20}$ & $3.4\times10^{-20}$ \\
$|d^Z_\tau|$ [$e\,{\rm cm}$]         & $1.1\times10^{-18}$ & $2.8\times10^{-19}$ & $1.0\times10^{-19}$ & $5.3\times10^{-20}$ \\
\hline
\end{tabular}
\caption{Projected 95\% CL  constraints on the electromagnetic and weak dipole moments of the $\tau$ lepton from the combination of all channels considered in this work. The constraints on $\Delta a^{(Z)}_\tau$ are derived assuming a purely real $\mathcal{C}_{\tau V}$, while those on $d^{(Z)}_\tau$ assume a purely imaginary $\mathcal{C}_{\tau V}$.}
\label{tab:constraints_muC}
\end{table}

\begin{figure}[htb]
\includegraphics[width=0.47\textwidth]{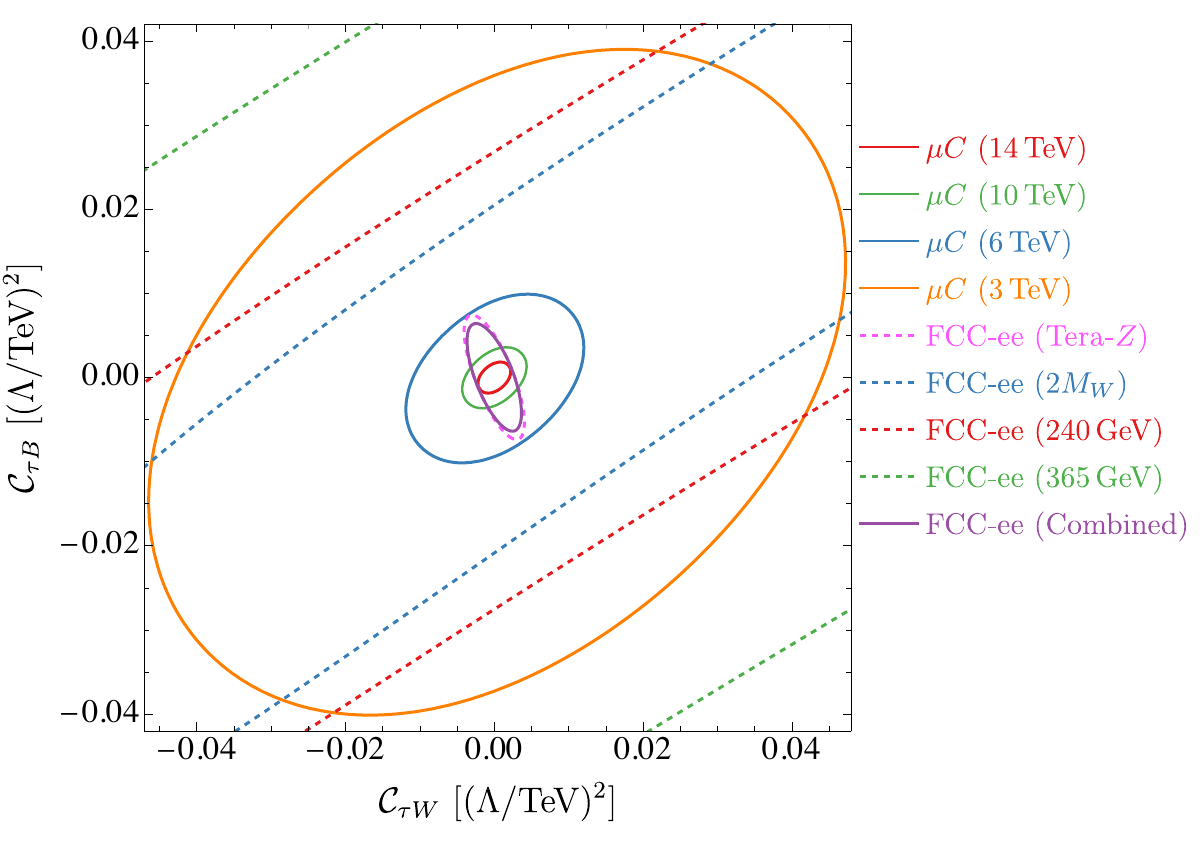}
\includegraphics[width=0.465\textwidth]{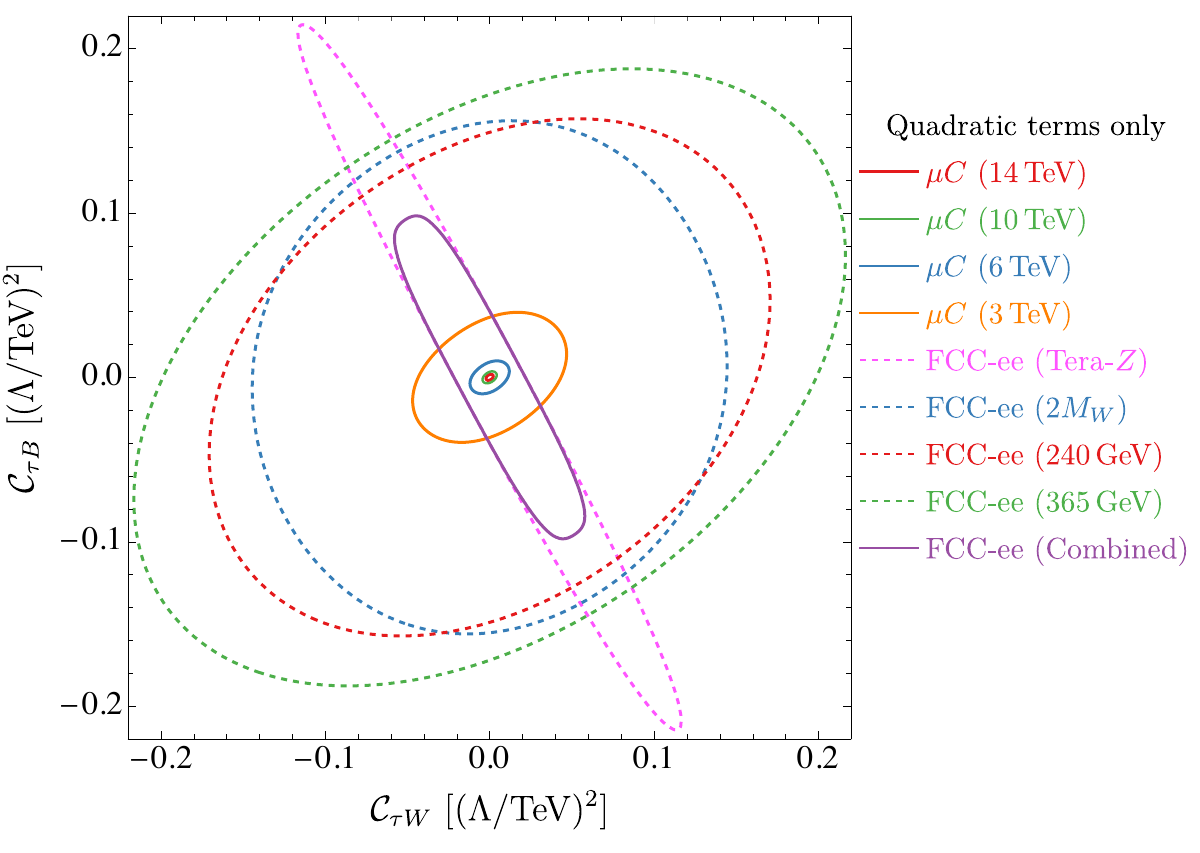}
\includegraphics[width=0.47\textwidth]{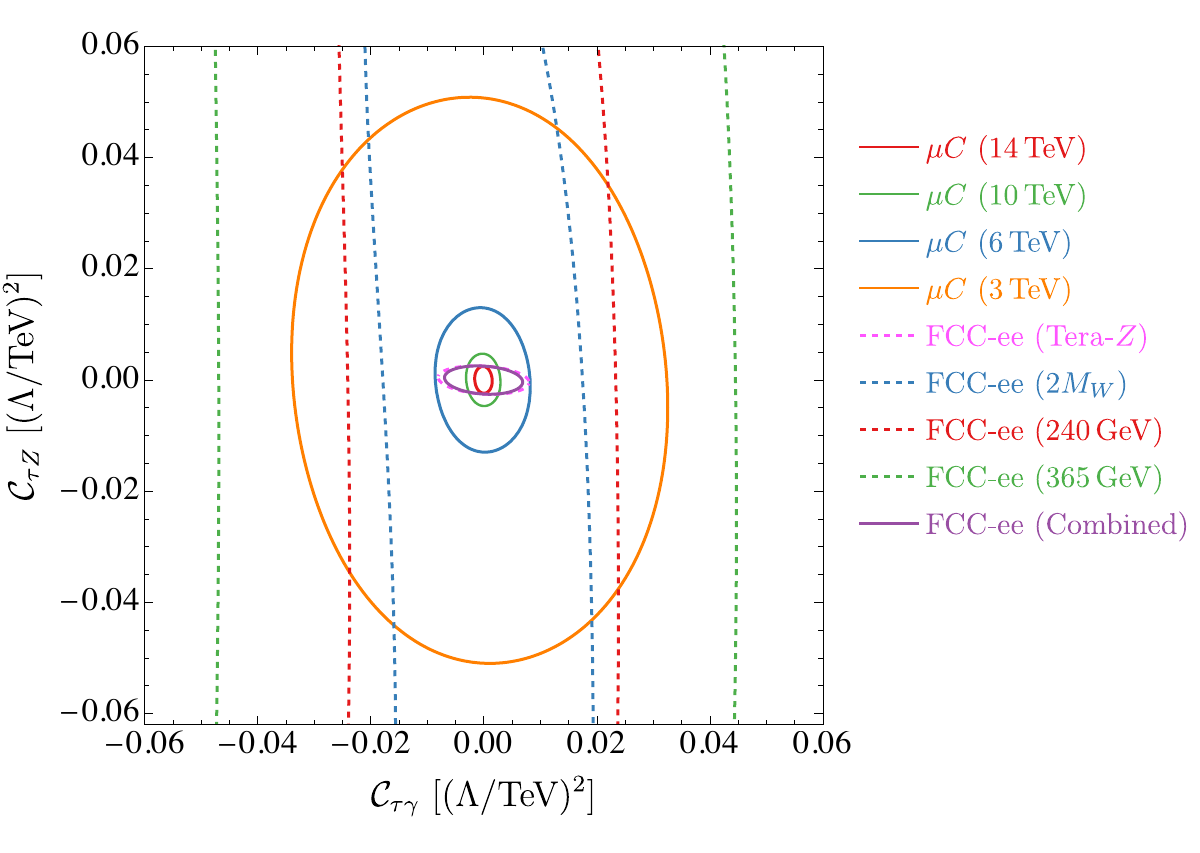}
\includegraphics[width=0.465\textwidth]{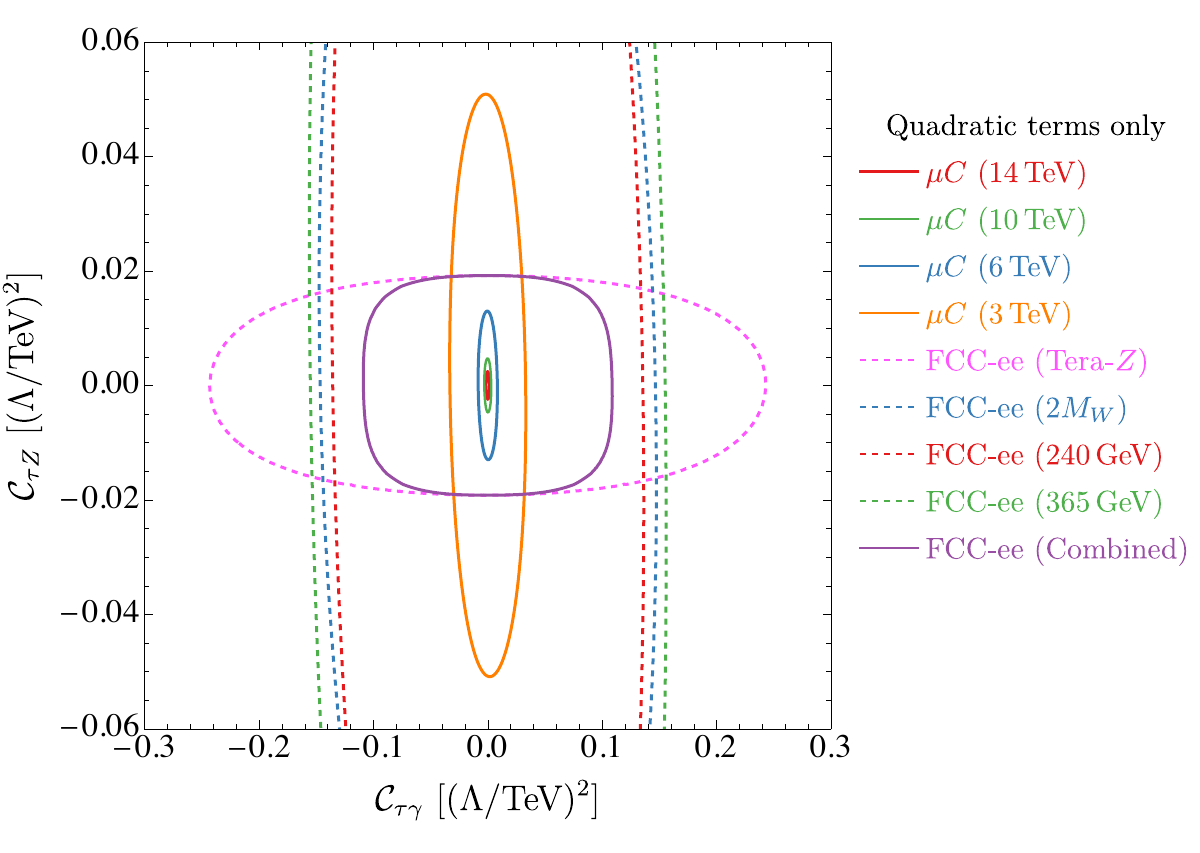}
\caption{Projected 95\% CL contours for the $\tau$ dipole operators at FCC-ee and future muon colliders. The upper row uses the gauge-invariant basis $(\mathcal C_{\tau W},\mathcal C_{\tau B})$, and the lower row the mass-eigenstate basis $(\mathcal C_{\tau\gamma},\mathcal C_{\tau Z})$. The left column includes the SM--dimension-six interference and the quadratic dimension-six terms, relevant for real (CP-even) coefficients; the right column retains only the quadratic terms, as appropriate for the total-rate sensitivity to imaginary (CP-odd) coefficients. Dashed contours show the individual FCC-ee running stages, with the FCC-ee combination in solid purple; the other solid contours correspond to muon collider energies of $3$, $6$, $10$, and $14$~TeV. The statistical and reconstruction assumptions differ between the two collider studies, as detailed in the respective analyses.}
\label{fig:contour_all}
\end{figure}

Fig.~\ref{fig:contour_all} makes the comparison between the two collider programs explicit in both the gauge-invariant and mass-eigenstate operator bases. The distinction between the columns is essential: the left-hand panels retain the interference with the SM, whereas the right-hand panels display the quadratic contribution alone. For inclusive rate measurements, the latter also determines the sensitivity to CP-odd dipole couplings, whose interference with the CP-even SM amplitude vanishes. These plots therefore compare two different assumptions about the dipole coefficients, rather than independently measuring their CP properties.

The FCC-ee and muon-collider contours exhibit different geometries and energy dependences. At FCC-ee, the high-statistics scan around the $Z$ pole constrains combinations involving the weak dipole interaction, while photon-induced $\tau$-pair production probes the electromagnetic operator efficiently, particularly through interference effects in the appropriate kinematic region~\cite{Buttazzo:2026ulg}. Consequently, the FCC-ee contours can change appreciably when the linear contribution is removed. By contrast, the muon-collider contours are much less affected by this change. Their approximate symmetry about the SM point reflects the predominance of quadratic dimension-six contributions in the multi-TeV channels studied here. The orientations of the contours further illustrate why combining the neutral and charged-current channels can help disentangle $\mathcal C_{\tau B}$ from $\mathcal C_{\tau W}$, even when one channel supplies most of the total statistical sensitivity.

A quantitative comparison is provided by the one-parameter 95\% CL projections. The combined FCC-ee study gives $|\Delta a_\tau|<2.8\times10^{-5}$, $|d_\tau|<2.5\times10^{-18}\,e\,{\rm cm}$, $|\Delta a_\tau^Z|<1.0\times10^{-5}$, and $|d_\tau^Z|<4.4\times10^{-19}\,e\,{\rm cm}$~\cite{Buttazzo:2026ulg}. At $\sqrt{s}=10$~TeV, Table~\ref{tab:constraints_muC} gives stronger projected bounds on the electromagnetic magnetic and electric dipole moments and on the weak electric dipole moment, whereas the projected weak magnetic bound remains less restrictive than the FCC-ee result. In particular, the electromagnetic EDM reach improves by nearly a factor of forty in this comparison. At $14$~TeV the projected muon-collider constraints become slightly stronger for the weak magnetic moment as well. These numerical comparisons refer to one-parameter limits and should not be read as a direct measurement of the areas or degeneracy directions of the two-dimensional contours in Fig.~\ref{fig:contour_all}.

The two machines thus gain sensitivity in different ways. FCC-ee exploits high luminosity, resonant $Z$ production, and interference-sensitive observables at comparatively low energies; the muon collider tests the same dipole interactions through energy-enhanced production amplitudes, most notably in $\mu^+\mu^-\to\tau^+\tau^-H$. The latter mechanism is especially advantageous for CP-odd moments when the sensitivity is extracted from CP-even event rates, although dedicated CP-odd observables could further improve the constraints. Finally, the reaches quoted here are nominal statistical projections based on the respective reconstruction and kinematic selection assumptions. The muon-collider values use the counting statistic and reference thresholds in Sec.~\ref{sec:analysis_framework}, without a dedicated coverage study or independently verified event-category correlations; the comparison with FCC-ee is therefore indicative rather than a harmonized likelihood combination. Their interpretation within the SMEFT is conditional on the dipole-only benchmark and additionally requires the perturbative-unitarity and EFT-validity considerations discussed above. A scale-consistent comparison of low-energy moments would also require confirming the treatment of RG evolution in the fitted coefficients. These projections should not be regarded as detector-level forecasts or as universally valid for arbitrary ultraviolet completions.

\clearpage
\section{Summary}
\label{sec:summary}

We have examined the prospects for determining the electromagnetic and weak dipole moments of the $\tau$ lepton at future multi-TeV muon colliders. Within the dimension-six SMEFT, we have studied direct $\mu^+\mu^-\to\tau^+\tau^-$ production, inclusive $\tau$-pair production with additional muons or neutrinos, the electroweak-boson-associated channels $\tau^+\tau^-H$, $\tau^+\tau^-Z$ and $W^\pm\tau^\mp\nu_\tau$, and radiative Higgs decays $H\to\tau^+\tau^-\gamma$. Their distinct production mechanisms provide access to different combinations of the two gauge-invariant dipole operators and allow us to assess the gain in sensitivity as the collision energy increases.

A characteristic feature of the multi-TeV regime is the importance of terms quadratic in the dimension-six dipole coefficients. In direct $\tau$-pair annihilation, helicity suppression limits the interference with the SM, while the quadratic contribution remains approximately constant at high energy and the SM cross section decreases as $1/s$. With the benchmark luminosity assumption $\mathcal L\propto s$, this leads to an approximate $s^{-1/2}$ scaling of the projected limits on the dipole moments. In electroweak-boson-associated production, the Higgs and longitudinal-gauge-boson interactions generated by the dipole operators instead yield energy-growing quadratic contributions, giving an approximately $s^{-1}$ scaling under the same assumptions. These behaviours are subject to the event selections and backgrounds considered in the analysis, rather than being universal properties of muon-collider sensitivities.

Among the processes investigated,
$\mu^+\mu^-\to\tau^+\tau^-H$ provides the strongest
individual sensitivity over much of the multi-TeV
range, owing to its energy-enhanced contact
interaction and relatively small SM background.
The $\tau^+\tau^-Z$ channel probes the same dipole
operators with weaker sensitivity due to its larger
SM background, while charged-current $W\tau\nu$
production offers complementary sensitivity
to the weak dipole moment.
At higher energies, the inclusive vector-boson-scattering
sample becomes competitive with direct
$\mu^+\mu^-\to\tau^+\tau^-$ production.
Radiative Higgs decay provides a distinct probe
through its linear interference sensitivity to the
CP-even electromagnetic dipole interaction,
although its projected bounds remain weaker
than those of the leading production channels.
Combining the channels as discussed in
Sec.~\ref{sec:analysis_framework}, we obtain the following one-parameter projections for a $10$~TeV muon collider with $10\,\mathrm{ab}^{-1}$:
\begin{equation}
\begin{aligned}
|\Delta a_\tau| &< 1.2\times10^{-5}, &
|d_\tau| &< 6.6\times10^{-20}\,e\,\mathrm{cm},\\
|\Delta a_\tau^Z| &< 1.8\times10^{-5}, &
|d_\tau^Z| &< 1.0\times10^{-19}\,e\,\mathrm{cm}.
\end{aligned}
\end{equation}
Within this statistical benchmark and the dipole-only assumption, these projected sensitivities improve substantially on existing experimental constraints. Compared with the projected combined FCC-ee reach~\cite{Buttazzo:2026ulg}, the $10$~TeV muon collider yields stronger one-parameter bounds on $\Delta a_\tau$ and $d_\tau$, with an improvement of nearly a factor of forty for $d_\tau$. The FCC-ee projection remains more restrictive for $\Delta a_\tau^Z$ at this energy, illustrating the complementarity of high-statistics $Z$-pole measurements and energy-enhanced multi-TeV production. The corresponding two-coefficient contours provide additional information about correlations that is not captured by these one-dimensional limits.

Our results provide estimates of statistical sensitivity within the specified SMEFT and collider benchmarks. Establishing the experimental reach will require realistic simulations of $\tau$ decays and reconstruction, beam-induced and reducible backgrounds, systematic uncertainties, and higher-order
electroweak effects, as well as further checks of EFT validity and perturbative unitarity. Our findings nevertheless highlight the potential
of multi-TeV muon colliders to complement the FCC-ee precision program by probing $\tau$ dipole interactions through energy-enhanced processes at substantially higher collision energies.

\begin{acknowledgments}
We thank B. Batell, I. Brivio, C. Degrande, G. Durieux, T. Han, M. Hoferichter, F. Jaffredo, J. Liu, O. Mattelaer, M. Passera, and X. Wang for useful discussions. 
This research is partially supported by the IISN convention 4.4517.08, ``Theory of fundamental interactions.'' Y. Ma acknowledges the support as a Postdoctoral Fellow of the Fond de la Recherche Scientifique de Belgique (F.R.S.-FNRS), Belgium.
Computational resources have been provided by the supercomputing facilities of the Universit\'e catholique de Louvain (CISM/UCL) and the Consortium des \'Equipements de Calcul Intensif en 
F\'ed\'eration Wallonie Bruxelles (C\'ECI) funded by the Fond de la Recherche Scientifique de Belgique (F.R.S.-FNRS) under convention 2.5020.11 and by the Walloon Region.
G. Levati gratefully acknowledges financial support by the Swiss National Science Foundation (Project No.\ 
TMCG-2\_213690). Z.Q. Wang acknowledges support from the China Scholarship Council (CSC)- UCLouvain Co‑funding PhD Fellowship (Grant No.\ 202206040017).
P. Paradisi received funding by the European Union’s Horizon 2020 research and innovation program under the Marie Sklodowska-Curie grant agreements n.~860881~-~HIDDeN, n.~101086085~-~ASYMMETRY, by the Italian MUR Departments of Excellence grant 2023-2027 ``Quantum Frontiers'' and by the European Union - Next Generation EU and by the Italian Ministry of University and Research (MUR) via the PRIN 2022 project n. 2022K4B58X - AxionOrigins. D. Buttazzo is partially funded by the European Union - Next Generation EU and MUR through the grant PRIN 202289JEW4. We also acknowledge support from the COMETA COST Action CA22130.
\end{acknowledgments}

\appendix

\section{Renormalization-group evolution of the tau dipole coefficients}
\label{app:RGE}
The dipole moments in Eq.~\eqref{eq:adtau_final} are defined at low scales, whereas multi-TeV scattering probes the associated operators at much higher energies. Here we estimate the renormalization-group (RG) evolution between a reference high scale $\Lambda$, the electroweak scale, and $m_\tau$. This exercise assumes that only the two dipole operators in Eq.~\eqref{eq:SMEFT} have nonzero Wilson coefficients at $\Lambda$. In particular, contributions induced through mixing from other dimension-six operators are not included. The semileptonic tensor operator $\mathcal O^{(3)}_{lequ}$, for example, may generate additional terms enhanced by the top Yukawa coupling~\cite{Jenkins:2013wua}.

For the evolution, it is convenient to remove the explicit gauge couplings from the dipole-moment matching. Following the normalization of Ref.~\cite{Alonso:2013hga}, we introduce
\begin{equation}
\label{eq:C_hat_definition}
\hat{\mathcal C}_{\tau\gamma} \equiv \frac{\mathcal C_{\tau B}}{g_1}-\frac{\mathcal C_{\tau W}}{g_2}=\frac{\mathcal C_{\tau\gamma}}{e},
\qquad
\hat{\mathcal C}_{\tau Z} \equiv -\frac{\mathcal C_{\tau B}}{g_2}-\frac{\mathcal C_{\tau W}}{g_1}=\frac{\mathcal C_{\tau Z}}{e},
\end{equation}
where $g_1=e/c_W$ and $g_2=e/s_W$ are the hypercharge and weak-isospin gauge couplings, respectively. The relations to the physical moments can then be expressed as
\begin{equation}
\label{eq:adtau_hat}
\Delta a_\tau^{(Z)}=\frac{2\sqrt{2}\,m_\tau v}{\Lambda^2}\,\mathrm{Re}\,\hat{\mathcal C}_{\tau\gamma(Z)},
\qquad
\frac{d_\tau^{(Z)}}{e}=\frac{\sqrt{2}\,v}{\Lambda^2}\,\mathrm{Im}\,\hat{\mathcal C}_{\tau\gamma(Z)}.
\end{equation}
The photon coefficient is evaluated at $m_\tau$ and the $Z$ coefficient at $M_Z$. This rescaling is also useful because the scale dependence of the gauge couplings is included in the definition of the hatted coefficients.

Above the electroweak threshold, the one-loop anomalous dimensions of the dipoles in the gauge basis~\cite{Jenkins:2013wua,Alonso:2013hga}, together with the scale-dependent transformation in Eq.~\eqref{eq:C_hat_definition}, yield
\begin{equation}
\label{eq:RGE_matrix}
16\pi^2\frac{d}{d\ln\mu}
\begin{pmatrix}
\hat{\mathcal C}_{\tau\gamma}\\
\hat{\mathcal C}_{\tau Z}
\end{pmatrix}
=\left[3y_t^2\,\mathbf{1}+
\begin{pmatrix}
\gamma_{\gamma\gamma}&\gamma_{\gamma Z}\\
\gamma_{Z\gamma}&\gamma_{ZZ}
\end{pmatrix}\right]
\begin{pmatrix}
\hat{\mathcal C}_{\tau\gamma}\\
\hat{\mathcal C}_{\tau Z}
\end{pmatrix},
\end{equation}
with
\begin{equation}
\label{eq:gamma_matrix}
\begin{aligned}
\gamma_{\gamma\gamma}&=e^2\left(12-\frac94\csc^2\theta_W+\frac14\sec^2\theta_W\right),\\
\gamma_{\gamma Z}&=12e^2\cot 2\theta_W,\qquad
\gamma_{Z\gamma}=-\frac23e^2\left(2\cot\theta_W+31\tan\theta_W\right),\\
\gamma_{ZZ}&=\frac{e^2}{12}\left(151\tan^2\theta_W-92-11\cot^2\theta_W\right).
\end{aligned}
\end{equation}
The derivative of the weak mixing angle, implicit in the transformation to the $(\gamma,Z)$ basis, must be included alongside the gauge-basis evolution. Consequently, the two off-diagonal entries in Eq.~\eqref{eq:gamma_matrix} are not equal.

At leading-logarithmic order, and treating the couplings as fixed over the interval, the coefficients at $M_Z$ are
\begin{align}
\label{eq:Cgamma_MZ}
\hat{\mathcal C}_{\tau\gamma}(M_Z)\simeq{}&
\hat{\mathcal C}_{\tau\gamma}(\Lambda)-\frac{\ln(\Lambda/M_Z)}{16\pi^2}
\left[\big(3y_t^2+\gamma_{\gamma\gamma}\big)\hat{\mathcal C}_{\tau\gamma}(\Lambda)+\gamma_{\gamma Z}\hat{\mathcal C}_{\tau Z}(\Lambda)\right],\\
\label{eq:CZ_MZ}
\hat{\mathcal C}_{\tau Z}(M_Z)\simeq{}&\hat{\mathcal C}_{\tau Z}(\Lambda)-\frac{\ln(\Lambda/M_Z)}{16\pi^2}
\left[\big(3y_t^2+\gamma_{ZZ}\big)\hat{\mathcal C}_{\tau Z}(\Lambda)+\gamma_{Z\gamma}\hat{\mathcal C}_{\tau\gamma}(\Lambda)\right].
\end{align}

Below the electroweak threshold, the heavy electroweak states and the top quark are integrated out. In the restricted dipole setup considered here, the remaining electromagnetic coefficient evolves according to the one-loop QED equation
\begin{equation}
\label{eq:LEFT_dipole_RGE}
16\pi^2\frac{d\hat{\mathcal C}_{\tau\gamma}}{d\ln\mu}=10e^2\hat{\mathcal C}_{\tau\gamma},
\end{equation}
where the normalization in Eq.~\eqref{eq:C_hat_definition} includes the contribution from the running of the electromagnetic coupling. Its leading-logarithmic solution is
\begin{equation}
\label{eq:Cgamma_mtau}
\hat{\mathcal C}_{\tau\gamma}(m_\tau)\simeq
\hat{\mathcal C}_{\tau\gamma}(M_Z)-
\frac{10e^2}{16\pi^2}\ln\frac{M_Z}{m_\tau}\,
\hat{\mathcal C}_{\tau\gamma}(M_Z).
\end{equation}

For illustration, we consider a muon collider operating at
$\sqrt{s}=10$~TeV and identify the reference high scale
with the collider energy, $\Lambda=10$~TeV.
Using the input parameters $s_W^2\simeq 0.23$, $\alpha \simeq\frac{1}{128}$, and $3y_t^2\simeq 2.9$, the
leading-logarithmic evolution in Eqs.~(\ref{eq:Cgamma_MZ}) and (\ref{eq:Cgamma_mtau}) gives
\begin{equation}
\begin{aligned}
\hat{\mathcal C}_{\tau\gamma}(m_\tau)
\simeq {}&
0.88\,\hat{\mathcal C}_{\tau\gamma}(\Lambda)
-0.02\,\hat{\mathcal C}_{\tau Z}(\Lambda).
\end{aligned}
\end{equation}
Thus, neglecting the weak dipole coefficient at the
high scale, the electromagnetic dipole coefficient
decreases by about $12\%$ when evolved down to
$m_\tau$.

By contrast, for a dipole coefficient specified at the electroweak scale, as relevant to the FCC-ee $Z$-pole program, only the QED evolution below $M_Z$ needs to be considered. Equation~(\ref{eq:Cgamma_mtau}) yields
\begin{equation}
\hat{\mathcal C}_{\tau\gamma}(m_\tau)
\simeq
0.98\,\hat{\mathcal C}_{\tau\gamma}(M_Z),
\end{equation}
corresponding to a correction of about $2\%$.
These estimates illustrate that RG effects become
more relevant over the larger scale separation
probed by multi-TeV muon colliders.
In particular, for FCC-ee, the resulting RG corrections 
are expected to have a limited impact on the projected
sensitivities~\cite{Buttazzo:2026ulg}, compared with the 
anticipated experimental uncertainties.

\bibliographystyle{utphys}
\bibliography{ref}

\end{document}